\documentclass[aps,prx,twocolumn,superscriptaddress]{revtex4-2}
\usepackage[english]{babel}
\usepackage{amssymb}
\usepackage{amsmath}
\usepackage{mathtools}
\usepackage{txfonts}
\usepackage{mathdots}
\usepackage[normalem]{ulem}
\usepackage{array}
\usepackage{amsfonts}
\usepackage{mathrsfs}
\usepackage{booktabs}
\usepackage{multirow}
\usepackage[dvips]{graphicx}
\usepackage{epsfig}
\usepackage{subfigure}
\usepackage{threeparttable}
\usepackage{chngpage}
\usepackage{float}
\usepackage[dvipsnames,svgnames,x11names,hyperref]{xcolor}
\usepackage{bm}
\usepackage[toc]{appendix}
\usepackage{tabularx}
\usepackage{adjustbox}
\usepackage{comment}
\usepackage{hyperref}
\usepackage{todonotes}
\usepackage{lineno}

\hypersetup{
	unicode=false,     
	pdftoolbar=false,  
	pdfmenubar=true,   
	pdffitwindow=false, 
	pdfstartview={FitH},
	pdftitle={},    
	pdfauthor={Authors},     
	pdfsubject={},   
	pdfcreator={},   
	pdfproducer={}, 
	pdfnewwindow=true,
	colorlinks=true,
	linkcolor=RoyalBlue,
	citecolor=blue, 
	filecolor=magenta,
	urlcolor=blue
}

\newcommand{\bfA}{{\bf a}}
\newcommand{\bfB}{{\bf b}}
\newcommand{\bfC}{{\bf c}}
\newcommand{\bfD}{{\bf d}}
\newcommand{\bfE}{{\bf e}}
\newcommand{\bfF}{{\bf f}}

\newcommand{\ssz}{L}

\begin{document}	
\title{Fock space prethermalization and time-crystalline order on a quantum processor}
\author{Zehang Bao}
\author{Zitian Zhu}
\affiliation{School of Physics, ZJU-Hangzhou Global Scientific and Technological Innovation Center, and Zhejiang Key Laboratory of Micro-nano Quantum Chips and Quantum Control, Zhejiang University, Hangzhou, China}
\author{Yang-Ren Liu}
\affiliation{Kavli Institute for Theoretical Sciences, University of Chinese Academy of Sciences, Beijing 100190, China}
\author{Zixuan Song}
\author{Feitong Jin}
\author{Xuhao Zhu}
\author{Yu Gao}
\author{Chuanyu Zhang}
\author{Ning Wang}
\author{Yiren Zou}
\author{Ziqi Tan}
\author{Aosai Zhang}
\author{Zhengyi Cui}
\author{Fanhao Shen}
\author{Jiarun Zhong}
\author{Yiyang He}
\author{Han Wang}
\author{Jia-Nan Yang}
\author{Yanzhe Wang}
\author{Jiayuan Shen}
\author{Gongyu Liu}
\author{Yihang Han}
\author{Yaozu Wu}
\author{Jinfeng Deng}
\author{Hang Dong}
\author{Pengfei Zhang}
\affiliation{School of Physics, ZJU-Hangzhou Global Scientific and Technological Innovation Center, and Zhejiang Key Laboratory of Micro-nano Quantum Chips and Quantum Control, Zhejiang University, Hangzhou, China}
\author{Hekang Li}
\affiliation{School of Physics, ZJU-Hangzhou Global Scientific and Technological Innovation Center, and Zhejiang Key Laboratory of Micro-nano Quantum Chips and Quantum Control, Zhejiang University, Hangzhou, China}
\author{Zhen Wang}
\affiliation{School of Physics, ZJU-Hangzhou Global Scientific and Technological Innovation Center, and Zhejiang Key Laboratory of Micro-nano Quantum Chips and Quantum Control, Zhejiang University, Hangzhou, China}
\author{Chao Song}
\affiliation{School of Physics, ZJU-Hangzhou Global Scientific and Technological Innovation Center, and Zhejiang Key Laboratory of Micro-nano Quantum Chips and Quantum Control, Zhejiang University, Hangzhou, China}

\author{Chen Cheng}
\affiliation{Lanzhou Center for Theoretical Physics, Key Laboratory of Theoretical Physics of Gansu Province, Key Laboratory of Quantum Theory and Applications of MoE, Gansu Provincial Research Center for Basic Disciplines of Quantum Physics, Lanzhou University, Lanzhou 730000, China}

\author{Rubem Mondaini}
\affiliation{Department of Physics,
University of Houston, Houston, Texas 77004}
\affiliation{Texas Center for Superconductivity, University of Houston, Houston, Texas 77004, USA}

\author{Qiujiang Guo}
\email{qguo@zju.edu.cn}
\affiliation{School of Physics, ZJU-Hangzhou Global Scientific and Technological Innovation Center, and Zhejiang Key Laboratory of Micro-nano Quantum Chips and Quantum Control, Zhejiang University, Hangzhou, China}

\author{Biao Huang}
\email{phys.huang.biao@gmail.com}
\affiliation{Kavli Institute for Theoretical Sciences, University of Chinese Academy of Sciences, Beijing 100190, China}

\author{H. Wang}
\affiliation{School of Physics, ZJU-Hangzhou Global Scientific and Technological Innovation Center, and Zhejiang Key Laboratory of Micro-nano Quantum Chips and Quantum Control, Zhejiang University, Hangzhou, China}
\affiliation{State Key Laboratory for Extreme Photonics and Instrumentation, Zhejiang University, Hangzhou, China}
\begin{abstract}
    {
    Periodically driven quantum many-body systems exhibit a wide variety of exotic nonequilibrium phenomena and provide a promising pathway for quantum applications.     
    A fundamental challenge for stabilizing and harnessing these highly entangled states of matter is system heating by energy absorption from the drive.     
    Here, we propose and demonstrate a disorder-free mechanism, dubbed Fock space prethermalization (FSP), to suppress heating. This mechanism divides the Fock-space network into linearly many sparse sub-networks, thereby prolonging the thermalization timescale even for initial states at high energy densities.
    Using ${72}$ superconducting qubits, we  
    observe an FSP-based time-crystalline order that persists over $120$ cycles for generic initial Fock states. 
    The underlying kinetic constraint of approximately conserved domain wall (DW) numbers is identified by measuring site-resolved correlators. Further, we perform finite-size scaling analysis for DW and Fock-space dynamics by varying system sizes, which reveals size-independent regimes for FSP-thermalization crossover and links the dynamical behaviors to the eigenstructure of the Floquet unitary.   
    Our work establishes FSP as a robust mechanism for breaking ergodicity, and paves the way for exploring novel nonequilibrium quantum matter and its applications.
    }
\end{abstract}
\maketitle

Understanding the thermalization of interacting quantum systems and its breakdown plays an essential role in bridging reversible quantum mechanics with irreversible thermodynamics \cite{Srednicki1994PRE, Nandkishore2015, Luca2016}. 
Intriguingly, the synergy of ergodicity breaking, many-body interactions, and Floquet driving gives rise to unexpected nonequilibrium phenomena -- exemplified by time-crystalline orders that are forbidden in equilibrium~\cite{Wilczek2012,Wilczek2013,Bruno2012,Watanabe2015,Khemani2016,Else2016,Yao2017,Zhang2017,Choi2017,Rovny2018,Pal2018,Mi2022,Randall2021,Frey2022,Kyprianidis2021,Bao2024NC,Zaletel2023RMP,Khemani2019b,Else2020,Sacha2017}. 
These exotic many-body phenomena also serve as versatile platforms for protecting quantum information \cite{Bao2024NC} and enhancing quantum metrology \cite{Choi2017,Ding2022NP,Iemini2024PRA}. Nevertheless, a fundamental challenge arises, as the interplay between interactions and driving typically heats quantum systems to infinite temperature, erasing all dynamical signatures~\cite{Deutsch2018}. 
Thus, looking for new pathways to suppress heating is foundational to exploring novel quantum matter and developing compatible applications in quantum computation and quantum sensing.

To date, several mechanisms have been identified to suppress runaway heating. 
Two prominent examples are many-body localization (MBL) induced by quenched disorders~\cite{Abanin2019,Sierant2024}, and Floquet prethermalization~\cite{Ho2023,Abanin2015,Abanin2017} that relies on high-frequency periodic driving. 
While detrimental quantum avalanches can devastate MBL systems, such a destabilization process takes a predictably long time until very different Fock configurations are hybridized, which underpins the concept of prethermal MBL~\cite{Morningstar2022a,Long2023}.
In addition, nonergodic behavior can emerge in specific models with kinetic constraints, where the Hilbert space is approximately decoupled into various sectors, with the non-thermal ones leading to many-body scars or Hilbert space fragmentation behaviors~\cite{Chandran2022,Moudgalya2022}. 

\begin{figure*}[t]
    \includegraphics[width=17.4cm]{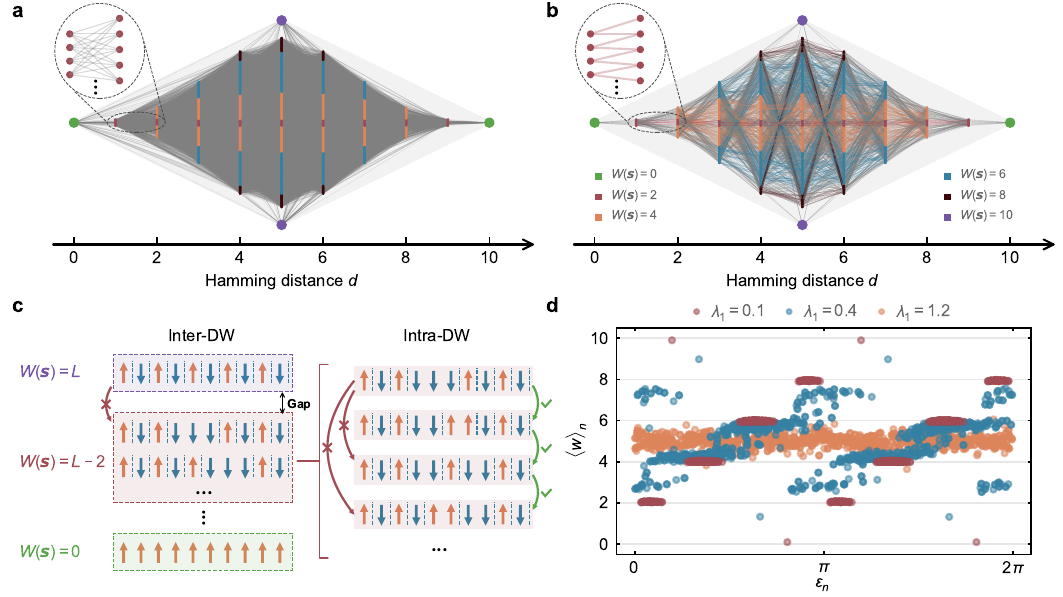}
    \caption{\label{fig1} 
    {\bf Schematic illustration of Fock space prethermalization (FSP).}
    {\bf a-b,} Fock space network in the thermal ({\bfA}) and FSP ({\bfB}) regimes. The thickness of lines denotes the relative hopping strength $|\mathcal{T}_{\boldsymbol{s}_1, \boldsymbol{s}_2}|$ of different bonds in each case, which is obtained numerically for the Floquet unitary without the perfect $\pi$-pulse, i.e., $ e^{-{\rm i} J \sum_j \tilde{\sigma}^z_j(\lambda_2)\tilde{\sigma}^z_{j+1}(\lambda_2)}U_{\rm p}(\varphi_1,\lambda_1,\varphi_2)$. 
    Bonds with strength weaker than $4\%$ of the maximum values are not plotted. Fock-space sites $\boldsymbol{s}$ are colored according to their total domain wall numbers $W(\boldsymbol{s})$. In ({\bfA}), with $\lambda_1=1.2$, qubits can flip freely, rendering a densely connected Fock-space network. In ({\bfB}), with $\lambda_1=0.1$, dominant Ising interactions enforce the approximate domain wall conservation and result in a sparse network, with four scarred states (marked by big green and purple dots) nearly disconnected.
    {\bf c,} Kinetic constraints in FSP regime enforced by strong Ising interactions ($J\gg\lambda_1$). At leading order, hopping among inter-DW sectors is suppressed by total DW conservation, while intra-DW hopping is only allowed if it locally conserves DW numbers, i.e., a spin can only flip if its neighbors are anti-parallel. A DW is denoted by a blue dashed line.
    {\bf d,} Illustrative numerical results of the quasienergy $\varepsilon_n$ and the averaged DW number $\langle w \rangle_n$ of each eigenstate $|\varepsilon_n\rangle$. We show $\lambda_1=0.1,0,4,1.2$ representing FSP, critical, and thermal regimes, respectively. Simulations for larger systems are presented in Supplementary Information section 2B.
    In all cases, we set $J=1$ for a 10-qubit ring, and fix $\lambda_1/\lambda_2=2$.
    }
\end{figure*}

In this work, we propose and experimentally demonstrate an unconventional mechanism for robustly suppressing heating, dubbed Fock space prethermalization (FSP), and observe the resulting time-crystalline behaviors in Fock space~\cite{Yao2023,DeTomasi2021,Mace2019,Ippoliti2021} using $L=72$ superconducting qubits. The concept of FSP can be intuitively illustrated in Fig.~\ref{fig1}{\bfA}. In Fock space, the time evolution unitary $U$ of an $L$-qubit many-body system can be expressed as a complex network spanned by Fock bases $|\boldsymbol{s}\rangle = |s^1\rangle \otimes \dots \otimes |s^L\rangle$, where each Fock-space site is denoted by a photon excitation configuration $\boldsymbol{s}$, with $s^j = 0$ or $1$ for qubit $j$ being in the $|0\rangle$ or $|1\rangle$ state. The connectivity between two sites $\boldsymbol{s}_1$ and $\boldsymbol{s}_2$ is defined by the corresponding hopping strength $\mathcal{T}_{\boldsymbol{s}_1, \boldsymbol{s}_2} = \langle \boldsymbol{s}_1 | U | \boldsymbol{s}_2 \rangle$ and the Hamming distance $D(\boldsymbol{s}_1, \boldsymbol{s}_2) = \sum_{j=1}^L |s_1^j - s_2^j|$ between them. In general scenarios (Fig.~\ref{fig1}\bfA), the dynamics of quantum states obeys the eigenstate thermalization hypothesis~\cite{Deutsch1991PRA, Rigol2008Nature} and rapidly explores the entire Fock space via extensive hoppings $\{\mathcal{T}_{\boldsymbol{s}_1, \boldsymbol{s}_2}\}$. However, for many-body systems with large Ising interaction energy gaps that are immune to generic perturbations, the dense Fock-space network is approximately decoupled into a series of sparse, isolated sub-networks (Fig.~\ref{fig1}\bfB), leading to FSP that significantly prolongs the time required to reach ergodicity, even for initial states at high energy densities.

While reminiscent of Hilbert space fragmentation~\cite{Sala2020PRX, Khemani2020PRB}, which hosts exponentially many disconnected subspaces, FSP only consists of a linear number of sub-networks. In particular, its Fock-space network is highly organized and can withstand generic perturbations. FSP also sharply differs from other well-established prethermal phenomena. Its interaction energy scale is comparable to the driving frequency, distinguishing itself from Floquet prethermalization~\cite{Else2017,Machado2020,Kyprianidis2021}; FSP can sustain spatial inhomogeneity in observables up to prethermal time, in contrast to the prethermalization induced by approximate $U(1)$ symmetry~\cite{Ho2017,Luitz2020,Ho2020,Beatrez2023,Stasiuk2023,He2025} accommodating homogeneous patterns only~\cite{Luitz2020, Ippoliti2021}.
Furthermore, as a disorder-free mechanism, FSP is inherently distinct from stable or prethermal MBL~\cite{Randall2021,Mi2022,Frey2022}. Finally, unlike the weak ergodicity breaking observed in quantum many-body scars, e.g., in PXP models, FSP structures the entire Hilbert space rather than just a few scarred eigenstates~\cite{Turner2018,Bluvstein2021,Maskara2021,Huang2022,Huang2023,Bao2024NC}. Indeed, scarred subspaces can coexist as a marginal part of the structured Fock-space network in FSP regimes. 

\begin{figure*}[t]
    \includegraphics[width=17.4cm]{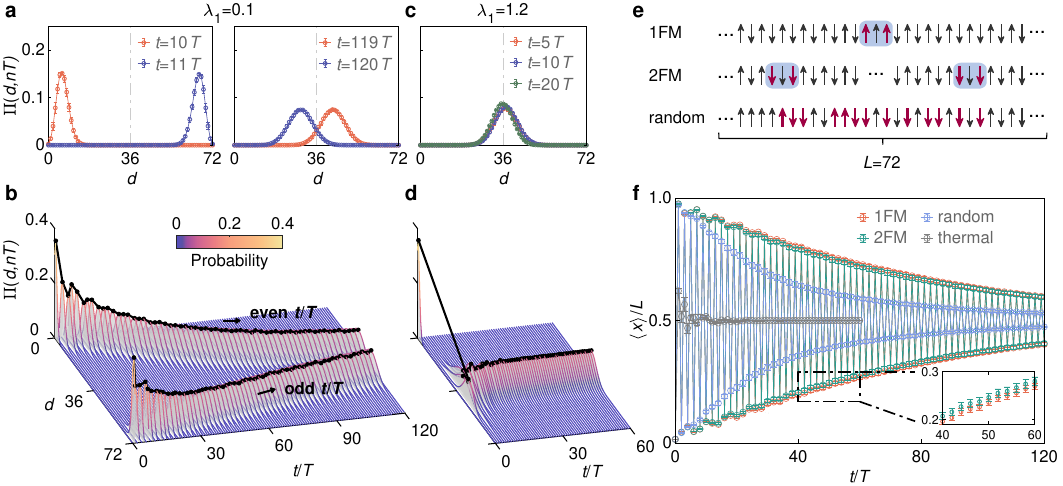}
    \caption{\label{fig2}
        {\bf FSP-induced time-crystalline wave packet dynamics.}
        {\bf a-d,} Measured dynamics of radial probability distribution $\Pi(d,nT)$ in FSP regime at $\lambda_1=2\lambda_2=0.1$ ({\bf a}, {\bf b}) and thermal regime at $\lambda_1=2\lambda_2=1.2$ ({\bf c}, {\bf d}). The initial state $|\boldsymbol{s}_0\rangle$ is prepared to the ``1FM" pattern.
        {\bf e,} Illustration of initial Fock states investigated in this work. $\uparrow, \downarrow$ denote $s^j = 1$ and $0$, respectively. ``1FM" and ``2FM" patterns differ from an overall AFM pattern by one or two mini ferromagnetic regions highlighted by the blue shades. To mitigate the unequal qubit errors in experiments, we average all results over five repeated experiments where the initial patterns are globally shifted along the physical qubit ring. The “random” state exemplifies one of five randomly sampled patterns, and the experimental results are averaged over them. The qubits satisfying DW constraints to flip are denoted by red colors.
        {\bf f,} Measured normalized mean Hamming distance $\langle x \rangle/L$ in FSP regime ($\lambda_1=2\lambda_2=0.1$) for ``1FM", ``2FM", and ``random" initial states, and in thermal regime at $\lambda_1=2\lambda_2=1.2$ for the ``1FM" initial state. Error bars, when present, are derived from 10 random samples of $\varphi_2$ across all figures in the text.
    }
\end{figure*}

\vspace{4mm}
\noindent\textbf{\large{}Model and FSP Mechanism}
\vspace{1mm}

To illustrate the physics of FSP in bulk, we consider a periodically driven Ising chain with periodic boundary conditions, whose Floquet unitary $U_{\rm F}$ for each drive cycle $T$ is given by 
\begin{align} \label{eq:uf}
    U_{\rm F} = \underbrace{e^{-{\rm i} J \sum_j \tilde{\sigma}^z_j(\lambda_2)\tilde{\sigma}^z_{j+1}(\lambda_2)}}_{\text{perturbed Ising interaction}}\underbrace{U_{\rm p}(\varphi_1,\lambda_1,\varphi_2)e^{-{\rm i}\pi\sum_j \sigma^x_j/2}}_{\text{single-qubit driving}}.
\end{align}
Here $\sigma_j^{x,y,z}$ are Pauli operators acting on qubit $j$, and the drive period is set as $T=1$. The model parameters are carefully designed to faithfully demonstrate FSP. 
To distinguish FSP from high-frequency ($JT\ll1$) Floquet prethermal behaviors, we set the Ising interaction strength $J=1$. 
To break the integrability of transverse-field Ising models while excluding the accidental localization caused by non-interacting echoes~\cite{Huang2023,Bao2024NC,Luitz2020}, we introduce generic on-site perturbations $U_{\rm p}(\varphi_1,\lambda_1,\varphi_2) = \prod_j e^{-{\rm i} \varphi_1 \sigma^z_j/2} e^{{\rm i}\lambda_1 \sigma^y_j/2} e^{-{\rm i} \varphi_2 \sigma^z_j/2}$ and average over 10 samples of different longitudinal uniform fields $\varphi_1, \varphi_2\in(0,2\pi)$ (Supplementary Information section 2A).
Moreover, to accelerate the early-time transient relaxation, two-qubit perturbation $\tilde{\sigma}^z_j(\lambda_2) = \cos(\lambda_2) \sigma^z_j + \sin(\lambda_2) \sigma^x_j$ is encoded into the Ising interaction $\tilde{\sigma}_j^z \tilde{\sigma}^z_{j+1}$. In our experiments, we evaluate the robustness of FSP and its breakdown by controlling the perturbation strength $\lambda_1, \lambda_2$, and always fix their relation to $\lambda_1/\lambda_2=2$. 

At the anchor point $\lambda_1=\lambda_2=0$, the Floquet unitary $U_{\rm F}$ only connects pairwise Fock states, as 
the single-qubit drive $e^{-{\rm i}\pi\sum_j \sigma^x_j/2}$ induces perfect Fock-state transitions $|\boldsymbol{s}\rangle\leftrightarrow |\boldsymbol{\bar{s}}\rangle$ ($\bar{s}^j=1-s^j$) while the Ising interaction and longitudinal field, $e^{-{\rm i} \sum_j [\sigma^z_j \sigma^z_{j+1} + (\varphi_1+\varphi_2)\sigma^z_j/2]}$, purely accumulate phases without altering the Fock-space connectivity. 
As $\lambda_1$ increases, an extensive Fock state network emerges, leading to eventual thermalization. 
However, when the Ising interaction is dominant ($J\gg\lambda_1$), the Ising energy $E_{\text{Ising}} = J(L-2w)$ introduces large gaps $2J|w-w'| \propto J$ among sectors $ \{ \boldsymbol{s} \} $ defined by different domain wall (DW) numbers $ w=W(\boldsymbol{s}) = 2\sum_{j} s^j(1-s^{j+1})$, where each DW is visualized by a blue dashed line in Fig.~\ref{fig1}\bfC. 
This splits the Fock-space network into distinct DW sectors (marked by different colors in Fig.~\ref{fig1}\bfB), with inter-DW hoppings suppressed by the Ising gap (see the left panel of Fig.~\ref{fig1}\bfC). 
Moreover, even within a DW sector of fixed $W(\boldsymbol{s})$, only Fock states differing by a qubit sandwiched between anti-parallel neighbors are directly connected (see right panel of Fig.~\ref{fig1}\bfC), which is the leading process conserving local DW number, and henceforth reducing the Fock bases connectivity inside a $W(\boldsymbol{s})$ subspace (see the inset of Fig.~\ref{fig1}\bfB). Together, these inter- and intra-DW constraints disentangle the dense Fock-space network in Fig.~\ref{fig1}\bfA\, into sparse networks in Fig.~\ref{fig1}\bfB, postponing any initial states to reach ergodicity in such FSP regimes.

The DW conservation central to FSP can be quantified by the underlying eigenstructure of the system~\cite{Huang2023} (Supplementary Information section 2). In Fig.~\ref{fig1}\bfD, we numerically compute the averaged DW number $ \langle w\rangle_n \equiv \sum_{\boldsymbol{s}} W(\boldsymbol{s}) |\langle \boldsymbol{s}| \varepsilon_n\rangle|^2 $ for each Floquet eigenstate $| \varepsilon_n\rangle$ of a ten-qubit system, where $\varepsilon_n$ denotes the quasienergy.
At weak perturbation ($\lambda_1=0.1$), the regime where FSP emerges, eigenstates fall into $L/2+1$ distinct DW sectors characterized by narrow distributions of $\langle w\rangle_n$ around even integers. Notably, each value of $ \langle w\rangle_n $ involves two eigenstate clusters separated roughly by quasienergy $\pi$, resulting in prethermal time-crystalline dynamics that will be shown later. 
In the opposite limit of strong perturbation ($\lambda_1=1.2$), the system obeys the eigenstate thermalization hypothesis, where all eigenstates yield similar $\langle w\rangle_n$.
These two extreme cases are bridged by a critical regime near $\lambda_1=0.4$, featuring wide distributions of $\langle w\rangle_n$ with significant DW sector mixing.
\begin{figure*}[t]
    \includegraphics[width=17.4cm]{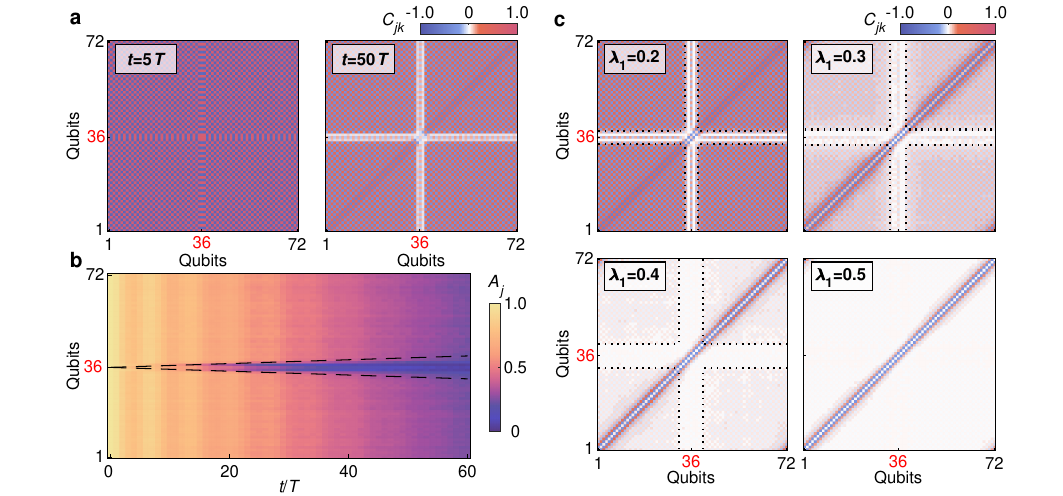}
    \caption{\label{fig3}{\bf Site-resolved dynamics of equal-time correlators.}
        {\bf a,} Snapshots of the measured equal-time correlator $C_{jk}(t)$ at $\lambda_1=0.1$ for the ``1FM" initial state, with the flipped qubit $Q_{36}$ highlighted in red.
        {\bf b,} Measured dynamics of $A_j(t)$ for the ``1FM" initial state. Black dashed lines represent analytical predictions for light-cone propagation speed.
        {\bf c,} Measured snapshots of $C_{jk}(t)$ at $t=30T$ for $\lambda_1=0.2, 0.3, 0.4$, and $0.5$. Dashed lines indicate the light-cone boundary. $\lambda_1/\lambda_2$ is fixed to 2 during these experiments.
    }
\end{figure*}

\vspace{4mm}
\noindent\textbf{\large{}Time-crystalline dynamics in Fock space}
\vspace{1mm}

\noindent 
Our experiments start with measuring wave packet dynamics in Fock space, which reveals the associated time-crystalline signatures. In doing so, we employ digital quantum circuits to implement the kicked Ising ring $U_{\rm F}$  on a two-dimensional chip with up to $L=72$ qubits. The chip features long coherence time of $T_1\approx 118~\mu$s, and high-fidelity operations of 99.95\% and 99.57\% for single- and two-qubit gates, respectively (Supplementary Information section 1A). The system is initialized to a Fock state $|\boldsymbol{s}_0\rangle$, as illustrated in Fig.~\ref{fig2}\bfE, and then periodically driven by the Floquet unitary $U_{\rm F}$. After $t=nT$ drive cycles, the many-body wavefunction can be characterized by the dynamical radial probability distribution~\cite{DeTomasi2021,Yao2023}
\begin{align} \label{eq:Pi}
    \Pi(d,nT) &= \sum_{\boldsymbol{s} \in \{ D(\boldsymbol{s}, \boldsymbol{s}_0) = d \}} 
    \left| 
    \langle \boldsymbol{s} | U_{\rm F}^{n} | \boldsymbol{s}_0 \rangle 
    \right|^2.
\end{align}

To observe slow thermalization behaviors caused by FSP, we prepare an initial state with ``1FM" pattern (see Fig.~\ref{fig2}\bfE), which belongs to the DW sector of $W(\boldsymbol{s}) = 70$ and consists of a single 3-qubit mini ferromagnetic (FM) region out of an overall anti-ferromagnetic (AFM) background. At small perturbations ($\lambda_1=0.1$) hosting FSP, the initial state connects to other Fock states in a fashion analogous to the inset of Fig.~\ref{fig1}\bfB. In this scenario, as shown in Fig.~\ref{fig2}\bfB, experimentally measured wave packet $\Pi(d, nT)$ remains localized away from $d=L/2$ throughout the entire $120T$ duration. In particular, its center exhibits persistent subharmonic oscillations between $d\rightarrow0$ and $L$ at even and odd periods, respectively~\cite{Ippoliti2021} (see Fig.~\ref{fig2}\bfA ~for snapshots). This is a hallmark feature of a discrete time crystal (DTC), which rigidly breaks the time translation symmetry $T$ of the underlying drive, as physical observables oscillate with periods that are integer multiples of $T$ under generic perturbations~\cite{Zaletel2023RMP,Else2020,Khemani2019b,Sacha2017}.
On the contrary, deep in the thermal regime ($\lambda_1=1.2$) with dense Fock-space networks, we see sharp differences in $\Pi(d,nT)$ dynamics in Fig.~\ref{fig2}\bfD. The wave packet rapidly equilibrates from its initial peak at $d=0$ into a thermal distribution centering around $d=L/2$. Snapshots in Fig.~\ref{fig2}\bfC~confirm that $\Pi(d,nT)$ already approaches a Gaussian shape by the early time $t=5T$, consistent with an idealized ergodic wave packet that reflects the density of states at each $d$~(Supplementary Information section 2D). 

DTC behaviors induced by  FSP can also manifest for different initial states, each probing a distinct region of the Fock-space network. In Fig.~\ref{fig2}\bfF, we present the measured mean Hamming distance $\langle x \rangle(t) = \sum_{d=0}^L d \cdot \Pi(d, t)$, a quantity that characterizes the motion of the wave packet's center. Owing to the inter- and intra-DW constraints detailed in Fig.~\ref{fig1}\bfC, qubit flips are only possible for qubits with anti-parallel neighbors (e.g., red arrows in Fig.~\ref{fig2}\bfE). Thus, initial states with more mini-FM regions, such as the ``2FM" pattern, should allow for more qubit flips, and consequently higher connectivity among Fock states. As expected, we observe a slight acceleration of thermalization for the ``2FM" initial state relative to ``1FM" state. Following this trend, we further explore more generic spots in the Fock space by randomly sampling the initial configuration. For example, the ``random" pattern in Fig.~\ref{fig2}\bfE~contains more qubits (denoted by red) that are allowed to flip, leading to a significantly faster decay of $\langle x\rangle/L$. In all three cases (``1FM", ``2FM", ``random"), however, the DTC oscillation outlives the full $120$ drive periods. This stands in sharp contrast to the thermal scenarios, where $\langle x\rangle/L$ equilibrates to $0.5$ within just 10 periods. These nonequilibrium dynamics of $\Pi(d,nT)$ thus confirm that the Fock-space network in the FSP regime is generically sparser than the thermal one across the entire Fock space.
	
\begin{figure*}[t]
    \includegraphics[width=17.4cm]{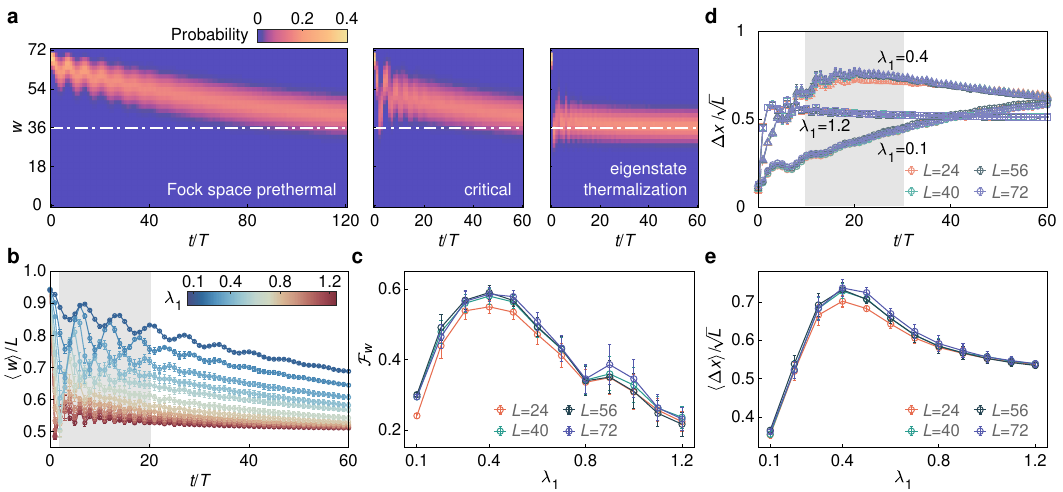}
    \caption{\label{fig4}{\bf Finite-size scaling of FSP-thermalization crossover.}
    {\bf a,} Measured DW probability distribution $\mathcal{D}(w,nT)$ in the Fock space prethermal ($\lambda_1=0.1$), critical ($\lambda_1=0.4$), and thermal ($\lambda_1=1.2$) regimes.
    {\bf b,} Measured dynamics of the normalized averaged DW number $\langle w\rangle/L$ as a function of perturbation strength $\lambda_1$.
    {\bf c,} Maximum amplitude of the Fourier spectra of DW dynamics for system sizes $L=24, 40, 56,$ and $72$. Fourier transformation is performed over the time window $t\in[2T,20T]$ (gray regime in {\bfB}).
    {\bf d,} Measured dynamics of normalized wave packet width $\Delta x/\sqrt{L}$ at $\lambda_1=0.1$ (circles), $0.4$ (triangles), and $1.2$ (squares). For each $\lambda_1$, experimental results are shown for system sizes $L=24, 40, 56,$ and $72$.
   {\bf e,} Time-averaged $\Delta x/\sqrt{L}$ (averaged over the time window $t\in[10T,30T]$, i.e., gray regime in {\bfD}) as a function of $\lambda_1$ for different system sizes $L$. Error bars in  {\bfB}-{\bfE} stem from 10 random samples of $\varphi_2$.
		}
\end{figure*}

\vspace{4mm}
\noindent\textbf{\large{}Site-resolved correlators in real space}
\vspace{1mm}

\noindent
To unveil the underlying DW constraints that give rise to FSP, we observe the site-resolved microscopic dynamics in real space by measuring the equal-time correlator
\begin{align}
    C_{jk}(t) = \langle \psi(t)| \sigma^z_j \sigma^z_k |\psi(t)\rangle 
\end{align}
for qubit pairs $\{Q_j, Q_k\}$, where  $|\psi\left(t=nT\right)\rangle = U_{\rm F}^n |\boldsymbol{s}_0\rangle$, $j,k=1,2,\dots,L$. 
For a perfectly localized Fock state, such as the initial state $|\boldsymbol{s}_0\rangle$, $C_{jk}$ reaches maximal magnitudes: $+1$ if $s_0^j$ and $s_0^k$ are parallel, $-1$ if they are anti-parallel. In contrast, local thermalization at a specific qubit $Q_{j_0}$ implies a superposition of various Fock states, which tends to take random values for $s^{j_0}$ and erase correlations between $Q_{j_0}$ and all other qubits so that $C_{j_0k}\rightarrow0$. Thus, $C_{jk}(t)$ provides site-resolved insights into the local thermalization process. 

In Fig.~\ref{fig3}{\bfA}, we show experimentally measured snapshots of $C_{jk}(t)$ at $\lambda_1=0.1$ for the ``1FM" initial state with a single mini FM region centered at $Q_{36}$. It is clear that at early time $t=5T$, $C_{jk}\rightarrow\pm1$ with the sign reflecting the mini FM region surrounding $Q_{36}$ out of an overall AFM pattern. According to Fig.~\ref{fig2}\bfE, only $Q_{35}$ and $Q_{37}$ satisfy the local DW constraints for flipping, which trigger local thermalization. Consequently, at late time $t=50T$, we observe a salient ``{\bf +}"-shaped spatial structure in $|C_{jk}(t)|$, whose values are small, indicating reduced correlations of $Q_{35}$ and $Q_{37}$ with all other qubits. However, AFM correlations, $C_{j,k}\approx - C_{j,k\pm1}$, persist in the area away from the mini FM region. Interestingly, while the mini FM region is created by flipping $Q_{36}$ in a global AFM pattern, our high-fidelity quantum circuits faithfully reveal that it is the neighboring qubits ($Q_{35}$ and $Q_{37}$), instead, exhibit the fastest thermalization. This observation aligns precisely with the predictions of DW constraints.

A full landscape of thermalizing dynamics is captured by the correlation magnitude $A_{j}(t)=\frac{1}{\ssz-1}\sum_{k\neq j} \left\lvert C_{jk}(t)\right\rvert$, with $A_{j}\to0$ in the fully thermalized limit. As shown in Fig.~\ref{fig3}\bfB, a clear light-cone stems from the mini FM region for the measured $A_{j}(t)$ dynamics. Notably, a bifurcation of the tip emerges and circumvents the central qubit $Q_{36}$ at early times, which further underscores the role of DW constraints as the underlying mechanism.
Based on the conservation of local DW number, we perform a perturbative calculation to extract the butterfly velocity $v_{\rm B}\approx 0.074$, which characterizes the propagation speed of the observed light-cone (Supplementary Information section 2B). As indicated by the dashed lines in Fig.~\ref{fig3}\bfB, the analytical prediction shows remarkable agreement with experimental results.  

With the increase of perturbation strengths, as shown in Fig.~\ref{fig3}\bfC, FSP gradually breaks down and is succeeded by conventional thermalization. For relatively small perturbations ($\lambda_1=0.2$ and $0.3$), the light-cone expands with increasing perturbation strength. Even so, DW constraints remain evident, indicated by the clear light-cone boundary and the retarded thermalization of the flipped qubit $Q_{36}$. Further going to  $\lambda_1=0.4$, the light-cone boundary obscures, finally giving way to the thermal case at $\lambda_1=0.5$ with short-range correlations alone.

\vspace{4mm}
\noindent\textbf{\large{}Scaling of FSP-thermal crossover}
\vspace{1mm}

\noindent
The observed spatial structure of $C_{jk}(t)$ in the 72-qubit system intuitively confirms the local DW constraints in the FSP regime. However, to verify that the slow thermalizing rate of FSP character is a many-body phenomenon that can persist in the thermodynamic limit, a finite-size scaling analysis is essential. To this end, we experimentally perform two types of scaling analysis: domain wall dynamics and wave packet properties. They yield consistent parameter estimations that characterize the perturbation-induced crossover from the FSP regime to the thermal regime.
        
Analogous to the wave packet $\Pi(d,nT)$, we introduce the DW probability distribution,
\begin{align}
    \mathcal{D}(w,nT) = \sum_{\boldsymbol{s}\in \{ W(\boldsymbol{s}) = w \}}  \left|\langle \boldsymbol{s} | U_{\rm F}^n |\boldsymbol{s}_0\rangle \right|^2,
\end{align}
to connect dynamical features of FSP to the defining eigenstructures of $U_{\rm F}$ (Fig.~\ref{fig1}\bfD). $\mathcal{D}(w,nT)$ characterizes how the wavefunction $|\psi(t=nT)\rangle = U^{n}_{\rm F}|\boldsymbol{s}_0\rangle$ spreads over Fock states belonging to distinct DW sectors. In Fig.~\ref{fig4}\bfA, we initialize the 72-qubit ring to the ``1FM" state with 70 DWs and measure representative $\mathcal{D}(w, nT)$ dynamics in three different regimes. In the FSP regime ($\lambda_1=0.1$) where the initial state has large overlaps with eigenstates of DW number $w\sim70$, we observe that, in early times ($t<30T$), $\mathcal{D}(w,nT)$ remains concentrated in exponentially small subspaces far from $L/2$, bouncing back and forth between these subspaces, consistent with the eigenstructure in Fig.~\ref{fig1}\bfD~(Note that the density of states at a DW subspace $w$ is $\propto e^{-\frac{(w-L/2)^2}{L/2}}$ for $L\gg1$, see Supplementary Information section 2D). In particular, $\mathcal{D}(w,nT)$ barely reach the region $w=L/2=36$ until $t\approx80 T$. In contrast, at $\lambda_1=0.4$ entering the critical regime (middle panel of Fig.~\ref{fig4}\bfA), $\mathcal{D}(w,nT)$ dynamics already exhibits large oscillations over $w$ within $t\approx2T$, penetrating all the way through $L/2$. This reflects the significant DW sector mixing in eigenstates (Fig.~\ref{fig1}\bfD). Further increasing the perturbation to $\lambda_1=1.2$ leads to an immediate jump of DW number from the initial $w\approx 70$ directly to an equilibrium value $w\approx L/2$, where the density of states is highest. These characteristic DW dynamics nicely validate the theoretical understanding of eigenstructure for FSP and its breaking mechanism in Fig.~\ref{fig1}\bfD.
 	
To diagnose the critical regime, we take a step further to investigate DW dynamics across varying perturbation strengths. In Fig.~\ref{fig4}\bfB, we present the measured dynamics of the averaged DW number, defined as $\langle w\rangle(t) = \sum_{w=0}^L w\cdot \mathcal{D}(w,t)$, for a 72-qubit system. We find that, in early times ($t\lesssim 20T$), where the dynamics is dominated by unitary evolution, $\langle w\rangle$ exhibits anomalously large oscillation amplitudes under intermediate perturbation strengths near $\lambda_1=0.4$. By contrast, in the FSP and thermal limits, these oscillations remain small. To quantify these amplitudes, we perform a Fourier analysis of the DW dynamics over $t\in[2T, 20T]$. Figure~\ref{fig4}\bfC~shows the maximum value of Fourier spectra $\mathcal{F}_{w}$ as a function of perturbation strength for system sizes $L=24, 40, 56$, and $72$. Crucially, all system sizes consistently identify the critical regime around $\lambda_1\sim0.4$. More importantly, for large system sizes ($L=40, 56$, and $72$), where finite-size effects are negligible, the results exhibit nearly perfect data collapse. This provides strong evidence that FSP is not a finite-size effect but a robust phenomenon that can exist in large-scale systems.  
		
The scaling behaviors of DW dynamics above can be confirmed by an independent observation of wave function fluctuations in Fock space, $\Delta x(t) = \sqrt{\sum_{d=0}^\ssz (d-\langle x\rangle)^2 \Pi(d,t)}$. It quantifies the width of wave packet $\Pi(d)$ and has been shown to be effective in characterizing the crossover from MBL to thermal regimes in finite-size systems~\cite{DeTomasi2021, Yao2023,liu2025}. Due to noise effects, $\Delta x$ predominantly grows with $\sqrt{L}$ experimentally (Supplementary Information section 2D), while the magnitudes $\Delta x/\sqrt{L}$ distinguish perturbation parameters $\lambda_1, \lambda_2$ in different regimes. Figure~\ref{fig4}\bfD~presents experimentally measured $\Delta x/\sqrt{L}$ dynamics in three different regimes: In the FSP regime ($\lambda_1=0.1$), $\Delta x/\sqrt{L}$ grows extremely slowly but its value exceed 0.5 in late times, an explicit signature of slow thermalization behaviors; in the critical regime, an anomalously large fluctuation emerges rapidly in early times (e.g., $t=20T$), arising from the wide distribution of DW numbers in eigenstates; in the thermal regime, $\Delta x/\sqrt{L}$ surges up and saturates to the value around $0.5$, in line with the prediction of thermal ensembles. Notably, like the DW dynamics, satisfactory data collapse is observed for different system sizes. For finite-size scaling analysis, we further measure the time-averaged $\langle \Delta x\rangle/\sqrt{L}$ over the time window $t\in[10T,30T]$, during which sharp distinctions for different regimes emerge, and present it as a function of perturbation strength in Fig.~\ref{fig4}\bfE. Remarkably, this analysis also identifies a system-size independent critical regime near $\lambda_1=0.4$, consistent with DW dynamics scaling in Fig.~\ref{fig4}\bfC. This provides independent validation of FSP and its breaking mechanism, from the perspective of wave packet dynamics.

\vspace{4mm}
\noindent\textbf{\large{}Conclusions}
\vspace{1mm}
	
\noindent In summary, we have proposed and experimentally demonstrated Fock space prethermalization as a robust mechanism against heating, which, intriguingly, works under strong interactions comparable to driving frequency that defies conventional Floquet prethermal conditions. The local DW constraints that lead to a sparse Fock-space network only require strong Ising interactions, and therefore can be compatible with additional structures to construct exotic non-equilibrium phases of quantum matter, such as the DTC exemplified in this work.
	
Our experiment also demonstrates the viability of large-scale superconducting qubits for directly revealing the Fock-space structure of many-body systems. This capability holds utility for several promising research directions of wide interest, which are currently under intensive investigation. For instance, the dynamics of Fock-space wave packets could shed light on the eigenstate resonance due to hybridized Fock states in prethermal MBL~\cite{Morningstar2022a,Ha2023,Long2023}. Meanwhile, recent work has shown that strong bottlenecks in Fock space can protect Hilbert space fragmentation 
against avalanche induced by edge thermal bath~\cite{Han2024}, therefore it will be of interest to explore the stability of FSP against coupling to environments. Furthermore, a notable feature of FSP is the non-uniform spatial structures, where thermalization occurs at drastically different rates inside and outside the well-shaped light cones. It could provide potential controllability for the thermalization rate of many-body systems by initializing different spatial patterns, in stark contrast to conventional prethermalization, i.e., in U(1) prethermal DTCs at high frequency~\cite{Luitz2020,Stasiuk2023} (Supplementary Information section 2C), that usually thermalize uniformly in space. Finally, it is of interest to extend the FSP scheme to non-periodically driven systems that potentially host discrete time quasi-crystals~\cite{Beatrez2023,He2025,He2023,Else2020a}.

\vspace{.1cm}
\noindent\textbf{\large Data availability}
\\
The data generated in this study are available on GitHub at \href{https://github.com/ZJU-zbao/FSP-72q-data}{https://github.com/ZJU-zbao/FSP-72q-data}.

\vspace{.1cm}
\noindent\textbf{\large Code availability}
\\
The simulation codes used in this study are available on GitHub at \href{https://github.com/ZJU-zbao/FSP-72q}{https://github.com/ZJU-zbao/FSP-72q}.

\vspace{.1cm}
\noindent\textbf{\large Acknowledgements} 
\\
The device was fabricated at the Micro-Nano Fabrication Center of Zhejiang University. 
Numerical simulation of quantum circuits was performed using MindSpore Quantum and Qiskit framework. 
We acknowledge the support from the Zhejiang Provincial Natural Science Foundation of China (Grant Nos. LR24A040002 and LDQ23A040001) and the National Natural Science Foundation of China (Grant Nos. 92065204, 12274368, 12174389, U20A2076, 12174342, 12274367, 12322414, 12404570, 12404574, 12247101). R.M.~acknowledges support from the T$_c$SUH Welch Professorship Award. Z.B.~acknowledges support from CPS-Yangtze Delta Region Industrial Innovation Center of Quantum and Information Technology-MindSpore Quantum Open Fund.

\vspace{.1cm}
\noindent\textbf{\large Author contributions}
\\
B.H. and Q.G. conceived the idea; Z.B. carried out the experiments and analyzed the experimental data under the supervision of Q.G. and H. Wang; H.L. fabricated the device supervised by H. Wang; B.H., Q.G., Z.B., and H. Wang co-wrote the manuscript. H. Wang, Q.G., C.S., Z.W., Z.B., Z.Z., Z.S., F.J., X.Z., Y.G., C.Z., N.W., Y.Z., Z.T., A.Z., Z.C., F.S., J. Zhong, Y. He, Han Wang, J.-N.Y., Y. Wang, J.S., G.L., Y. Han, Y. Wu, J.D., H.D., and P.Z. contributed to the experimental setup. All authors contributed to the discussions of the results and the writing of the manuscript.

\vspace{.1cm}
\noindent\textbf{\large Competing interests} 
\\
The authors declare no competing interests.
\bibliography{TC}

@article{Arute2019Nature,
	title = {Quantum supremacy using a programmable superconducting processor},
	volume = {574},
	issn = {0028-0836, 1476-4687},
	doi = {10.1038/s41586-019-1666-5},
	number = {7779},
	urldate = {2022-02-11},
	journal = {Nature},
	author = {Arute, Frank and Arya, Kunal and Babbush, Ryan and Bacon, Dave and Bardin, Joseph C. and Barends, Rami and Biswas, Rupak and Boixo, Sergio and Brandao, Fernando G. S. L. and Buell, David A. and Burkett, Brian and Chen, Yu and Chen, Zijun and Chiaro, Ben and Collins, Roberto and Courtney, William and Dunsworth, Andrew and Farhi, Edward and Foxen, Brooks and Fowler, Austin and Gidney, Craig and Giustina, Marissa and Graff, Rob and Guerin, Keith and Habegger, Steve and Harrigan, Matthew P. and Hartmann, Michael J. and Ho, Alan and Hoffmann, Markus and Huang, Trent and Humble, Travis S. and Isakov, Sergei V. and Jeffrey, Evan and Jiang, Zhang and Kafri, Dvir and Kechedzhi, Kostyantyn and Kelly, Julian and Klimov, Paul V. and Knysh, Sergey and Korotkov, Alexander and Kostritsa, Fedor and Landhuis, David and Lindmark, Mike and Lucero, Erik and Lyakh, Dmitry and Mandrà, Salvatore and McClean, Jarrod R. and McEwen, Matthew and Megrant, Anthony and Mi, Xiao and Michielsen, Kristel and Mohseni, Masoud and Mutus, Josh and Naaman, Ofer and Neeley, Matthew and Neill, Charles and Niu, Murphy Yuezhen and Ostby, Eric and Petukhov, Andre and Platt, John C. and Quintana, Chris and Rieffel, Eleanor G. and Roushan, Pedram and Rubin, Nicholas C. and Sank, Daniel and Satzinger, Kevin J. and Smelyanskiy, Vadim and Sung, Kevin J. and Trevithick, Matthew D. and Vainsencher, Amit and Villalonga, Benjamin and White, Theodore and Yao, Z. Jamie and Yeh, Ping and Zalcman, Adam and Neven, Hartmut and Martinis, John M.},
	month = oct,
	year = {2019},
	pages = {505--510}
}

@article{Vidal2003PRL,
	title = {Efficient Classical Simulation of Slightly Entangled Quantum Computations},
	author = {Vidal, Guifr\'e},
	journal = {Phys. Rev. Lett.},
	volume = {91},
	issue = {14},
	pages = {147902},
	numpages = {4},
	year = {2003},
	month = {Oct},
	publisher = {American Physical Society},
	doi = {10.1103/PhysRevLett.91.147902}
}

@Article{Song2019Science,
	author  = {Chao Song and Kai Xu and Hekang Li and Yu-Ran Zhang and Xu Zhang and Wuxin Liu and Qiujiang Guo and Zhen Wang and Wenhui Ren and Jie Hao and Hui Feng and Heng Fan and Dongning Zheng and Da-Wei Wang and H. Wang and Shi-Yao Zhu},
	journal = {Science},
	title   = {Generation of multicomponent atomic {S}chrödinger cat states of up to 20 qubits},
	year    = {2019},
	number  = {6453},
	pages   = {574-577},
	volume  = {365},
	doi     = {10.1126/science.aay0600},
}

@article{xu2018prl,
  title = {Emulating Many-Body Localization with a Superconducting Quantum Processor},
  author = {Xu, Kai and Chen, Jin-Jun and Zeng, Yu and Zhang, Yu-Ran and Song, Chao and Liu, Wuxin and Guo, Qiujiang and Zhang, Pengfei and Xu, Da and Deng, Hui and Huang, Keqiang and Wang, H. and Zhu, Xiaobo and Zheng, Dongning and Fan, Heng},
  journal = {Phys. Rev. Lett.},
  volume = {120},
  issue = {5},
  pages = {050507},
  numpages = {6},
  year = {2018},
  month = {Feb},
  publisher = {American Physical Society},
  doi = {10.1103/PhysRevLett.120.050507},
  url = {https://link.aps.org/doi/10.1103/PhysRevLett.120.050507}
}

@article{Morvan2024Nature,
	title = {Phase transitions in random circuit sampling},
	volume = {634},
	issn = {0028-0836, 1476-4687},
	url = {https://www.nature.com/articles/s41586-024-07998-6},
	doi = {10.1038/s41586-024-07998-6},
	number = {8033},
	urldate = {2025-02-03},
	journal = {Nature},
	author = {Morvan, A. and Villalonga, B. and Mi, X. and Mandr\.a, S. and Bengtsson, A. and Klimov, P. V. and Chen, Z. and Hong, S. and Erickson, C. and Drozdov, I. K. and Chau, J. and Laun, G. and Movassagh, R. and Asfaw, A. and Brand\~ao, L. T. A. N. and Peralta, R. and Abanin, D. and Acharya, R. and Allen, R. and Andersen, T. I. and Anderson, K. and Ansmann, M. and Arute, F. and Arya, K. and Atalaya, J. and Bardin, J. C. and Bilmes, A. and Bortoli, G. and Bourassa, A. and Bovaird, J. and Brill, L. and Broughton, M. and Buckley, B. B. and Buell, D. A. and Burger, T. and Burkett, B. and Bushnell, N. and Campero, J. and Chang, H.-S. and Chiaro, B. and Chik, D. and Chou, C. and Cogan, J. and Collins, R. and Conner, P. and Courtney, W. and Crook, A. L. and Curtin, B. and Debroy, D. M. and Barba, A. Del Toro and Demura, S. and Paolo, A. Di and Dunsworth, A. and Faoro, L. and Farhi, E. and Fatemi, R. and Ferreira, V. S. and Burgos, L. Flores and Forati, E. and Fowler, A. G. and Foxen, B. and Garcia, G. and Genois, \'E. and Giang, W. and Gidney, C. and Gilboa, D. and Giustina, M. and Gosula, R. and Dau, A. Grajales and Gross, J. A. and Habegger, S. and Hamilton, M. C. and Hansen, M. and Harrigan, M. P. and Harrington, S. D. and Heu, P. and Hoffmann, M. R. and Huang, T. and Huff, A. and Huggins, W. J. and Ioffe, L. B. and Isakov, S. V. and Iveland, J. and Jeffrey, E. and Jiang, Z. and Jones, C. and Juhas, P. and Kafri, D. and Khattar, T. and Khezri, M. and Kieferov\'a, M. and Kim, S. and Kitaev, A. and Klots, A. R. and Korotkov, A. N. and Kostritsa, F. and Kreikebaum, J. M. and Landhuis, D. and Laptev, P. and Lau, K.-M. and Laws, L. and Lee, J. and Lee, K. W. and Lensky, Y. D. and Lester, B. J. and Lill, A. T. and Liu, W. and Livingston, W. P. and Locharla, A. and Malone, F. D. and Martin, O. and Martin, S. and McClean, J. R. and McEwen, M. and Miao, K. C. and Mieszala, A. and Montazeri, S. and Mruczkiewicz, W. and Naaman, O. and Neeley, M. and Neill, C. and Nersisyan, A. and Newman, M. and Ng, J. H. and Nguyen, A. and Nguyen, M. and Niu, M. Yuezhen and O'Brien, T. E. and Omonije, S. and Opremcak, A. and Petukhov, A. and Potter, R. and Pryadko, L. P. and Quintana, C. and Rhodes, D. M. and Rocque, C. and Rosenberg, E. and Rubin, N. C. and Saei, N. and Sank, D. and Sankaragomathi, K. and Satzinger, K. J. and Schurkus, H. F. and Schuster, C. and Shearn, M. J. and Shorter, A. and Shutty, N. and Shvarts, V. and Sivak, V. and Skruzny, J. and Smith, W. C. and Somma, R. D. and Sterling, G. and Strain, D. and Szalay, M. and Thor, D. and Torres, A. and Vidal, G. and Heidweiller, C. Vollgraff and White, T. and Woo, B. W. K. and Xing, C. and Yao, Z. J. and Yeh, P. and Yoo, J. and Young, G. and Zalcman, A. and Zhang, Y. and Zhu, N. and Zobrist, N. and Rieffel, E. G. and Biswas, R. and Babbush, R. and Bacon, D. and Hilton, J. and Lucero, E. and Neven, H. and Megrant, A. and Kelly, J. and Roushan, P. and Aleiner, I. and Smelyanskiy, V. and Kechedzhi, K. and Chen, Y. and Boixo, S.},
	month = oct,
	year = {2024},
	pages = {328--333}
}

@Article{Zhang2017,
	author       = {Zhang, J and Hess, P W and Kyprianidis, A and Becker, P and Lee, A and Smith, J and Pagano, G and Potirniche, I-D and Potter, A C and Vishwanath, A and Yao, N Y and Monroe, C},
	title        = {Observation of a discrete time crystal.},
	journal      = {Nature},
	year         = {2017},
	volume       = {543},
	pages        = {217--220},
	month        = mar,
	issn         = {1476-4687},
	completed    = {2017-07-18},
	country      = {England},
	doi          = {10.1038/nature21413},
	issn-linking = {0028-0836},
	issue        = {7644},
	nlm-id       = {0410462},
	pii          = {nature21413},
	pmid         = {28277505},
	pubmodel     = {Print},
	pubstatus    = {ppublish},
	revised      = {2017-07-18},
}

@Article{Yao2017,
	author       = {Yao, N. Y. and Potter, A. C. and Potirniche, I.-D. and Vishwanath, A.},
	journal      = {Phys. Rev. Lett.},
	title        = {Discrete Time Crystals: Rigidity, Criticality, and Realizations.},
	year         = {2017},
	issn         = {1079-7114},
	month        = jan,
	pages        = {030401},
	volume       = {118},
	completed    = {2018-02-05},
	country      = {United States},
	doi          = {10.1103/PhysRevLett.118.030401},
	issn-linking = {0031-9007},
	issue        = {3},
	nlm-id       = {0401141},
	pmid         = {28157355},
	pubmodel     = {Print-Electronic},
	pubstatus    = {ppublish},
	revised      = {2018-02-05},
}

@Article{Else2016,
	author       = {Else, Dominic V and Bauer, Bela and Nayak, Chetan},
	title        = {Floquet Time Crystals.},
	journal      = {Phys. Rev. Lett.},
	year         = {2016},
	volume       = {117},
	pages        = {090402},
	month        = aug,
	issn         = {1079-7114},
	completed    = {2018-01-24},
	country      = {United States},
	doi          = {10.1103/PhysRevLett.117.090402},
	issn-linking = {0031-9007},
	issue        = {9},
	nlm-id       = {0401141},
	pmid         = {27610834},
	pubmodel     = {Print-Electronic},
	pubstatus    = {ppublish},
	revised      = {2018-01-24},
}

@Article{Choi2017,
	author  = {Soonwon Choi and Joonhee Choi and Renate Landig and Georg Kucsko and Hengyun Zhou and Junichi Isoya and Fedor Jelezko and Shinobu Onoda and Hitoshi Sumiya and Vedika Khemani and Curt von Keyserlingk and Norman Y. Yao and Eugene Demler and Mikhail D. Lukin},
	title   = {Observation of discrete time-crystalline order in a disordered dipolar many-body system},
	journal = {Nature},
	year    = {2017},
	volume  = {543},
	pages   = {221-225},
	issn    = {0028-0836},
	doi     = {10.1038/nature21426},
}

@Article{Ho2017,
	author       = {Wen Wei Ho and Soonwon Choi and Mikhail D. Lukin and Dmitry A. Abanin},
	title        = {Critical Time Crystals in Dipolar Systems},
	journal      = {Phys. Rev. Lett.},
	year         = {2017},
	volume       = {119},
	pages        = {010602},
	date         = {2017-03-14},
	doi          = {10.1103/PhysRevLett.119.010602},
	eprintclass  = {cond-mat.dis-nn},
	eprinttype   = {arXiv},
	journaltitle = {Phys. Rev. Lett. 119, 010602 (2017)},
}

@Article{Khemani2016,
	author    = {Vedika Khemani and Achilleas Lazarides and Roderich Moessner and S. L. Sondhi},
	journal   = {Phys. Rev. Lett.},
	title     = {Phase Structure of Driven Quantum Systems},
	year      = {2016},
	month     = {jun},
	number    = {25},
	pages     = {250401},
	volume    = {116},
	doi       = {10.1103/physrevlett.116.250401},
	publisher = {American Physical Society ({APS})},
}

@Article{Wilczek2012,
	author    = {Wilczek, Frank},
	title     = {Quantum time crystals},
	journal   = {Phys. Rev. Lett.},
	year      = {2012},
	volume    = {109},
	number    = {16},
	pages     = {160401},
	doi       = {10.1103/physrevlett.109.160401},
	publisher = {APS},
}

@Article{Wilczek2013,
	author    = {Wilczek, Frank},
	title     = {Superfluidity and space-time translation symmetry breaking},
	journal   = {Phys. Rev. Lett.},
	year      = {2013},
	volume    = {111},
	number    = {25},
	pages     = {250402},
	doi       = {10.1103/physrevlett.111.250402},
	publisher = {APS},
}

@Article{Bruno2012,
	author  = {Bruno, Patrick},
	title   = {Comment on “Quantum Time Crystals”},
	journal = {Phys. Rev. Lett.},
	year    = {2012},
	volume  = {110},
	pages   = {118901},
	doi     = {10.1103/PhysRevLett.110.118901},
}

@Article{Watanabe2015,
	author    = {Watanabe, Haruki and Oshikawa, Masaki},
	title     = {Absence of quantum time crystals},
	journal   = {Phys. Rev. Lett.},
	year      = {2015},
	volume    = {114},
	number    = {25},
	pages     = {251603},
	doi       = {10.1103/physrevlett.114.251603},
	publisher = {APS},
}

@Article{Bluvstein2021,
	author    = {D. Bluvstein and A. Omran and H. Levine and A. Keesling and G. Semeghini and S. Ebadi and T. T. Wang and A. A. Michailidis and N. Maskara and W. W. Ho and S. Choi and M. Serbyn and M. Greiner and V. Vuleti{\'{c}} and M. D. Lukin},
	journal   = {Science},
	title     = {Controlling quantum many-body dynamics in driven Rydberg atom arrays},
	year      = {2021},
	month     = {feb},
	number    = {6536},
	pages     = {1355--1359},
	volume    = {371},
	doi       = {10.1126/science.abg2530},
	publisher = {American Association for the Advancement of Science ({AAAS})},
}

@Article{Rovny2018,
	author    = {Jared Rovny and Robert L. Blum and Sean E. Barrett},
	journal   = {Phys. Rev. Lett.},
	title     = {Observation of Discrete-Time-Crystal Signatures in an Ordered Dipolar Many-Body System},
	year      = {2018},
	month     = {may},
	number    = {18},
	pages     = {180603},
	volume    = {120},
	doi       = {10.1103/physrevlett.120.180603},
	fjournal  = {Physical Review Letters},
	publisher = {American Physical Society ({APS})},
}

@Article{Randall2021,
	author    = {J. Randall and C. E. Bradley and F. V. van der Gronden and A. Galicia and M. H. Abobeih and M. Markham and D. J. Twitchen and F. Machado and N. Y. Yao and T. H. Taminiau},
	journal   = {Science},
	title     = {Many-body{\textendash}localized discrete time crystal with a programmable spin-based quantum simulator},
	year      = {2021},
	month     = {dec},
	number    = {6574},
	pages     = {1474--1478},
	volume    = {374},
	doi       = {10.1126/science.abk0603},
	publisher = {American Association for the Advancement of Science ({AAAS})},
}

@Article{Kyprianidis2021,
	author    = {A. Kyprianidis and F. Machado and W. Morong and P. Becker and K. S. Collins and D. V. Else and L. Feng and P. W. Hess and C. Nayak and G. Pagano and N. Y. Yao and C. Monroe},
	journal   = {Science},
	title     = {Observation of a prethermal discrete time crystal},
	year      = {2021},
	month     = {jun},
	number    = {6547},
	pages     = {1192--1196},
	volume    = {372},
	doi       = {10.1126/science.abg8102},
	publisher = {American Association for the Advancement of Science ({AAAS})},
}

@article{Mi2022,
  title = {Time-Crystalline Eigenstate Order on a Quantum Processor},
  year = {2022},
  month = jan,
  journal = {Nature},
  volume = {601},
  number = {7894},
  pages = {531--536},
  issn = {0028-0836, 1476-4687},
  doi = {10.1038/s41586-021-04257-w},
  langid = {english},
  author = {Mi, Xiao and Ippoliti, Matteo and Quintana, Chris and Greene, Ami and Chen, Zijun and Gross, Jonathan and Arute, Frank and Arya, Kunal and Atalaya, Juan and Babbush, Ryan and Bardin, Joseph C. and Basso, Joao and Bengtsson, Andreas and Bilmes, Alexander and Bourassa, Alexandre and Brill, Leon and Broughton, Michael and Buckley, Bob B. and Buell, David A. and Burkett, Brian and others}
}

@Article{Else2020,
	author    = {Dominic V. Else and Christopher Monroe and Chetan Nayak and Norman Y. Yao},
	journal   = {Annu. Rev. Condens. Matter Phys.},
	title     = {Discrete Time Crystals},
	year      = {2020},
	month     = {mar},
	number    = {1},
	pages     = {467--499},
	volume    = {11},
	doi       = {10.1146/annurev-conmatphys-031119-050658},
	fjournal  = {Annual Review of Condensed Matter Physics, Vol 2},
	publisher = {Annual Reviews},
}

@Article{Sacha2017,
	author    = {Krzysztof Sacha and Jakub Zakrzewski},
	journal   = {Rep. Progr. Phys.},
	title     = {Time crystals: a review},
	year      = {2017},
	month     = {nov},
	number    = {1},
	pages     = {016401},
	volume    = {81},
	doi       = {10.1088/1361-6633/aa8b38},
	publisher = {{IOP} Publishing},
}

@Article{Abanin2019,
	author    = {Dmitry A. Abanin and Ehud Altman and Immanuel Bloch and Maksym Serbyn},
	journal   = {Rev. Mod. Phys.},
	title     = {Colloquium : Many-body localization, thermalization, and entanglement},
	year      = {2019},
	month     = {may},
	number    = {2},
	pages     = {021001},
	volume    = {91},
	doi       = {10.1103/revmodphys.91.021001},
	fjournal  = {Reviews of Modern Physics},
	publisher = {American Physical Society ({APS})},
}

@Article{Abanin2015,
	author    = {Dmitry A. Abanin and Wojciech De Roeck and Fran{\c{c}}ois Huveneers},
	journal   = {Phys. Rev. Lett.},
	title     = {Exponentially Slow Heating in Periodically Driven Many-Body Systems},
	year      = {2015},
	month     = {dec},
	number    = {25},
	pages     = {256803},
	volume    = {115},
	doi       = {10.1103/physrevlett.115.256803},
	fjournal  = {Physical Review Letters},
	publisher = {American Physical Society ({APS})},
}

@Article{Else2017,
	author    = {Dominic V. Else and Bela Bauer and Chetan Nayak},
	journal   = {Phys. Rev. X},
	title     = {Prethermal Phases of Matter Protected by Time-Translation Symmetry},
	year      = {2017},
	month     = {mar},
	number    = {1},
	pages     = {011026},
	volume    = {7},
	doi       = {10.1103/physrevx.7.011026},
	publisher = {American Physical Society ({APS})},
}

@Article{Luitz2020,
	author    = {David J. Luitz and Roderich Moessner and S. L. Sondhi and Vedika Khemani},
	journal   = {Phys. Rev. X},
	title     = {Prethermalization without Temperature},
	year      = {2020},
	month     = {may},
	number    = {2},
	pages     = {021046},
	volume    = {10},
	doi       = {10.1103/physrevx.10.021046},
	publisher = {American Physical Society ({APS})},
}

@Misc{Ho2020,
	author      = {Wen Wei Ho and Wojciech De Roeck},
	title       = {A Rigorous Theory of Prethermalization without Temperature},
	date        = {2020-11-30},
        year={2020},
	eprint      = {arXiv:2011.14583},
	primaryClass = {quant-ph},
	eprinttype  = {arXiv},
}

@Article{Turner2018,
	author    = {C. J. Turner and A. A. Michailidis and D. A. Abanin and M. Serbyn and Z. Papi{\'{c}}},
	journal   = {Nat. Phys.},
	title     = {Weak ergodicity breaking from quantum many-body scars},
	year      = {2018},
	month     = {may},
	number    = {7},
	pages     = {745--749},
	volume    = {14},
	doi       = {10.1038/s41567-018-0137-5},
	fjournal  = {Nature Physics},
	publisher = {Springer Science and Business Media {LLC}},
}

@Article{Pal2018,
	author    = {Soham Pal and Naveen Nishad and T. S. Mahesh and G. J. Sreejith},
	journal   = {Phys. Rev. Lett.},
	title     = {Temporal Order in Periodically Driven Spins in Star-Shaped Clusters},
	year      = {2018},
	month     = {may},
	number    = {18},
	pages     = {180602},
	volume    = {120},
	doi       = {10.1103/physrevlett.120.180602},
	fjournal  = {Physical Review Letters},
	publisher = {American Physical Society ({APS})},
}

@Article{Frey2022,
	author    = {Philipp Frey and Stephan Rachel},
	journal   = {Sci. Adv.},
	title     = {Realization of a discrete time crystal on 57 qubits of a quantum computer},
	year      = {2022},
	month     = {mar},
	number    = {9},
	pages     = {eabm7652},
	volume    = {8},
	doi       = {10.1126/sciadv.abm7652},
	fjournal  = {Science Advances},
	publisher = {American Association for the Advancement of Science ({AAAS})},
}

@Article{Huang2022,
	author    = {Biao Huang and Tsz-Him Leung and Dan M. Stamper-Kurn and W. Vincent Liu},
	journal   = {Phys. Rev. Lett.},
	title     = {Discrete Time Crystals Enforced by {F}loquet-{B}loch Scars},
	year      = {2022},
	month     = {sep},
	number    = {13},
	pages     = {133001},
	volume    = {129},
	doi       = {10.1103/physrevlett.129.133001},
	fjournal  = {Physical Review Letters},
	publisher = {American Physical Society ({APS})},
}

@Article{Chandran2022,
	author    = {Anushya Chandran and Thomas Iadecola and Vedika Khemani and Roderich Moessner},
	journal   = {Annu. Rev. Condens. Matter Phys.},
	title     = {Quantum Many-Body Scars: A Quasiparticle Perspective},
	year      = {2023},
	volume    = {14},
        pages     = {443-468},
        doi = {https://doi.org/10.1146/annurev-conmatphys-031620-101617},
	fjournal  = {Annual Review of Condensed Matter Physics},
	publisher = {Annual Reviews},
}

@Misc{Khemani2019b,
	author        = {Vedika Khemani and Roderich Moessner and S. L. Sondhi},
	title         = {A Brief History of Time Crystals},
	month         = nov,
	archiveprefix = {arXiv},
        year={2019},
	eprint        = {1910.10745},
        primaryClass = {quant-ph}
}

@Article{Beatrez2023,
  author    = {William Beatrez and Christoph Fleckenstein and Arjun Pillai and Erica de Leon Sanchez and Amala Akkiraju and Jesus Diaz Alcala and Sophie Conti and Paul Reshetikhin and Emanuel Druga and Marin Bukov and Ashok Ajoy},
  journal   = {Nat. Phys.},
  title     = {Critical prethermal discrete time crystal created by two-frequency driving},
  year      = {2023},
  month     = {jan},
  pages     = {407},
  volume    = {19},
  doi       = {10.1038/s41567-022-01891-7},
  fjournal  = {Nature Physics},
  publisher = {Springer Science and Business Media {LLC}},
}

@Article{Machado2020,
	author    = {Francisco Machado and Dominic V. Else and Gregory D. Kahanamoku-Meyer and Chetan Nayak and Norman Y. Yao},
	journal   = {Phys. Rev. X},
	title     = {Long-Range Prethermal Phases of Nonequilibrium Matter},
	year      = {2020},
	month     = {feb},
	number    = {1},
	pages     = {011043},
	volume    = {10},
	doi       = {10.1103/physrevx.10.011043},
	fjournal  = {Physical Review X},
	publisher = {American Physical Society ({APS})},
}

@Article{Ippoliti2021,
	author    = {Matteo Ippoliti and Kostyantyn Kechedzhi and Roderich Moessner and S.L. Sondhi and Vedika Khemani},
	journal   = {{PRX} Quantum},
	title     = {Many-Body Physics in the {NISQ} Era: Quantum Programming a Discrete Time Crystal},
	year      = {2021},
	month     = {sep},
	number    = {3},
	pages     = {030346},
	volume    = {2},
	doi       = {10.1103/prxquantum.2.030346},
	publisher = {American Physical Society ({APS})},
}

@article{Zaletel2023RMP,
	title = {Colloquium: Quantum and classical discrete time crystals},
	author = {Zaletel, Michael P. and Lukin, Mikhail and Monroe, Christopher and Nayak, Chetan and Wilczek, Frank and Yao, Norman Y.},
	journal = {Rev. Mod. Phys.},
	volume = {95},
	issue = {3},
	pages = {031001},
	numpages = {34},
	year = {2023},
	month = {Jul},
	publisher = {American Physical Society},
	doi = {10.1103/RevModPhys.95.031001}
}

@Article{Nandkishore2015,
	author    = {Rahul Nandkishore and David A. Huse},
	journal   = {Annu. Rev. Condens. Matter Phys.},
	title     = {Many-Body Localization and Thermalization in Quantum Statistical Mechanics},
	year      = {2015},
	month     = {mar},
	number    = {1},
	pages     = {15--38},
	volume    = {6},
	doi       = {10.1146/annurev-conmatphys-031214-014726},
	fjournal  = {Annual Review of Condensed Matter Physics},
	publisher = {Annual Reviews},
}

@Article{Morningstar2022a,
	author    = {Alan Morningstar and Luis Colmenarez and Vedika Khemani and David J. Luitz and David A. Huse},
	journal   = {Phys. Rev. B},
	title     = {Avalanches and many-body resonances in many-body localized systems},
	year      = {2022},
	month     = {may},
	number    = {17},
	pages     = {174205},
	volume    = {105},
	doi       = {10.1103/physrevb.105.174205},
	fjournal  = {Physical Review B},
	publisher = {American Physical Society ({APS})},
}

@Article{Ha2023,
	author    = {Hyunsoo Ha and Alan Morningstar and David A. Huse},
	journal   = {Phys. Rev. Lett.},
	title     = {Many-Body Resonances in the Avalanche Instability of Many-Body Localization},
	year      = {2023},
	month     = {jun},
	number    = {25},
	pages     = {250405},
	volume    = {130},
	doi       = {10.1103/physrevlett.130.250405},
	fjournal  = {Physical Review Letters},
	publisher = {American Physical Society ({APS})},
}

@Article{Stasiuk2023,
  author    = {Stasiuk, Andrew and Cappellaro, Paola},
  journal   = {Phys. Rev. X},
  title     = {Observation of a Prethermal ${U}(1)$ Discrete Time Crystal},
  year      = {2023},
  issn      = {2160-3308},
  month     = oct,
  number    = {4},
  pages     = {041016},
  volume    = {13},
  doi       = {10.1103/physrevx.13.041016},
  fjournal  = {Physical Review X},
  publisher = {American Physical Society (APS)},
}

@Article{Deutsch2018,
  author    = {Deutsch, Joshua M},
  journal   = {Rep. Progr. Phys.},
  title     = {Eigenstate thermalization hypothesis},
  year      = {2018},
  issn      = {1361-6633},
  month     = jul,
  number    = {8},
  pages     = {082001},
  volume    = {81},
  doi       = {10.1088/1361-6633/aac9f1},
  fjournal  = {Reports on Progress in Physics},
  publisher = {IOP Publishing},
}

@Article{Huang2023,
  author    = {Huang, Biao},
  journal   = {Phys. Rev. B},
  title     = {Analytical theory of cat scars with discrete time-crystalline dynamics in {F}loquet systems},
  year      = {2023},
  issn      = {2469-9969},
  month     = sep,
  number    = {10},
  pages     = {104309},
  volume    = {108},
  doi       = {10.1103/physrevb.108.104309},
  fjournal  = {Physical Review B},
  publisher = {American Physical Society (APS)},
}

@Article{Maskara2021,
  author    = {Maskara, N. and Michailidis, A. A. and Ho, W. W. and Bluvstein, D. and Choi, S. and Lukin, M. D. and Serbyn, M.},
  journal   = {Phys. Rev. Lett.},
  title     = {Discrete Time-Crystalline Order Enabled by Quantum Many-Body Scars: Entanglement Steering via Periodic Driving},
  year      = {2021},
  issn      = {1079-7114},
  month     = aug,
  number    = {9},
  pages     = {090602},
  volume    = {127},
  doi       = {10.1103/physrevlett.127.090602},
  fjournal  = {Physical Review Letters},
  publisher = {American Physical Society (APS)},
}

@article{Long2023,
  title = {Phenomenology of the Prethermal Many-Body Localized Regime},
  author = {Long, David M. and Crowley, Philip J. D. and Khemani, Vedika and Chandran, Anushya},
  journal = {Phys. Rev. Lett.},
  volume = {131},
  issue = {10},
  pages = {106301},
  numpages = {7},
  year = {2023},
  month = {Sep},
  publisher = {American Physical Society},
  doi = {10.1103/PhysRevLett.131.106301},
  //url = {https://link.aps.org/doi/10.1103/PhysRevLett.131.106301}
}

@article{Sala2020PRX,
  title = {Ergodicity Breaking Arising from {H}ilbert Space Fragmentation in Dipole-Conserving Hamiltonians},
  author = {Sala, Pablo and Rakovszky, Tibor and Verresen, Ruben and Knap, Michael and Pollmann, Frank},
  journal = {Phys. Rev. X},
  volume = {10},
  issue = {1},
  pages = {011047},
  numpages = {19},
  year = {2020},
  month = {Feb},
  publisher = {American Physical Society},
  doi = {10.1103/PhysRevX.10.011047},
  //url = {https://link.aps.org/doi/10.1103/PhysRevX.10.011047}
}

@article{Khemani2020PRB,
  title = {Localization from {H}ilbert space shattering: From theory to physical realizations},
  author = {Khemani, Vedika and Hermele, Michael and Nandkishore, Rahul},
  journal = {Phys. Rev. B},
  volume = {101},
  issue = {17},
  pages = {174204},
  numpages = {17},
  year = {2020},
  month = {May},
  publisher = {American Physical Society},
  doi = {10.1103/PhysRevB.101.174204},
  //url = {https://link.aps.org/doi/10.1103/PhysRevB.101.174204}
}

@Article{Mace2019,
  author    = {Macé, Nicolas and Alet, Fabien and Laflorencie, Nicolas},
  journal   = {Phys. Rev. Lett.},
  title     = {Multifractal Scalings Across the Many-Body Localization Transition},
  year      = {2019},
  issn      = {1079-7114},
  month     = oct,
  number    = {18},
  pages     = {180601},
  volume    = {123},
  doi       = {10.1103/physrevlett.123.180601},
  fjournal  = {Physical Review Letters},
  publisher = {American Physical Society (APS)},
}

@Article{Sierant2024,
title = {Many-body localization in the age of classical computing},
volume = {88},
number = {2},
journal = {Rep. Prog. Phys.},
fjournal   = {Reports on Progress in Physics},
doi = {10.1088/1361-6633/ad9756},
author = {Sierant, Piotr and Lewenstein, Maciej and Scardicchio, Antonello and Vidmar, Lev and Zakrzewski, Jakub},
year = {2025},
month = jan,
publisher = {IOP Publishing},
pages = {026502},
//url = {https://doi.org/10.1088/1361-6633/ad9756},
}

@Article{Yao2023,
  author    = {Yao, Yunyan and Xiang, Liang and Guo, Zexian and Bao, Zehang and Yang, Yong-Feng and Song, Zixuan and Shi, Haohai and Zhu, Xuhao and Jin, Feitong and Chen, Jiachen and Xu, Shibo and Zhu, Zitian and Shen, Fanhao and Wang, Ning and Zhang, Chuanyu and Wu, Yaozu and Zou, Yiren and Zhang, Pengfei and Li, Hekang and Wang, Zhen and Song, Chao and Cheng, Chen and Mondaini, Rubem and Wang, H. and You, J. Q. and Zhu, Shi-Yao and Ying, Lei and Guo, Qiujiang},
  journal   = {Nat. Phys.},
  title     = {Observation of many-body {F}ock space dynamics in two dimensions},
  year      = {2023},
  issn      = {1745-2481},
  month     = jul,
  number    = {10},
  pages     = {1459--1465},
  volume    = {19},
  doi       = {10.1038/s41567-023-02133-0},
  publisher = {Springer Science and Business Media LLC},
}

@Article{DeTomasi2021,
  author    = {De Tomasi, Giuseppe and Khaymovich, Ivan M. and Pollmann, Frank and Warzel, Simone},
  journal   = {Phys. Rev. B},
  title     = {Rare thermal bubbles at the many-body localization transition from the {F}ock space point of view},
  year      = {2021},
  issn      = {2469-9969},
  month     = jul,
  number    = {2},
  pages     = {024202},
  volume    = {104},
  doi       = {10.1103/physrevb.104.024202},
  publisher = {American Physical Society (APS)},
}

@Article{Ho2023,
  author    = {Ho, Wen Wei and Mori, Takashi and Abanin, Dmitry A. and Dalla Torre, Emanuele G.},
  journal   = {Ann. Phys.},
  title     = {Quantum and classical {F}loquet prethermalization},
  year      = {2023},
  issn      = {0003-4916},
  month     = jul,
  pages     = {169297},
  volume    = {454},
  doi       = {10.1016/j.aop.2023.169297},
  publisher = {Elsevier BV},
}

@Article{He2023,
  author    = {He, Guanghui and Ye, Bingtian and Gong, Ruotian and Liu, Zhongyuan and Murch, Kater W. and Yao, Norman Y. and Zu, Chong},
  journal   = {Phys. Rev. Lett.},
  title     = {Quasi-{F}loquet Prethermalization in a Disordered Dipolar Spin Ensemble in Diamond},
  year      = {2023},
  issn      = {1079-7114},
  month     = sep,
  number    = {13},
  pages     = {130401},
  volume    = {131},
  doi       = {10.1103/physrevlett.131.130401},
  publisher = {American Physical Society (APS)},
}

@Article{Else2020a,
  author    = {Else, Dominic V. and Ho, Wen Wei and Dumitrescu, Philipp T.},
  journal   = {Phys. Rev. X},
  title     = {Long-Lived Interacting Phases of Matter Protected by Multiple Time-Translation Symmetries in Quasiperiodically Driven Systems},
  year      = {2020},
  issn      = {2160-3308},
  month     = may,
  number    = {2},
  pages     = {021032},
  volume    = {10},
  doi       = {10.1103/physrevx.10.021032},
  publisher = {American Physical Society (APS)},
}

@Article{Abanin2017,
  author    = {Abanin, Dmitry A. and De Roeck, Wojciech and Ho, Wen Wei and Huveneers, François},
  journal   = {Phys. Rev. B},
  title     = {Effective {H}amiltonians, prethermalization, and slow energy absorption in periodically driven many-body systems},
  year      = {2017},
  issn      = {2469-9969},
  month     = jan,
  number    = {1},
  pages     = {014112},
  volume    = {95},
  doi       = {10.1103/physrevb.95.014112},
  publisher = {American Physical Society (APS)},
}

@Article{Moudgalya2022,
  author    = {Moudgalya, Sanjay and Bernevig, B Andrei and Regnault, Nicolas},
  journal = {Rep. Prog. Phys.},
  fjournal   = {Reports on Progress in Physics},
  title     = {Quantum many-body scars and {H}ilbert space fragmentation: a review of exact results},
  year      = {2022},
  issn      = {1361-6633},
  month     = jul,
  number    = {8},
  pages     = {086501},
  volume    = {85},
  doi       = {10.1088/1361-6633/ac73a0},
  publisher = {IOP Publishing},
}

@Misc{Han2024,
  author    = {Han, Yiqiu and Chen, Xiao and Lake, Ethan},
      title={Exponentially slow thermalization and the robustness of {H}ilbert space fragmentation}, 
      author={Yiqiu Han and Xiao Chen and Ethan Lake},
      year={2024},
      eprint={2401.11294},
      primaryClass = {quant-ph},
      archivePrefix={arXiv},
}

@article{Xu2023CPL,
doi = {10.1088/0256-307X/40/6/060301},
year = {2023},
month = {jun},
publisher = {Chinese Physical Society and IOP Publishing Ltd},
volume = {40},
number = {6},
pages = {060301},
author = {Shibo Xu and Zheng-Zhi Sun and Ke Wang and Liang Xiang and Zehang Bao and Zitian Zhu and Fanhao Shen and Zixuan Song and Pengfei Zhang and Wenhui Ren and Xu Zhang and Hang Dong and Jinfeng Deng and Jiachen Chen and Yaozu Wu and Ziqi Tan and Yu Gao and Feitong Jin and Xuhao Zhu and Chuanyu Zhang and Ning Wang and Yiren Zou and Jiarun Zhong and Aosai Zhang and Weikang Li and Wenjie Jiang and Li-Wei Yu and Yunyan Yao and Zhen Wang and Hekang Li and Qiujiang Guo and Chao Song and H. Wang and Dong-Ling Deng},
title = {{Digital Simulation of Projective Non-Abelian Anyons with 68 Superconducting Qubits}},
fjournal = {Chinese Phys. Lett.},
journal = {Chin. Phys. Lett.},
}

@article{Bao2024NC,
	title = {Creating and controlling global {Greenberger}-{Horne}-{Zeilinger} entanglement on quantum processors},
	volume = {15},
	issn = {2041-1723},
	doi = {10.1038/s41467-024-53140-5},
	number = {1},
	urldate = {2024-10-21},
	journal = {Nat. Commun.},
	author = {Bao, Zehang and Xu, Shibo and Song, Zixuan and Wang, Ke and Xiang, Liang and Zhu, Zitian and Chen, Jiachen and Jin, Feitong and Zhu, Xuhao and Gao, Yu and Wu, Yaozu and Zhang, Chuanyu and Wang, Ning and Zou, Yiren and Tan, Ziqi and Zhang, Aosai and Cui, Zhengyi and Shen, Fanhao and Zhong, Jiarun and Li, Tingting and Deng, Jinfeng and Zhang, Xu and Dong, Hang and Zhang, Pengfei and Liu, Yang-Ren and Zhao, Liangtian and Hao, Jie and Li, Hekang and Wang, Zhen and Song, Chao and Guo, Qiujiang and Huang, Biao and Wang, H.},
	month = oct,
	year = {2024},
	pages = {8823},
	//url = {https://www.nature.com/articles/s41467-024-53140-5},
}

@article{Ding2022NP,
   title={Enhanced metrology at the critical point of a many-body Rydberg atomic system},
   volume={18},
   ISSN={1745-2481},
   DOI={10.1038/s41567-022-01777-8},
   number={12},
   journal={Nat. Phys.},
   publisher={Springer Science and Business Media LLC},
   author={Ding, Dong-Sheng and Liu, Zong-Kai and Shi, Bao-Sen and Guo, Guang-Can and Mølmer, Klaus and Adams, Charles S.},
   year={2022},
   month=oct, pages={1447–1452},
  //url={http://dx.doi.org/10.1038/s41567-022-01777-8},
}

@inproceedings{Mlmer1993MonteCarlo,
author = {K. M{\o}lmer and K. Berg-S{\o}rensen and Y. Castin and J. Dalibard},
booktitle = {J Opt Soc Am},
journal = {Optical Society of America Annual Meeting},
pages = {MFF1},
title = {A {M}onte {C}arlo wave function method in quantum optics},
year = {1992},
doi = {10.1364/OAM.1992.MFF1},
}

@misc{xu2024mindspore,
      title={MindSpore Quantum: A User-Friendly, High-Performance, and {AI}-Compatible Quantum Computing Framework},
      author={Xusheng Xu and Jiangyu Cui and Zidong Cui and Runhong He and Qingyu Li and Xiaowei Li and Yanling Lin and Jiale Liu and Wuxin Liu and Jiale Lu and others},
      year=2024,
      eprint={2406.17248},
      primaryClass = {quant-ph},
      archivePrefix={arXiv},
}

@misc{qiskit2024,
      title={Quantum computing with {Q}iskit},
      author={Javadi-Abhari, Ali and Treinish, Matthew and Krsulich, Kevin and Wood, Christopher J. and Lishman, Jake and Gacon, Julien and Martiel, Simon and Nation, Paul D. and Bishop, Lev S. and Cross, Andrew W. and Johnson, Blake R. and Gambetta, Jay M.},
      year={2024},
      eprint={2405.08810},
      primaryClass = {quant-ph},
      archivePrefix={arXiv}
}

@misc{liu2025,
      title={On the criticality of the configuration-space statistical geometry}, 
      author={Yu-Jing Liu and Wen-Yu Su and Yong-Feng Yang and Nvsen Ma and Chen Cheng},
      year={2025},
      eprint={2508.00787},
      primaryClass={cond-mat},
      archivePrefix={arXiv},
}

@article{Srednicki1994PRE,
  title = {Chaos and quantum thermalization},
  author = {Srednicki, Mark},
  journal = {Phys. Rev. E},
  volume = {50},
  issue = {2},
  pages = {888--901},
  numpages = {0},
  year = {1994},
  month = {Aug},
  publisher = {American Physical Society},
  doi = {10.1103/PhysRevE.50.888},
  //url = {https://link.aps.org/doi/10.1103/PhysRevE.50.888}
}

@article{Luca2016,
	doi = {10.1080/00018732.2016.1198134},
	year = 2016,
	month = {may},
	publisher = {Informa {UK} Limited},
	volume = {65},
	number = {3},
	pages = {239--362},
	author = {Luca D'Alessio and Yariv Kafri and Anatoli Polkovnikov and Marcos Rigol},
	title = {From quantum chaos and eigenstate thermalization to statistical mechanics and thermodynamics},
        journal = {Adv. Phys.},
	fjournal = {Advances in Physics},
 //url = {https://doi.org/10.1080%2F00018732.2016.1198134},
}

@article{Iemini2024PRA,
  title = {Floquet time crystals as quantum sensors of ac fields},
  author = {Iemini, Fernando and Fazio, Rosario and Sanpera, Anna},
  journal = {Phys. Rev. A},
  volume = {109},
  issue = {5},
  pages = {L050203},
  numpages = {6},
  year = {2024},
  month = {May},
  publisher = {American Physical Society},
  doi = {10.1103/PhysRevA.109.L050203},
  //url = {https://link.aps.org/doi/10.1103/PhysRevA.109.L050203}
}

@Article{He2025,
  author    = {He, Guanghui and Ye, Bingtian and Gong, Ruotian and Yao, Changyu and Liu, Zhongyuan and Murch, Kater W. and Yao, Norman Y. and Zu, Chong},
  journal   = {Phys. Rev. X},
  title     = {Experimental Realization of Discrete Time Quasicrystals},
  year      = {2025},
  issn      = {2160-3308},
  month     = mar,
  number    = {1},
  pages     = {011055},
  volume    = {15},
  doi       = {10.1103/physrevx.15.011055},
  publisher = {American Physical Society (APS)},
}

@Article{Porter1956,
  author    = {Porter, C. E. and Thomas, R. G.},
  journal   = {Phys. Rev.},
  fjournal   = {Physical Review},
  title     = {Fluctuations of Nuclear Reaction Widths},
  year      = {1956},
  issn      = {0031-899X},
  month     = oct,
  number    = {2},
  pages     = {483--491},
  volume    = {104},
  doi       = {10.1103/physrev.104.483},
  publisher = {American Physical Society (APS)},
}

@Article{Boixo2018,
  author    = {Boixo, Sergio and Isakov, Sergei V. and Smelyanskiy, Vadim N. and Babbush, Ryan and Ding, Nan and Jiang, Zhang and Bremner, Michael J. and Martinis, John M. and Neven, Hartmut},
  journal   = {Nat. Phys.},
  fjournal   = {Nature Physics},
  title     = {Characterizing quantum supremacy in near-term devices},
  year      = {2018},
  issn      = {1745-2481},
  month     = apr,
  number    = {6},
  pages     = {595--600},
  volume    = {14},
  doi       = {10.1038/s41567-018-0124-x},
  publisher = {Springer Science and Business Media LLC},
}

@Article{Claeys2025,
  author    = {Claeys, Pieter W. and De Tomasi, Giuseppe},
  journal   = {Phys. Rev. Lett.},
  title     = {Fock-Space Delocalization and the Emergence of the Porter-Thomas Distribution from Dual-Unitary Dynamics},
  year      = {2025},
  issn      = {1079-7114},
  month     = feb,
  number    = {5},
  pages     = {050405},
  volume    = {134},
  doi       = {10.1103/physrevlett.134.050405},
  publisher = {American Physical Society (APS)},
}

@article{Deutsch1991PRA,
  title = {Quantum statistical mechanics in a closed system},
  author = {Deutsch, J. M.},
  journal = {Phys. Rev. A},
  volume = {43},
  issue = {4},
  pages = {2046--2049},
  numpages = {0},
  year = {1991},
  month = {Feb},
  publisher = {American Physical Society},
  doi = {10.1103/PhysRevA.43.2046},
  //url= {https://link.aps.org/doi/10.1103/PhysRevA.43.2046}
}

@ARTICLE{Rigol2008Nature,
       author = {{Rigol}, Marcos and {Dunjko}, Vanja and {Olshanii}, Maxim},
        title = "{Thermalization and its mechanism for generic isolated quantum systems}",
      journal = {Nature},
         year = 2008,
        month = apr,
       volume = {452},
       number = {7189},
        pages = {854-858},
          doi = {10.1038/nature06838}
}
	
\end{document}


\nolinenumbers	
\title{Supplementary Information for\\
``Fock space prethermalization and time-crystalline order on a quantum processor''
}
\maketitle

\tableofcontents
\beginsupplement

\section{Device characterization and benchmarking}
\begin{figure}[t]
    \begin{center}
   \includegraphics[width=0.95 \textwidth]{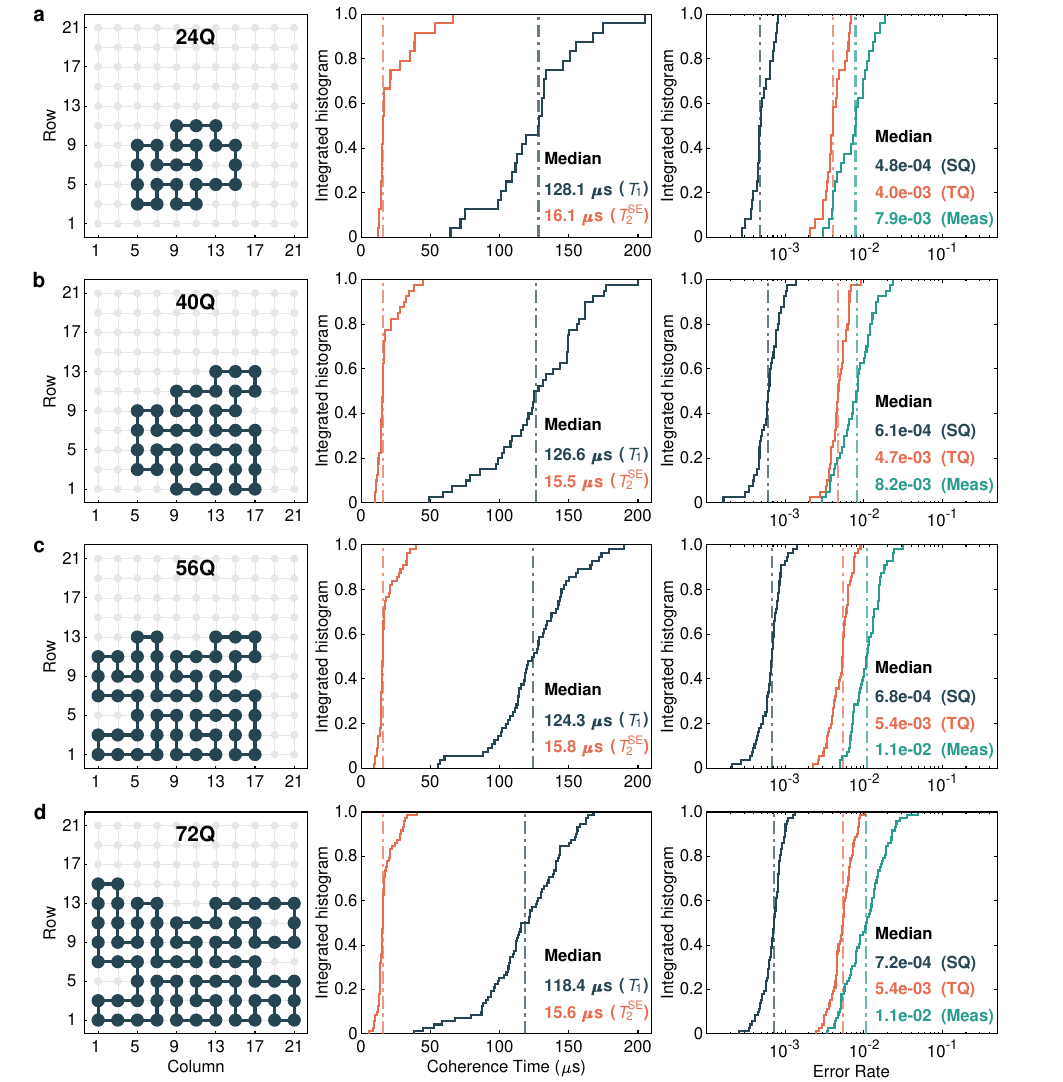} 
    \end{center}
    \caption{\label{fig:dev_info}{\bf Device performance for different system sizes.} {\bf a-d,} Qubit layout and characteristic performance for system sizes of $L=$ 24, 40, 56, and 72. Left panels: Illustration of qubit ring on our processor. Active qubits and couplers are represented by dark dots and lines, respectively. Center panels: Distributions of qubit energy relaxation times ($T_1$) and spin-echo dephasing times ($T_2^{\rm SE}$). Right panels: Distributions of single-qubit gate errors (SQ), two-qubit gate errors (TQ), and readout errors (Meas). Vertical dashed lines denote the median values.}
\end{figure}

\subsection{Device information}
The quantum processor used in our experiments comprises 121 frequency-tunable transmon qubits arranged in an $11\times 11$ square lattice. These qubits are connected by 220 tunable couplers, which are inserted between the neighboring qubits~\cite{Xu2023CPL}. To conduct scaling experiments, we construct four rings of varying system sizes ($L=24$, $40$, $56$, and $72$) on the processor, which are depicted in the left panels of Fig.~\ref{fig:dev_info}. Typical performance of these qubits is illustrated in the middle and right panels. Taking the largest system size (72 qubits, Fig.~\ref{fig:dev_info}{\bf d}) as an example, the median values of the energy relaxation time $T_1$ and the spin-echo dephasing time $T_2^{\rm SE}$ are $118.4 \, \mu$s and $15.6 \, \mu $s, respectively. 

Single-qubit operations are implemented using 20-ns microwave pulses. Error rates for these gates are calibrated via parallel cross-entropy benchmarking (XEB)~\cite{Arute2019Nature} with nine different depths, reaching up to 600 cycles. For two-qubit operations, we employ 44-ns controlled phase (CPhase) gates to realize ZZ interactions; additional experimental details can be found in Ref.~\cite{Bao2024NC}. Error rates for CPhase gates are measured through parallel XEB interleaved with CPhase layers used in the experiments, using seven different depths up to 300 cycles. All XEB results presented use 20 random circuit instances. For qubit readout, we apply an additional microwave pulse at the frequency of $|1\rangle \leftrightarrow|2\rangle$ transition before the measurement pulse. Readout errors are calibrated by preparing random bitstring states and measuring the probability of incorrect labeling for each qubit. Using the methods described above, we obtain median Pauli errors of $0.072\%$ for parallel single-qubit gates, $0.54\%$ for parallel two-qubit gates, and a readout error rate of $1.1\%$ in the 72-qubit system.

\subsection{Context-aware gate optimization}

\begin{figure}[t]
    \includegraphics[width=18cm]{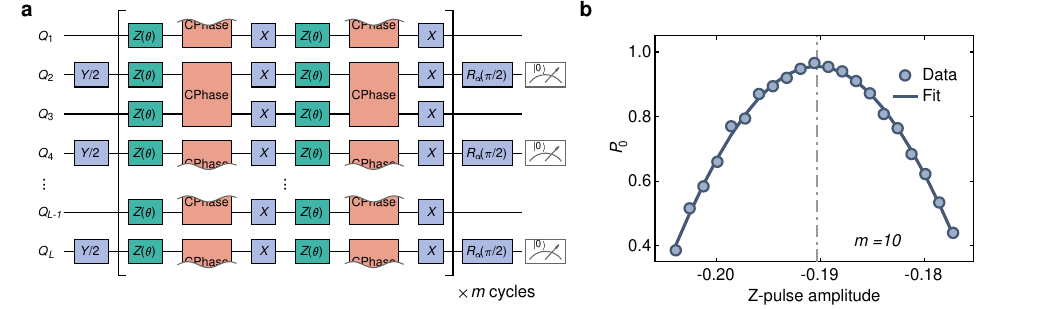}
    \caption{{\bf Calibration of conditional phases for CPhase gates.} 
    {\bf a,} Quantum circuit for calibrating conditional phases of CPhase gates. Z pulse amplitudes of real $Z(\theta)$ gates are the same as that in the other CPhase layer.
    {\bf b,} Example of experimental data displaying the $|0\rangle$-state probability of qubit $Q_2$ as a function of the Z-pulse amplitude applied to $Q_2$, with a fixed $m=10$ cycles. The gray dashed line denotes the optimized Z pulse amplitude after calibration.}
    \label{fig2}
\end{figure}

Although we have carefully calibrated parameters for the parallel CPhase gate with the Floquet calibration method~\cite{Mi2022,Bao2024NC}, we observe a systematic discrepancy between the results predicted by XEB fidelity measured for each individual circuit layer and the experimental outcome of the overall composite circuit. This phenomenon, recently reported in Ref.~\cite{Morvan2024Nature}, is attributed to distortions in qubit flux pulses that induce a slow settling response of qubit frequencies. Ref.~\cite{Morvan2024Nature} mitigates such an effect by inserting idling time between gates, but it is not a perfect solution as idling time introduces extra decoherence errors. In our experiments, we develop a context-aware calibration strategy aiming at considering the gate layers in the context of experimental circuits rather than focusing on individual gate operations in isolation.

There are two different CPhase layers for each Floquet cycle in our model. In addition to the conventional individual calibration for each CPhase layer, we further optimize the conditional phases of each CPhase layer under the context that the control pulses for qubits in the other CPhase layer are also applied. In detail, the calibration circuit is illustrated in Fig.~\ref{fig2}{\bf a}: Orange rectangles represent the CPhase gates in the target layer, while green rectangles denote real $Z(\theta)$ gates. These $Z(\theta)$ gates are implemented by executing the pulses of the CPhase gate from the other CPhase layer, but with all coupler pulses removed. For each cycle in Fig.~\ref{fig2}{\bf a}, the phases accumulated by the two $Z(\theta)$ layers are canceled out due to the inserted $X$-gate layer. Meanwhile, the target qubits initialized via $Y/2$ pulses acquire a conditional phase $\phi$ on their ground state component. After $m$-cycle evolution, a final $\pi/2$ rotation pulse [$R_\alpha(\pi/2)$] is applied to the target qubit. This pulse rotates the target qubit's state by $\pi/2$ around the axis with an angle $\alpha = -m\phi - \pi/2$ to the $x$ axis on the $xy$ plane. By sweeping the applied Z-pulse amplitude of the target qubit for realizing the CPhase gate and identifying its value that maximizes the measured $|0\rangle$-state probability for the target qubit, we can determine the optimized Z-pulse amplitude to realize the conditional phase $\phi$ in the context of our experimental circuits. Exemplary data from this sweep are shown in Fig.~\ref{fig2}{\bf b}.

\begin{figure}[h]
    \includegraphics[width=18cm]{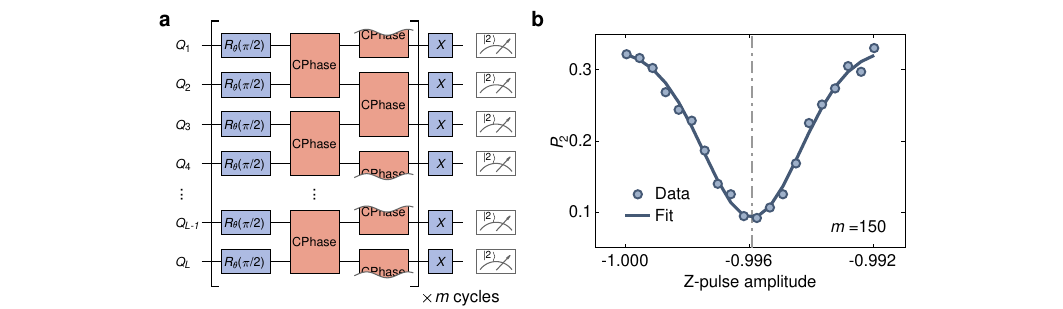}
    \caption{{\bf Calibration of 
    CPhase gates to minimize state leakage.} 
    {\bf a,} Quantum circuit to minimize the state leakage to higher energy levels.  $R_{\theta}(\pi/2)$ denote $\pi/2$ rotations around a random axis on the $xy$ plane with a $\theta$ angle to the $x$ axis.
    {\bf b,} Measured $P_2$ (the sum of the |2⟩-state probabilities of the two qubits involved in a CPhase gate) versus the coupler's Z-pulse amplitude with $m=150$ cycles. The calibrated amplitude, which minimizes leakage, is marked by the gray dashed line.}
    \label{fig3}
\end{figure}

After calibrating the conditional phases, we next suppress the state leakage to higher energy levels. The leakage calibration sequence (Fig.~\ref{fig3}{\bf a} is similar to the XEB circuit but incorporates the measurement of the $|2\rangle$ state. The circuit consists of $m$ cycles, each containing a layer of random single-qubit $\pi/2$ [$R_\theta(\pi/2)$] rotations, followed by two layers of CPhase gates. The $|2\rangle$-state probability is measured at the end of the sequence. For each CPhase gate, we sweep the Z-pulse amplitudes of couplers in this layer in parallel to minimize the state leakage. For each CPhase gate, its state leakage is quantified as the sum of the $|2\rangle$-state probabilities ($P_2$) of the two qubits involved in this gate. We repeat this optimization procedure iteratively for the two CPhase layers several times.
We note that after optimization, $P_2$ remains non-zero (example data are shown in Fig.~\ref{fig3}{\bf b} due to residual state leakage during the gate operation).

\begin{figure}[t]
    \includegraphics[width=18cm]{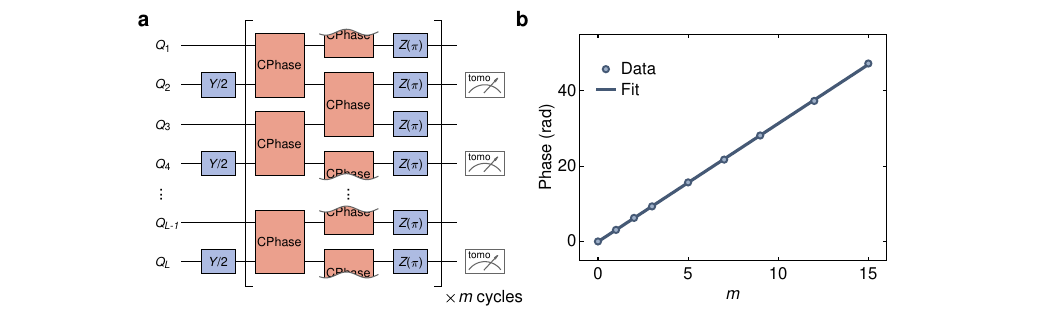}
    \caption{{\bf Calibration of single-qubit phases for CPhase gates.} 
    {\bf a,} Quantum circuit for calibrating single-qubit phases of CPhase gates. Only qubits initialized by $Y/2$ gates are applied tomographic pulses to obtain expectation values of $\langle X\rangle$ and $\langle Y\rangle$. The phase is obtained through the formula $\arctan(\langle Y\rangle/\langle X\rangle)$.
    {\bf b,} Example of experimental data showing the measured phase as a function of the cycle number $m$. The solid line represents a linear fit, and its slope determines the correction for the single-qubit phase.}
    \label{fig4}
\end{figure}

We iteratively calibrate the controlled phases and the state leakage using the above procedures until the parameters converge, then we perform a final calibration for the single-qubit phases of CPhase gates. To decouple the single-qubit phases from the conditional phases, we divide all qubits into two groups (an exemplary group is shown in Fig.~\ref{fig4}{\bf a}), and calibrate them separately. The target qubits in each group are initialized with $Y/2$ pulses and are applied state tomography pulses at the end of circuits. For more precise phase estimation, we measure up to 15 cycles using a the sparse set of circuit depths. Representative data from the calibrated correction of single-qubit gates is shown in Fig.~\ref{fig4}{\bf b}.

\begin{figure}[h]
    \includegraphics[width=18cm]{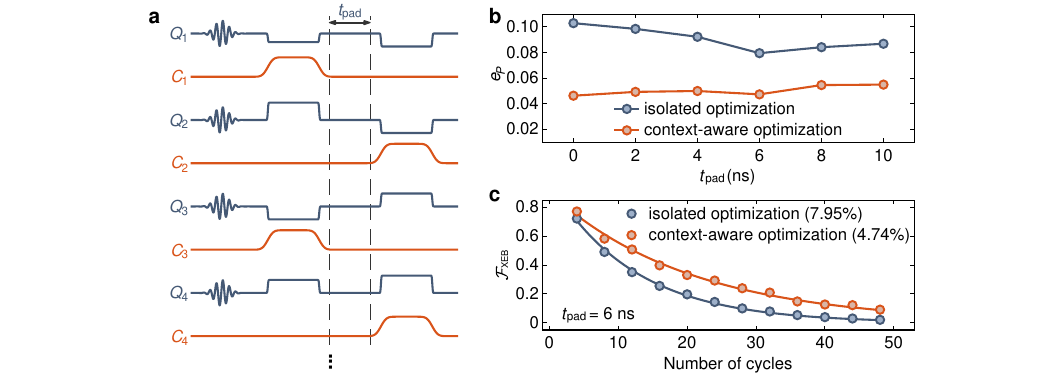}
    \caption{{\bf Experimental comparison between context-aware optimization and isolated optimization.} 
    {\bf a,} Pulse sequence for a single XEB cycle.
    {\bf b,} Measured Pauli error per cycle ($e_P$) as a function of padding time $t_{\rm pad}$.
    {\bf c,} Exemplary XEB data for a fixed pad time of 6 ns.
    }
    \label{fig5}
\end{figure}

We benchmark the effectiveness of our context-aware calibration protocol using multi-qubit XEB on a 12-qubit ring. The pulse sequence for a single XEB cycle is illustrated in Fig.~\ref{fig5}{\bf a}. It comprises one layer of single-qubit gates, followed by two successive CPhase layers with a padding time $t_{\text{pad}}$ inserted between them. We compare two optimization strategies, termed context-aware optimization and isolated optimization. The former corresponds to the method we introduce above, where the control parameters of one CPhase layer are optimized under the context of the experimental circuit (i.e., accounting for the influence of the other CPhase layer). The latter optimizes each CPhase layer in isolation, without considering any cross-layer interactions. Figure~\ref{fig5}{\bf b} and {\bf c} compare the measured Pauli error per cycle ($e_P$) for the two strategies. It is evident that context-aware optimization yields lower Pauli error and shows weaker dependence on the padding time $t_{\rm pad}$, key evidence of our context-aware calibration.

\section{Theoretical details}

\subsection{Theoretical model and the introduction of generic perturbation}

The model in our main text reads
\begin{align} \label{eq:uf}
	U_{\rm F} = \underbrace{e^{-{\rm i} J \sum_j \tilde{\sigma}^z_j(\lambda_2)\tilde{\sigma}^z_{j+1}(\lambda_2)}}_{\text{perturbed Ising interaction}}\underbrace{U_{\rm p}(\varphi_1,\lambda_1,\varphi_2)e^{-{\rm i}\pi\sum_j \sigma^x_j/2}}_{\text{single-qubit driving}},
\end{align}
where the single-qubit perturbation is represented by three Euler angles $(\varphi_1, \lambda_1, \varphi_2)$ for rotations
\begin{align} \label{eq:euler}
	U_{\rm p}(\varphi_1,\lambda_1,\varphi_2) = \prod_j e^{-{\rm i} \varphi_1 \sigma^z_j/2} e^{{\rm i} {\lambda_1 \sigma^y_j}/2} e^{-{\rm i} \varphi_2 \sigma^z_j/2},
\end{align}
and the two-qubit perturbation are $\sim \sigma^z_j\sigma^x_{j+1}, \sigma^x_j\sigma^z_{j+1}$ and $\sigma^x_j \sigma^x_{j+1}$ terms. These two-qubit terms are incorporated into the Ising interaction as
\begin{align} \label{eq:2qubitpert}
	e^{-{\rm i}J\sum_j \tilde{\sigma}^z_j \tilde{\sigma}^z_{j+1}} = 
	e^{-{\rm i}J\sum_j \left(\cos^2(\lambda_2) \sigma^z_j\sigma^z_{j+1} + \cos(\lambda_2){\sin(\lambda_2) (\sigma^z_j\sigma^x_{j+1} + \sigma^x_j \sigma^z_{j+1}) + \sin^2(\lambda_2) \sigma^x_j \sigma^x_{j+1} } \right) },
    \end{align}
where  $\tilde{\sigma}^z_j = \cos(\lambda_2) \sigma^z_j + \sin(\lambda_2) \sigma^x_j$.
There are 5 parameters in the model, namely, the Ising interaction strength $J$, longitudinal field strengths $\varphi_1, \varphi_2$, and the single- and two-qubit perturbation strength $\lambda_1$ and $\lambda_2$, respectively.

At the unperturbed anchor point $\lambda_{1}=\lambda_2 =0$,
\begin{align}\label{eq:uf1}
	U_{\rm F}^{(0)} = e^{-{\rm i}\sum_j (J\sigma^z_j \sigma^z_{j+1} + h\sigma^z_j)} e^{-{\rm i}\pi\sum_j \sigma^x_j/2},
	\qquad
	h = \frac{\varphi_1+\varphi_2}{2},
\end{align}
the Floquet unitary can be solved exactly,
\begin{align} \label{eq:uf0}
	U_{\rm F}^{(0)} |E(\ell, \boldsymbol{s})\rangle = e^{{\rm i} E(\ell,\boldsymbol{s})} |E(\ell, \boldsymbol{s}) \rangle,
	\qquad
	|E(\ell,\boldsymbol{s})\rangle &= \frac{1}{\sqrt{2}} \left(
	e^{-{\rm i}h\sum_j (s^j-1/2)}|\boldsymbol{s}\rangle 
	+ (-1)^\ell e^{{\rm i}h\sum_j (s^j-1/2)}|\bar{\boldsymbol{s}}\rangle
	\right),
	\\
	E(\ell, \boldsymbol{s}) &= J \sum_j (2s^j-1) (2s^{j+1}-1) + \pi\ell , 
	\qquad \ell = 0,1;\quad s^j = 0,1
\end{align}
These eigenstates are all made of pairwise cat states of different qubit patterns $|\boldsymbol{s}\rangle = |s^1\rangle \otimes |s^2\rangle \otimes\dots \otimes |s^L\rangle$, $|\bar{\boldsymbol{s}}\rangle = |1-s^1\rangle\otimes\dots\otimes|1-s^L\rangle$. Each pair of cat eigenstates is separated by a spectral gap $\pi$, specified by the quantum number $\ell$. Aside from the spectral pairing, the quasienergy $E(\ell, \boldsymbol{s})$ is determined by the total domain wall (DW) number $W(\boldsymbol{s})$ of a qubit pattern $\boldsymbol{s}$,
\begin{align} \label{eq:ws}
	W(\boldsymbol{s}) = \sum_j s^j(1-s^{j+1}) + (1-s^j)s^{j+1} = 2\sum_j s^j(1-s^{j+1}),
\end{align} 
which counts the number of neighboring qubits being anti-parallel to each other. Then, the quasienergy can be written as
\begin{align} \label{eq:unperturbedE}
	E(\ell,\boldsymbol{s}) = J[L-2W(\boldsymbol{s})] + \pi\ell,
	\qquad
	W=0,2,4,\dots,L.
\end{align}
Note that each total DW number $\{ W(\boldsymbol{s})=w\}$ contains a manifold of $2C_L^w $ different qubit patterns $\boldsymbol{s}$. Thus, the density of states is
\begin{align} \label{eq:dosW}
	\text{dos}(w) = \frac{2}{2^L}C_L^w = \frac{1}{2^{L-1}}\frac{L!}{w!(L-w)!} \xrightarrow{L\gg1}  \frac{2}{\sqrt{\pi L/2}} e^{-\frac{(w-L/2)^2}{L/2}},
\end{align}
which is highest at $w=L/2$.
The only exceptions are the four cat scars~\cite{Huang2023,Bao2024NC} having ferromagnetic (FM) or anti-ferromagnetic (AFM) patterns respectively,
\begin{align}
	|E(\ell,\text{FM})\rangle &= \frac{1}{\sqrt{2}} \left(
	e^{-{\rm i}hL/2} |111\dots1\rangle + (-1)^\ell e^{{\rm i}hL/2}|000\dots 0\rangle 
	\right),
	&&w = 0,
	\\
	|E(\ell,\text{AFM})\rangle &= \frac{1}{\sqrt{2}} \left(
	|1010\dots10 \rangle + (-1)^\ell |0101\dots01\rangle 
	\right),
	&&w = L.
\end{align}
They take extreme values for the DW number $w$ and are non-degenerate. It implies that the FM and AFM are locally stable patterns, and thermalization would chiefly be triggered on their borders.

We aim at characterizing the peculiar thermalization process of such a system. For this purpose, we next describe the design of a generic perturbation, which avoids fine-tuned (non-interacting) localization effects and would render a faithful benchmark for the Fock space prethermalization behaviors in a generic strongly interacting system. 

Starting from the unperturbed limit Eq.~\eqref{eq:uf1}, if we only include a perturbation $\lambda$ in single-qubit flips, the model becomes $ U_{\rm F} = e^{-{\rm i}\sum_j (J\sigma^z_j \sigma^z_{j+1} + h\sigma^z_j)} e^{-{\rm i}(\pi/2+\lambda)\sum_j \sigma^x_j}$. This model faces a dilemma in choosing the value of longitudinal field strength $h$. At $h=0$, it reduces to a \emph{kicked} transverse-field Ising chain, which is quadratic in the fermionic operators, satisfying $\{c_j, c_k^\dagger\}=\delta_{jk}$, via the Jordan-Wigner transformation
\begin{align}\nonumber
	\left.
	\begin{array}{l}
		\sigma^z_j \rightarrow (c_j + c_j^\dagger)e^{{\rm i}\pi\sum_{k=1}^{j-1}c_k^\dagger c_k},
		\\
		\sigma^x_j \rightarrow 2c_j^\dagger c_j - 1
	\end{array}
	\right\}
	\quad\Rightarrow\quad 
	U_{\rm F}(h=0) &= e^{-{\rm i}\sum_j J\sigma^z_j \sigma^z_{j+1}} e^{-{\rm i}(\pi/2+\lambda)\sum_j \sigma^x_j}
	\\ \label{eq:jw}
	&\rightarrow
	e^{-{\rm i}J\sum_j (c_jc_{j+1} + c_j^\dagger c_{j+1}^\dagger + c_jc_{j+1}^\dagger + c_j^\dagger c_{j+1})} e^{-{\rm i}(\pi/2+\lambda)\sum_j (2c_j^\dagger c_j -1)}.
\end{align}
Each factor is a Gaussian (quadratic) fermionic unitary; hence, their product is also Gaussian. After Fourier and Bogoliubov transformations, the Floquet evolution decouples into independent $(k,-k)$ sectors. Thus, the stroboscopic dynamics are \emph{non-interacting}: the integrals of motion are the occupations of Bogoliubov quasiparticles (not the bare $c_k$).
Consequently, a strong longitudinal field $h$ is needed to break the fine-tuned integrability. However, taking a longitudinal field $h\rightarrow\pi/2$   introduces a different fine-tuning: a near-perfect single-qubit echo. To see this, consider the on-site evolution over two Floquet periods, $U_{\rm F}^2$
\begin{align}
	\left(e^{-{\rm i}(\pi/2)\sigma^z_j} e^{-{\rm i}(\pi/2+\lambda)\sigma^x_j} \right)
	\left(e^{-{\rm i}(\pi/2)\sigma^z_j} e^{-{\rm i}(\pi/2+\lambda)\sigma^x_j} \right)
	= 
	\sigma^z_j \sigma^x_j e^{-{\rm i}\lambda\sigma^x_j} 
	\sigma^z_j \sigma^x_j e^{-{\rm i}\lambda\sigma^x_j} 
	= 
	- 
	e^{+{\rm i}\lambda\sigma^x_j} e^{-{\rm i}\lambda\sigma^x_j} = -1,
\end{align}
where we used $\sigma^\mu_j \sigma^\nu_j \sigma^\mu_j = -\sigma^\nu_j, \mu\neq\nu$. Up to a global phase, the two-period propagator becomes the identity, independent of $\lambda$. Hence, a strong longitudinal field $h$ serves as a single-qubit echo to cancel the effect of qubit-flip perturbation $\lambda$ every two periods~\cite{Khemani2019b, Luitz2020}. With only the field $h$ present, the system therefore lacks any parameter regime free from such fine-tuned, non-interacting or echo-type effects, complicating efforts to benchmark the robustness of FSP in genuinely interacting conditions.

\begin{figure}
	[h]
	\includegraphics[width=18cm]{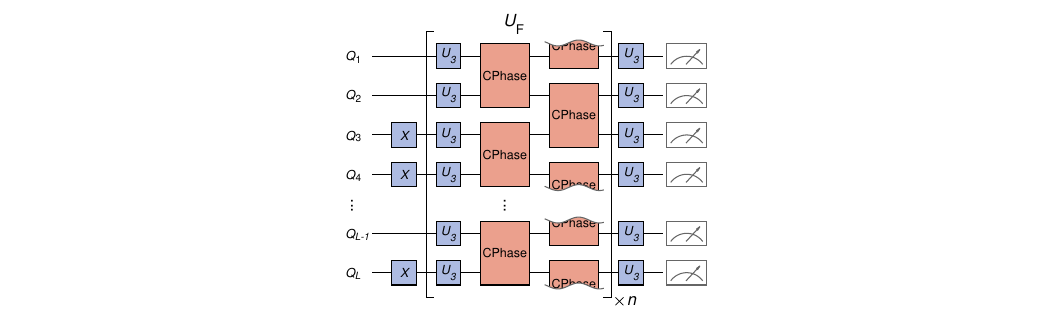}
	\caption{ \label{fig:circuit} Experimental quantum circuit designed for implementing Fock space prethermalization. The circuit consists of three main components: state initialization, Floquet unitary evolution, and measurement. The Floquet unitary $U_{\rm F}$ is constructed from two layers of CPhase gates, sandwiched by two layers of single-qubit gates (i.e., $U_3$ gates). To reduce the total number of circuit layers required in the experiment, we optimize the circuit layout for experimental implementation by merging the final single-qubit gate layer of the previous Floquet unitary with the initial layer of the subsequent one.}
\end{figure}

Such a dilemma is resolved by introducing a more generic single-qubit perturbation in Eq.~\eqref{eq:euler} parameterized by three Euler rotation angles. In such scenario, the single-qubit echo condition becomes
\begin{align}\label{eq:echo}
	\varphi_1=\varphi_2 \mod 2\pi.
\end{align}
With two parameters, we could choose large values for $\varphi_1, \varphi_2$ to break integrability, and make $|\varphi_1-\varphi_2|$ far from 0 to avoid single-qubit echoes. Finally, to distinguish the late-time steady state enforced by FSP from early time transient relaxation, we further introduce the two-qubit terms $\lambda_2$ in Eq.~\eqref{eq:2qubitpert}. Altogether, we could parameterize the strength of generic perturbation with $\lambda_1$ and $\lambda_2$. A schematic of the experimental quantum circuit is presented in Fig.~\ref{fig:circuit}. The protocol begins with state initialization using $\pi$-pulses. This is followed by the periodic application of the Floquet unitary $U_{\rm F}$, which is constructed from layers of parametrized single-qubit $U_3$ gates executed in parallel, interleaved with two CPhase gate layers.

\subsection{Mechanism of Fock space prethermalization (FSP)}

To investigate the FSP mechanism, we first demonstrate that the eigenstructure at small perturbation $\lambda_1, \lambda_2 \ll J$ exhibits DW separations, where each eigenstate approximately carries a good quantum number of the DW number $w$ with exponential accuracy, even under generic perturbation. Then, we next show that within each DW sector, thermalization occurs within finite-size light-cones, while qubits outside the light-cone are frozen. Such inter- and intra-DW constraints postpone the thermalization and give rise to the FSP phenomena.

\begin{figure}
	[h]
	\includegraphics[width=16cm]{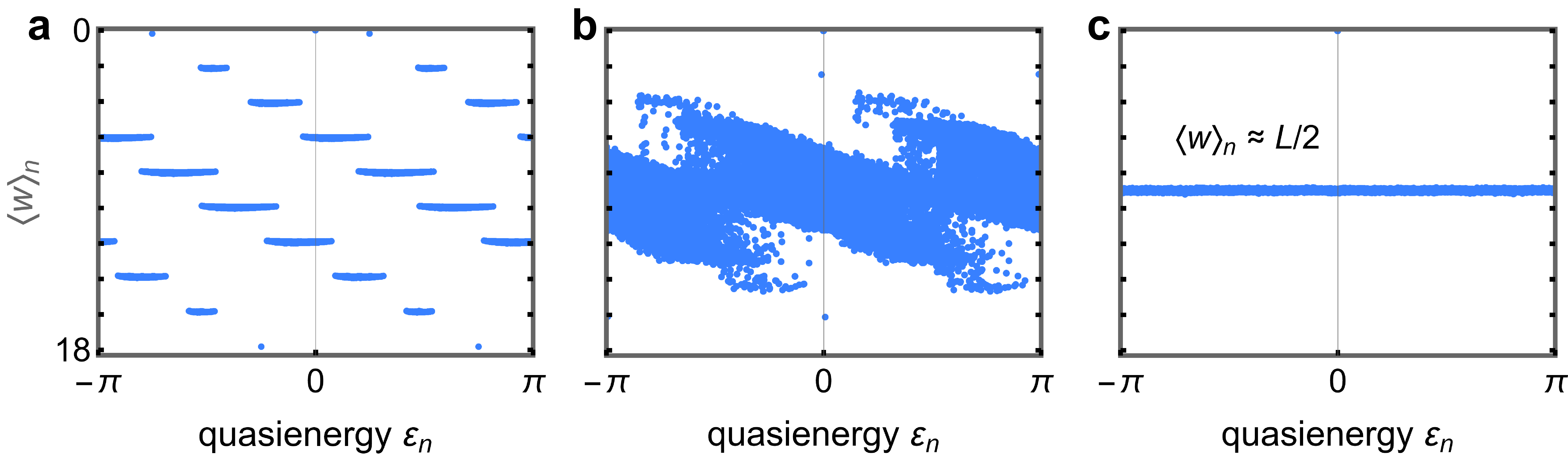}
	\caption{ \label{fig:eigs} Representative eigenstructures characterized by quasienergy $\varepsilon_n$ and the averaged domain wall number $\langle w\rangle_n$ in FSP (panel {\bf a}, $\lambda_1 = 2\lambda_2=0.1$), 
		critical (panel {\bf b}, $\lambda_1 = 2\lambda_2=0.4$), and 
		thermal (panel {\bf c}, $\lambda_1 = 2\lambda_2=1.2$) regimes. In all panels we take Ising interaction $J=1$, the longitudinal fields far detuned from the echos in Eq.~\eqref{eq:echo} ($\varphi_1 = -1.5701, \varphi_2= 0.9708 $). The system size $L=18$, which, under periodic boundary conditions, can host even integer numbers of DWs $W(\boldsymbol{s})=0,2,\dots,18$ for Fock states $|\boldsymbol{s}\rangle$. }
\end{figure}
FSP survives in the regime where the interaction strength $J$ is comparable to the driving frequency $\omega = 2\pi/T$. Under generic perturbation, its eigenstructure $U_{\rm F}|\varepsilon_n\rangle = e^{{\rm i}\varepsilon_n}|\varepsilon_n\rangle $ is shown in Fig.~\ref{fig:eigs}{\bf a}. We see that, unlike the conventional prethermal systems relying on high-frequency driving, the strong $J$ spreads the quasienergy across the whole frequency Brillouin zone $\varepsilon_n\in[-\pi, \pi]$. Meanwhile, for each eigenstate $|\varepsilon_n\rangle$, we calculate their averaged domain wall number 
\begin{align}
	\langle w\rangle_n = \sum_{\boldsymbol{s}} W(\boldsymbol{s})
	|\langle \boldsymbol{s} | \varepsilon_n\rangle|^2,
\end{align}
where $W(\boldsymbol{s})$ for Fock basis states $|\boldsymbol{s}\rangle$ is given in Eq.~\eqref{eq:ws}. As shown in Fig.~\ref{fig:eigs}{\bf a}, in the FSP regime, each eigenstate approximately carries a good quantum number $\langle w\rangle_n$ at an even integer (recall that we take periodic boundary conditions, so the domain wall number for any Fock state $|\boldsymbol{s}\rangle$ is always an even integer). Such an eigenstructure can be traced back to the unperturbed limit Eqs.~\eqref{eq:uf0} and \eqref{eq:unperturbedE}, where each domain wall number $w$ involves two clusters of eigenstates mutually separated by quasienergy $\pi$, while different $w$ sectors are gapped by $4J$. Under perturbation, eigenstates within each $w$ sector undergo degenerate hybridization and expand into a continuous band. However, we have eigenstates $|E(\ell,\boldsymbol{s})\rangle$ with the same domain wall number $W(\boldsymbol{s})=w$ hybridized strongly, such that the resulting perturbed eigenstates $|\varepsilon_n\rangle$ still approximately carries such a quantum number $\langle w\rangle_n\approx w$ from their parental eigenstates $|E(\ell,\boldsymbol{s})\rangle$. Hybridization among $w'\neq w$ sectors is suppressed by the large Ising gap $2J|w'-w|$.

With the increase of perturbation, the FSP system goes through a crossover to the thermal case. In Fig.~\ref{fig:eigs}{\bf b}, we increase the perturbation to a critical regime. As $\lambda_1, \lambda_2$ start to approach the Ising gap size $4J$, hybridization among different $w$ sectors becomes significant, which is reflected by large numbers of eigenstates having $\langle w\rangle_n$ between even integers. Finally, as $\lambda_1, \lambda_2$ becomes comparable to the Ising gap in Fig.~\ref{fig:eigs}{\bf c}, all domain wall sectors are fully mixed, resulting in the typical scenario predicted by the eigenstate thermalization hypothesis where all $\langle w\rangle_n$ are essentially the same.

\begin{figure}
	[h]
	\includegraphics[width=18cm]{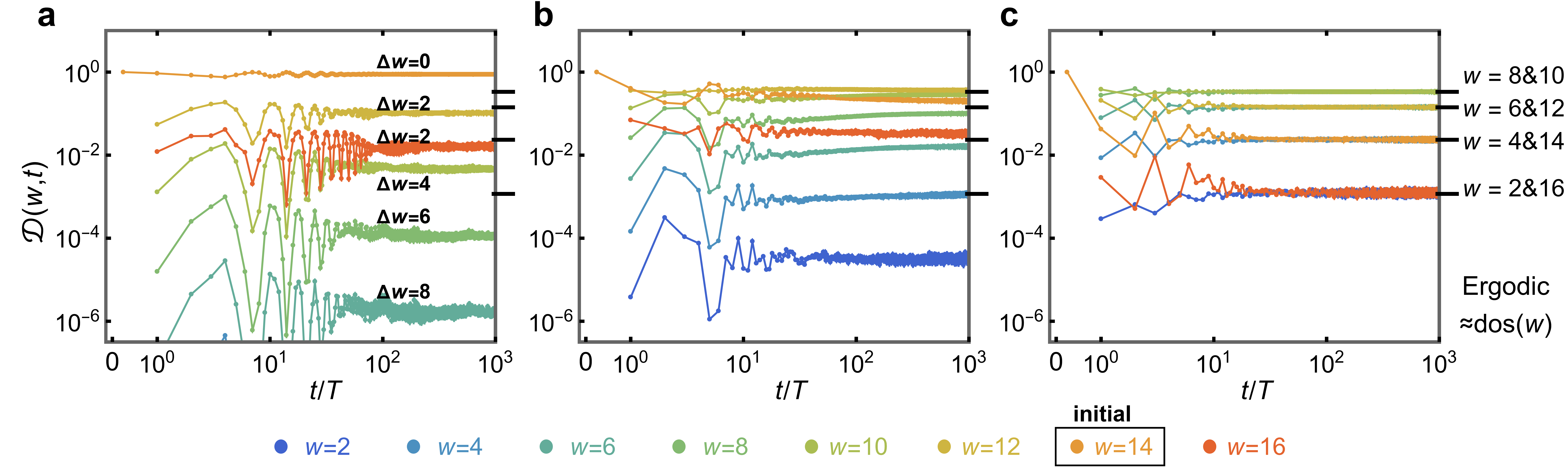}
	\caption{ \label{fig:dw_evo} 
		Evolution of wave function amplitudes projected onto different domain wall sectors. Here, the parameters are the same as those in Fig.~\ref{fig:eigs}, namely, the Ising interaction $J=1$, longitudinal fields $\varphi_1 = -1.5701, \varphi_2 = 0.9708$, and the perturbation strengths $\lambda_1 = 2\lambda_2=0.1$ for FSP ({\bf a}), 0.4 for critical regime ({\bf b}), and 1.2 for thermal regime ({\bf c}).
		For our system size $L=18$, the initial state $|\boldsymbol{s}_0\rangle = |001010100010101010\rangle$ under periodic boundary condition carries $W(\boldsymbol{s}_0)=w_0 = 14$ DWs. 
		We observe that after the oscillations at very early times, in the FSP regime (panel {\bf a}), $\mathcal{D}(w,t)$ exhibits a plateau for different values of $w$. The heights of these plateaus show an exponential localization of wave function onto the original DW sector with $\mathcal{D}(w,t) \sim e^{-\Delta w}, \Delta w = |w-w_0|$. 
		Contrarily, in the thermal regime (panel {\bf c}), such amplitude only reflects the density of states $\text{dos}(w)$ [see Eq.~\eqref{eq:dosW}] marked by black ticks on the right of each panel, indicating an ergodic wave function evenly distributed among all Fock bases.
		}
\end{figure}

The eigenstructure in the FSP regime enforces the exponentially accurate conservation of domain wall numbers for quantum states during evolution. In Fig.~\ref{fig:dw_evo}, we illustrate such a constraint by projecting the evolved state $|\psi(nT)\rangle = U_{\rm F}^{n}|\boldsymbol{s}_0\rangle $ into different domain wall sectors [see Eq.~(4) of the main text]
\begin{align}
	\mathcal{D}(w,nT) = \sum_{\boldsymbol{s}\in \{ W(\boldsymbol{s}) = w \}}  \left|\langle \boldsymbol{s} | U_{\rm F}^n |\boldsymbol{s}_0\rangle \right|^2.
\end{align}
We exemplify the process by evolving an initial state $|\boldsymbol{s}_0\rangle$ with $w_0=W(\boldsymbol{s}_0) = 14$ domain walls in a chain of $L=18$. In the FSP regime (Fig.~\ref{fig:dw_evo}{\bf a}), we see an exponential localization of the wave function onto the initial domain wall sector, where the amplitude shows the late time plateaus are exponentially proportional to the domain wall difference $\Delta w= |w-w_0|$. That is,  $\mathcal{D}(w,t) \sim O(e^{-\Delta w})$.
It is worth noting that the density of state is exponentially large at the domain wall sector $w = L/2 = 9$ [see Eq.~\eqref{eq:dosW}]. The localization onto $w=14$ sector with a tiny density of states is a clear signature for the exponentially suppressed hybridization in eigenstates for Fock bases from different domain wall sectors. Increasing $\lambda_1,\lambda_2$ in Fig.~\ref{fig:dw_evo}{\bf b} to the critical regime, we see that the amplitudes in different domain wall sector start to mix with the exponential localization in Fig.~\ref{fig:dw_evo}{\bf a} broken down at late time. Finally, in the thermal regime of Fig.~\ref{fig:dw_evo}{\bf c}, we see a quick convergence of $\mathcal{D}(w,t) \rightarrow \text{dos}(w)$ indicating fully thermalized, ergodic wave functions.

\begin{figure}
	[h]
	\includegraphics[width=18cm]{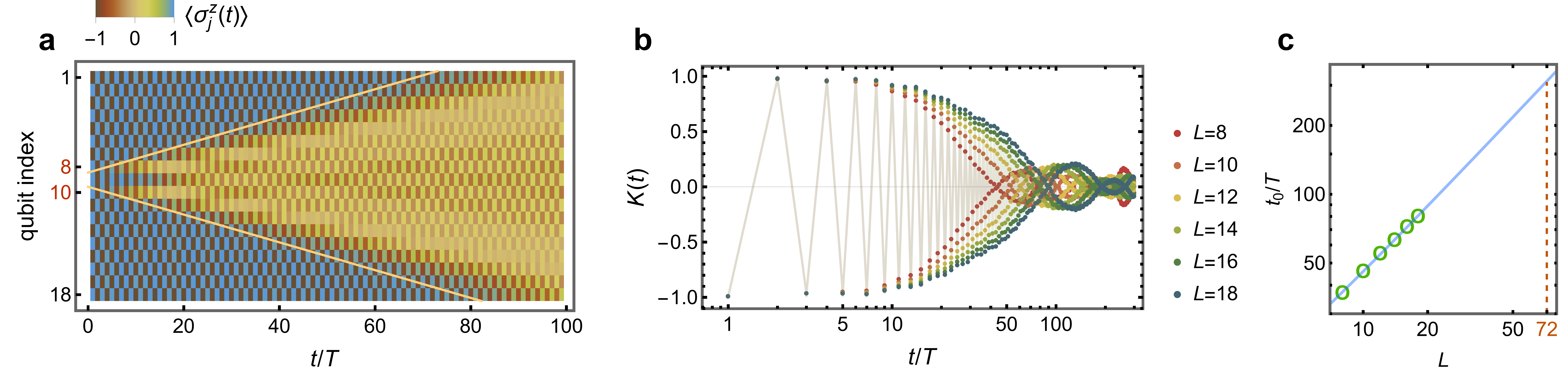}
	\caption{\label{fig:sz}Local magnetic order and light-cone spreading for DTC dynamics in the FSP regime. The initial state is $|\boldsymbol{s}_0\rangle = |10101010{\color{BrickRed}\boldsymbol{0}} 010101010\rangle$, where qubit $9$ is flipped with respect to an overall anti-ferromagnetic pattern.
		{\bf a,} Local magnetic order, where we see a light-cone spreading from the two qubits 8 and 10 neighboring the flipped qubit 9, consistent with the domain wall constraint. 
		Yellow lines are an analytical estimation of light-cone propagation with butterfly velocity given by Eq.~\eqref{eq:vb}.
		{\bf b,} Averaged temporal auto-correlator $K$ indicating both the period-doubled oscillation and the preservation of initial qubit pattern $|\boldsymbol{s}_0\rangle$.
		{\bf c,} The amplitude of $K(t)$ increases linearly with system size $L$, as the thermalizing light-cone propagates with parametrically controlled butterfly velocity, and needs a longer time to spread across the whole system. Circles represent critical times extracted from numerical data with a threshold of $|K(t_0)| < 10\%$, while the blue line, $t_0 = 2.9333 + 4.3L$, is the fit to the data.
	}
\end{figure}
Since the probability for an initial Fock state to ``leak'' from its original domain wall sector is exponentially small, the major dynamical process takes place in a way that conserves domain wall numbers. To visualize such a constraint, in Fig.~\ref{fig:sz} we numerically simulate the evolution of local magnetic order
\begin{align}\label{eq:mag}
	\langle\sigma^z_j(t)\rangle = \langle\boldsymbol{s}_0 |(U_{\rm F}^\dagger)^{t/T} \sigma^z_j U_{\rm F}^{t/T} |\boldsymbol{s}_0\rangle.
\end{align}
Here, since a global FM or AFM pattern resides in the non-degenerate scarred subspace, we flip one qubit (denoted by qubit 9 in Fig.~\ref{fig:sz}) out of an overall AFM background. Intriguingly, we see in Fig.~\ref{fig:sz}{\bf a} that it is the two neighbors (qubits 8 and 10) of the flipped one (qubit 9) that trigger the thermalizing lightcone. This is because the pattern of qubits [7,8,9,10,11] is [10001], which contains two domain walls. Then, qubits 8 and 10 can be freely flipped without changing the local domain wall number. In other words, the flipped qubit 9 creates a 3-qubit mini FM region surrounding it. Since FM and AFM are non-thermal cat scars~\cite{Huang2023, Bao2024NC}, only the qubit on the border of FM and AFM patterns will trigger thermalizing dynamics. The bifurcation of light-cone tips at qubit 9, therefore, underscores the effectiveness of the kinetic constraint preserving domain wall numbers.

We can further quantify the butterfly velocity of such a light-cone based on domain wall conservation~\cite{Bao2024NC},
\begin{align}\nonumber
	&v_{\rm B} = v_{\rm B}^{(1)} + v_{\rm B}^{(2)},
	\qquad
	v_{\rm B}^{(2)} = \sin(J(\beta_2^2+\gamma_2^2)) ,	
	\\ \nonumber			
	&
	v_{\rm B}^{(1)} = \frac{1}{2} \left|
	-(1-\cos(\lambda_1)) \sin(\lambda_2) \sin(\varphi_2-\varphi_1)   
	+ 
	\sin(\lambda_1) [\cos(\lambda_2) (\sin(\varphi_2) - \sin(\varphi_1))
	-{\rm i}(\cos(\varphi_2) - \cos(\varphi_1))] \right|,
	\\ \nonumber
	& 
	\beta_2 = \sin(\lambda_2) \cos(\lambda_2) \left[\sin(\varphi_2)\sin(\varphi_1) - \cos(\lambda_1)(1+\cos(\varphi_2)\cos(\varphi_1)) \right]
	-
	\sin(\lambda_1) \left[\cos^2 (\lambda_2) \cos(\varphi_1) - \sin^2 (\lambda_2) \cos(\varphi_2) \right],
	\\ \label{eq:vb}
	&
	\gamma_2 = -(\sin(\lambda_1)\cos(\lambda_2) + \cos(\lambda_1)\sin(\lambda_2) \cos(\varphi_2)) \sin(\varphi_1) 
	- 
	\sin(\lambda_2) \sin(\varphi_2) \cos(\varphi_1).
\end{align}
Such a butterfly velocity $v_{\rm B}$ is obtained by assuming that during the dynamical process, the domain wall number is conserved locally, so that $J\sigma^z_j\sigma^z_{j+1}$ can be replaced by a classical constant number, reducing $U_{\rm F}$ to single-qubit drivings. Furthermore, suppose we approximately assume that the qubit flip perturbation coherently accumulates over time, and that consecutive qubits must flip sequentially. In that case, we have the butterfly velocity $v_{\rm B} = 1/\tau$, where $\tau$ is the time it takes to completely flip a qubit by onsite perturbation. See Fig.~\ref{fig:c1j} for more benchmarks of the analytical butterfly velocity against results with parameters used in the experiments.

The butterfly velocity in Eq.~\eqref{eq:vb} is only controlled by model parameters. That means if we extend the system size $L$, the light-cone needs a longer time $\propto L$ to spread across the whole system, allowing for a longer lifetime of a discrete time crystal (DTC) with an inhomogeneous spatial pattern. To verify such an expectation, we calculate the temporal autocorrelator
\begin{align} \label{eq:autocorr}
    K(t) = \frac{1}{L} \sum_{j=1}^L \langle \boldsymbol{s}_0 | \sigma^z_j(t) \sigma^z_j |\boldsymbol{s}_0\rangle,
	\qquad
	\sigma^z_j(t) = (U_{\rm F}^\dagger)^{t/T} \sigma^z_j U_{\rm F}^{t/T}.
\end{align}
The amplitude $|K(t)|$ measures the fraction of the initial qubit pattern that are preserved during evolution. We observe in Fig.~\ref{fig:sz}{\bf b} that, indeed, with increasing system size $L$, a linear growth of the DTC lifetime is obtained, with the scaling confirmed in Fig.~\ref{fig:sz}{\bf c}.

It is of interest to compare the FSP we introduce here with conventional Floquet prethermal behaviors. Floquet prethermalization relying on high-frequency driving typically shows a {\it temporal} suppression of heating, in the sense that each photon from external driving has enormous energy $\omega \gg J$ compared with internal interaction strength. Then, it takes a considerable amount of time for the system to absorb energy from the drive. FSP, instead, exhibits a {\it spatial} suppression of heating, as the thermalizing light-cone can only be triggered at locations satisfying the domain wall constraint, and propagate with a finite butterfly velocity. Spacetime points outside the light cones are immune to perturbations and can show rigid DTC oscillations with inhomogeneous local magnetic orders. In the following section, we elaborate on the different phenomenological differences between FSP and conventional high-frequency Floquet prethermalization.

\subsection{Comparison with conventional prethermal mechanism}

In this section, we compare the DTC oscillations enforced by FSP in our work with the DTCs protected by conventional high-frequency Floquet prethermalization. To be concrete, we focus on the one-dimensional short-range interacting cases relevant to our experiments on superconducting qubits. In such scenarios, DTCs enforced by Landau's spontaneous symmetry breaking~\cite{Kyprianidis2021, Else2017} cannot occur for Ising models due to Mermin-Wagner types of no-go theorems. Nevertheless, it is possible for an alternative version, dubbed prethermal U(1) DTC~\cite{Stasiuk2023, Luitz2020}, to exist. We therefore focus on comparing the different behaviors of DTC dynamics in FSP and prethermal U(1) cases.

We adopt the prethermal U(1) DTC model from Ref.~\cite{Luitz2020}, which we rewrite in the following way to facilitate comparison,
\begin{align}\nonumber
	U_{\rm F} &= 
	e^{{\rm i} \theta_h \sum_j \sigma^z_j/2 }
	e^{-{\rm i}(\pi+\epsilon)\sum_j \sigma^x_j/2}
	e^{-{\rm i}JH_{\text{XXZ}}},
	\\ \label{eq:u1}
	H_{\text{XXZ}} &=  \sum_j \left[  
	1\times \left( 
	\frac{1}{2}(\sigma^x_j \sigma^x_{j+1} + \sigma^y_j\sigma^y_{j+1}) 
	-
	(\sigma^z_j\sigma^z_{j+1})
	\right) 
	+
	0.5\times \left( 
	\frac{1}{2} (\sigma^x_j \sigma^x_{j+2} + \sigma^y_j\sigma^y_{j+2}) 
	-
	(\sigma^z_j\sigma^z_{j+2})
	\right) 
	\right].
\end{align}
The controlling parameters $(J, \theta_h, \epsilon)$ above corresponds to the notations $(T_1 = 2\pi/\omega, h, \epsilon)$ in Ref.~\cite{Luitz2020} as
\begin{align}
	J = \frac{T_1}{2} = \frac{\pi}{\omega},\qquad
	\theta_h = h T_1,\qquad
	\epsilon = \epsilon.
\end{align}
There are two major differences between this model and ours. 
On the one hand, the high frequency regime $\omega\gg1$ in the above model corresponds to $J\ll1$, unlike our case with internal interactions comparable to the driving frequency $J\sim 1$. On the other hand, the interaction in Eq.~\eqref{eq:u1} is of XXZ type, which induces strong diffusion. In the Fock space perspective, such a diffusive interaction would induce extensive connections among Fock bases, unlike the Ising interaction, which is diagonal in the Fock bases.

The key mechanism for a prethermal U(1) DTC is an almost perfect qubit echo encoded in the longitudinal field
\begin{align}
	\theta_h \rightarrow \pi.
\end{align}
Then, in the limit of vanishing interaction $J\rightarrow0$, the qubit-flip imperfection $\epsilon$ is exactly canceled every two periods
\begin{align} \nonumber
	U_{\rm F}^2 &= 
	\left( e^{-{\rm i}\epsilon\sum_j \sigma^x_j/2} e^{{\rm i}\theta_h \sum_j \sigma^z_j/2}
	\left(\prod_j -{\rm i}\sigma^x_j \right) e^{-{\rm i}JH_{\text{XXZ}}} \right)
	\left( e^{-{\rm i}\epsilon\sum_j \sigma^x_j/2} e^{{\rm i}\theta_h \sum_j \sigma^z_j/2}
	\left(\prod_j -{\rm i}\sigma^x_j \right) e^{-{\rm i}JH_{\text{XXZ}}} \right)
	\\ \nonumber
	&= 
	e^{-{\rm i}\epsilon\sum_j \sigma^x_j/2} e^{{\rm i}\theta_h \sum_j \sigma^z_j/2}
	e^{-{\rm i}JH_{\text{XXZ}}} 
	e^{-{\rm i}\epsilon\sum_j \sigma^x_j/2} e^{-{\rm i}\theta_h \sum_j \sigma^z_j/2}
	e^{-{\rm i}JH_{\text{XXZ}}} 
	\\
	& \xrightarrow{J\rightarrow0, \theta_h\rightarrow \pi}
	e^{-{\rm i}\epsilon\sum_j \sigma^x_j/2}
	\left(\prod_j {\rm i}\sigma^z_j\right)
	e^{-{\rm i}\epsilon\sum_j \sigma^x_j/2}
	\left(\prod_j -{\rm i}\sigma^z_j\right)
	=
	e^{-{\rm i}\epsilon\sum_j \sigma^x_j/2}
	e^{+{\rm i}\epsilon\sum_j \sigma^x_j/2} = 1,
\end{align}
where we have repeatedly used $\sigma^\mu_j \sigma^\nu_j \sigma^\mu_j = -\sigma^\nu_j, \mu\neq \nu$. When $J\neq0$, the local qubit echo is destroyed. But the XXZ interaction preserves {\it global} U(1) symmetry, namely, the global magnetization
\begin{align}
	S^z = \sum_j  \sigma^z_j, 
	\qquad
	[H_{\text{XXZ}}, S^z]=0.
\end{align}
That means the global magnetization $S^z$ is approximately conserved under finite interaction, violated only by the qubit-flip imperfection $\epsilon\sum_j\sigma^x_j$. Then, generally any inhomogeneous initial qubit pattern $\langle\sigma^z_j\rangle$ will quickly diffuse into a homogeneous one contributed by large numbers of Fock states, while the total magnetization $S^z$ will undergo a relatively long-lived period-doubled oscillation for the prethermal U(1) DTC.

We would highlight two major difference in dynamical behavior for prethermal U(1) DTC from our case of FSP.

\begin{figure}
	[h]
	\includegraphics[width=18cm]{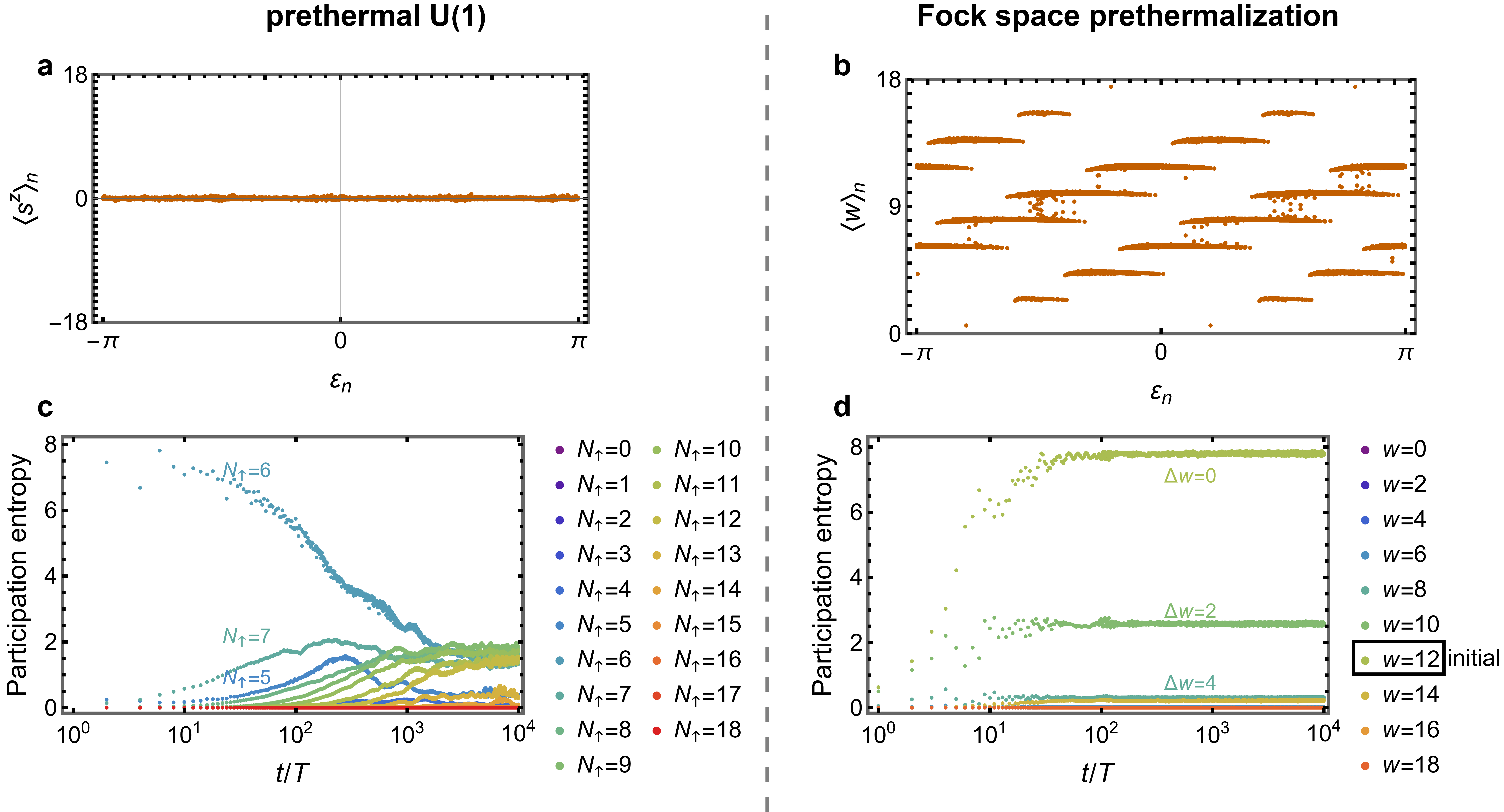}
	\caption{ \label{fig:u1_fsp_eigs} 
		Eigenstructure and participation entropy dynamics for the prethermal U(1) and FSP cases. 
		{\bf a, b,} Eigenstates plotted with their quasienergy $\varepsilon_n$ and the averaged total magnetization for prethermal U(1) case ({\bf a}), and the averaged DW number for FSP case ({\bf b}). 
		{\bf c, d,} Participation entropy dynamics for an evolved initial state $|\boldsymbol{s}_0\rangle = |001001001001001001\rangle$ projected onto the subspace of $N_{\uparrow}$ qubits being $1$ (panel {\bf c}, corresponding to total magnetization $2N_{\uparrow}-L$), and $w$ DW numbers ({\bf d}).
		For the prethermal U(1) case, we take high frequency $\omega=8$ ($J = \pi/\omega = \pi/8 $), longitudinal fields $90\%$ of an echo $\theta_h = 0.9\pi$, and perturbation strength $\epsilon=0.18$ similar to Fig.~3 in Ref.~\cite{Luitz2020}. For FSP situations, we take strong Ising interaction comparable to driving frequency $J=1$, longitudinal fields far-detuned from single-qubit echoes Eq.~\eqref{eq:echo} to showcase genuine many-body prethermalization $\varphi_1 = -1.5701, \varphi_2 = 0.9708$, and perturbation strengths $\lambda_1 = 2\lambda_2 = 0.18$. The simultaneous presence of single- and two-qubit perturbations $\lambda_1,  \lambda_2$ implies a stronger perturbation than $\epsilon$ in the U(1) case.
	}
\end{figure}

{\bf 1, The approximately conserved $S^z$ in U(1) DTC breaks down within timescale $\sim1/\epsilon^2$, while the domain wall number for DTC induced by FSP persists for exponentially long time.} We illustrate this point in Fig.~\ref{fig:u1_fsp_eigs}. 

From the perspective of eigenstructure, which stands for infinite-time response characters, we find that the prethermal U(1) case actually shows the feature consistent with eigenstate thermalization hypothesis, as shown in Fig.~\ref{fig:u1_fsp_eigs}{\bf a}. Here, we choose the parameters that have been shown in Ref.~\cite{Luitz2020} to be deep inside the U(1) DTC phase. Similar to $\langle w\rangle_n$ in FSP situations, for each Floquet eigenstate $U_{\rm F}|\varepsilon_n\rangle = e^{{\rm i}\varepsilon_n}|\varepsilon_n\rangle$ in prethermal U(1) cases we calculate its averaged total magnetization
\begin{align}
	\langle s^z\rangle_n = \langle\varepsilon_n |S^z |\varepsilon_n\rangle.
\end{align}
The fact that all eigenstates yield essentially the same $\langle s^z\rangle_n$ is in line with the ``thermal'' regime in our case as in Fig.~\ref{fig:eigs}{\bf c}. This is to be sharply contrasted with the FSP situation, plotted in Fig.~\ref{fig:u1_fsp_eigs}{\bf b}, where eigenstates separate into clusters with different averaged total domain wall numbers $\langle w\rangle_n$. 

Correspondingly, the timescales for preserving the $S^z$ symmetry in prethermal U(1) cases and domain wall numbers $w$ in FSP cases are qualitatively different. Here, we benchmark their difference using the participation entropy, as adopted in Ref.~\cite{Luitz2020}, for an evolved quantum state $|\psi(t)\rangle = U_{\rm F}^{t/T} |\boldsymbol{s}_0\rangle$,
\begin{align}
	\text{U(1):} \qquad &S[N_{\uparrow},t] = -\sum_{\boldsymbol{s}\in\{ \sum_j s^j = N_{\uparrow}\}} |\langle \boldsymbol{s}|\psi(t)\rangle|^2 \ln|\langle \boldsymbol{s}|\psi(t)\rangle|^2,
	\\
	\text{FSP:} \qquad &  S[w,t] = - \sum_{\boldsymbol{s}\in\{ W(\boldsymbol{s}) = w\}} |\langle \boldsymbol{s}|\psi(t)\rangle|^2 \ln|\langle\boldsymbol{s}|\psi(t)\rangle|^2.
\end{align}
The entropy $S[N_{\uparrow},t]$ concerns the subspace for prethermal U(1) DTC with $N_{\uparrow}$ qubits being $1$, corresponding to the subspace $S^z = N_{\uparrow} - (L-N_{\uparrow})$. Similarly, $S[w,t]$ concerns the total domain wall subspace $w$ for FSP. In Figs.~\ref{fig:u1_fsp_eigs}{\bf c} and \ref{fig:u1_fsp_eigs}{\bf d}, we see a sharp difference between the two. Here, we choose the same initial state $|\boldsymbol{s}_0\rangle = |001001001001001001\rangle$ with $L=18$ for both cases, similar to Fig.~3 in Ref.~\cite{Luitz2020}. For the prethermal U(1) situation in Fig.~\ref{fig:u1_fsp_eigs}{\bf c}, the initial state resides in the $N_{\uparrow} = 6$ subspace, while we see a rapid breakdown of the U(1) symmetry as $S[N_\uparrow=6,t]$ quickly decays, and merges with the others $N_\uparrow=5,7\dots$ in $t\sim 10^2 T$. This is to be contrasted against the FSP situation in Fig.~\ref{fig:u1_fsp_eigs}{\bf d}. Here, the same initial state belongs to the DW number $W(\boldsymbol{s}_0)=12$ subspace ($\Delta w = 0$). We note that initially $S[w=12,t]$ ramps up, implying the lightcone thermalization within the $w=12$ subspace. After $t\sim 10^2T$, $S[w=12,t]$ reaches a plateau for an exponentially long time, while other domain wall subspaces exhibit suppressed population as $S[w\neq 12,t]$ are much smaller than the $S[w=12,t]$ plateau. 

Results in Fig.~\ref{fig:u1_fsp_eigs}{\bf d} emphasize the long-term conservation of domain wall number $w$ in FSP systems, consistent with the eigenstate information in Fig.~\ref{fig:u1_fsp_eigs}{\bf b}. In contrast, the preservation of U(1) symmetry in Fig.~\ref{fig:u1_fsp_eigs}{\bf c} is only a transient process, whose signatures are absent from the eigenstructure in Fig.~\ref{fig:u1_fsp_eigs}{\bf a}. The U(1) case is consistent with our conventional understanding of Floquet prethermalization, as the prethermal phenomena only persist until the timescale $e^{\omega/J}$ ($\omega\gg J$), before which one can perform a Magnus expansion of $O(J/\omega)$ to obtain an effective Hamiltonian. Then, the eigenstates of the full Floquet operator should be thermal because they reflect the infinite-time responses. To the contrary, the FSP situation is enforced by a strong Ising interaction $J\sim \omega$ which structures the eigenstates of the {\it full Floquet operator} without any Magnus truncation. In summary, it is the different interaction strength $J$ compared with driving frequency $\omega$ for FSP and U(1) prethermal DTCs that results in the different timescale for $w$ and $S^z$ to be conserved.

\begin{figure}
	[h]
	\includegraphics[width=18cm]{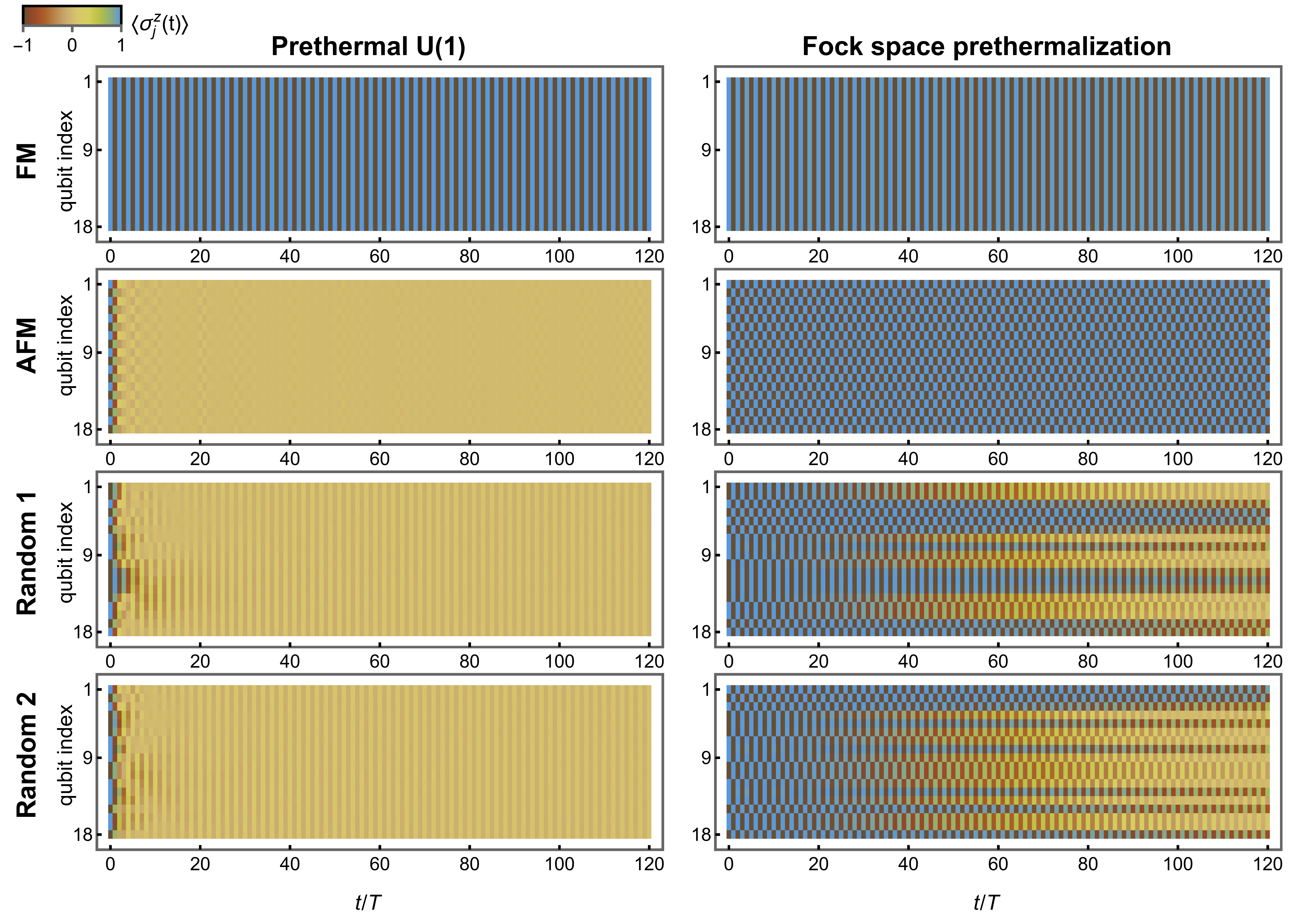}
	\caption{ \label{fig:u1_fsp_corr} 
		Comparison of the local magnetization dynamics.  
		For the prethermal U(1) case, we take high frequency $\omega = 12$ ($J=\pi/\omega=0.2618$) deep inside the phase, $90\%$ of a perfect echo in longitudinal fields, $\theta_h = 0.9\pi$, and perturbation strength $\epsilon=0.1$. To allow for a sensible comparison in early time dynamics, for the FSP case with $J=1$, we take a similar perturbation strength $\lambda_1=2\lambda_2=0.1$, longitudinal fields $\varphi_1=-0.5\pi, \varphi_2 = -0.6\pi$, similarly close to an echo. The system sizes are $L=18$ in both cases. }
\end{figure}
\begin{figure}
	[h]
	\includegraphics[width=16cm]{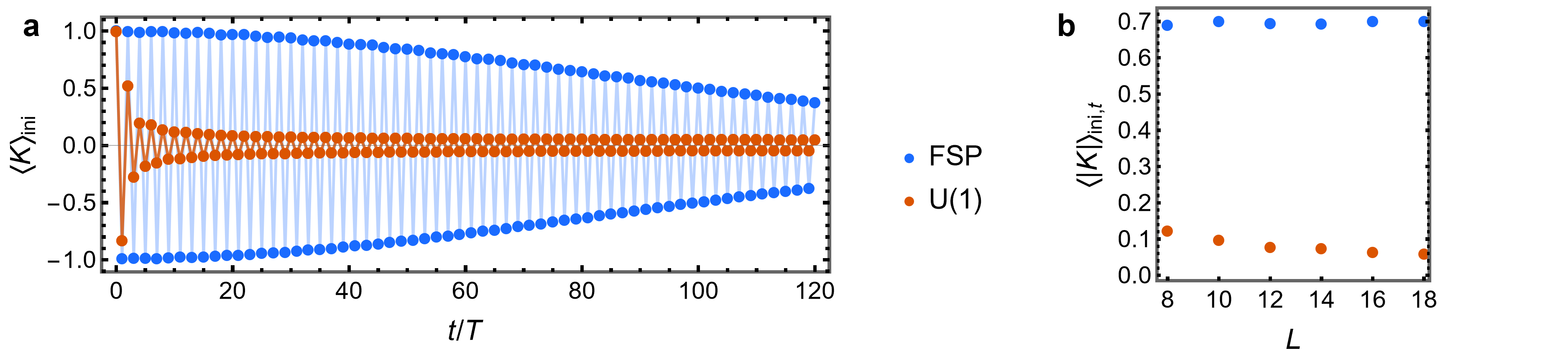}
	\caption{ \label{fig:u1_fsp_corr_mean} 
		{\bf a,} The auto-correlator averaged over random initial states. Here parameters $(J=\pi/\omega, \theta_h, \epsilon)$ for prethermal U(1) case and $(J,\varphi_1,\varphi_2,\lambda_1, \lambda_2)$ for FSP are the same as those in Fig.~\ref{fig:u1_fsp_corr}. For each system size, we average over 200 random initial states. {\bf b,} Averaged oscillation amplitudes within 120$T$ as a function of the system size $L$.}
\end{figure}

{\bf 2, DTC induced by FSP hosts inhomogeneous spatial patterns, while the U(1) DTC always relaxes quickly into a homogeneous pattern without spatial structure}. Comparisons are shown in Figs.~\ref{fig:u1_fsp_corr} and \ref{fig:u1_fsp_corr_mean}.

To reveal the real-space structure, we compute the local magnetic order $\langle\sigma^z_j(t)\rangle$ defined in Eq.~\eqref{eq:mag}.
Results for the two cases in Fig.~\ref{fig:u1_fsp_corr} intuitively illustrate their differences. For a completely homogeneous ferromagnetic (FM) initial state $|\boldsymbol{s}_0\rangle = |111\dots1\rangle$, the U(1) DTC and FSP induced DTC exhibit similar behaviors of subharmonic temporal responses. However, the dynamics for an anti-ferromagnetic (AFM) initial state $|\boldsymbol{s}_0\rangle = |1010\dots 10\rangle$ in two cases are completely opposite. Here, we see that the $\langle\sigma^z_j(t)\rangle$ for U(1) DTC quickly decays to zero within a few periods, as the total $ \langle\boldsymbol{s}_0| S^z|\boldsymbol{s}_0\rangle = 0$ in the initial state. Contrarily, the FSP-induced DTC shows rigid local period-doubled oscillations for $\langle \sigma^z_j(t)\rangle \sim (-1)^{t/T+j}$ preserving the initial maximally inhomogeneous spatial pattern. Further, for more generic initial states, we exemplify the results with two random patterns dubbed ``random 1'' and ``random 2''. In both cases, we note that the U(1) DTC quickly erases any spatial structure of the initial state within a few driving periods, while the global $\langle \boldsymbol{s}_0|S^z|\boldsymbol{s}_0\rangle$ shows subharmonic oscillations with homogeneous patterns. This is consistent with the major conclusion in Ref.~\cite{Luitz2020}. In contrast, the FSP-induced DTC only thermalizes within finite-size light-cones, outside of which the inhomogeneous spatial patterns are preserved in the period-doubled oscillations.

We further benchmark the fraction of the initial state in the entire Hilbert space that can show local DTC oscillations. For this purpose, we use the temporal auto-correlator $K(t)$ in Eq.~\eqref{eq:autocorr}.
A larger magnitude for such a quantity not only implies stronger temporal subharmonic oscillations, but also confirms that the oscillation can preserve initial state patterns $|\boldsymbol{s}_0\rangle$~\cite{Luitz2020}. 
In Fig.~\ref{fig:u1_fsp_corr_mean}{\bf a}, we average the result over 200 random initial states $|\boldsymbol{s}_0\rangle$ for both FSP-induced DTC and U(1) DTC. A sharp contrast appears, as the FSP shows large oscillations magnitudes with a decay timescale consistent with that for the thermalizing lightcone to spread across the whole system. In contrast, correlators in the U(1) DTC almost immediately decay over 10 initial periods to a small magnitude. We have further performed finite-size scaling for the averaged oscillation amplitudes of $|K(t)|$ over the first $120T$, as shown in Fig.~\ref{fig:u1_fsp_corr_mean}{\bf b}. 

The different capability of accommodating inhomogeneous patterns is related to the different types of interactions in the two cases. For prethermal U(1) DTC, the $\sigma^x_j\sigma^x_{k} + \sigma^y_j\sigma^y_k$ corresponds to a ``hopping'' term via Jordan-Wigner transformation Eq.~\eqref{eq:jw}. Thus, any inhomogeneity in $|\boldsymbol{s}_0\rangle$ would be smeared out by the XXZ interaction within the timescale $1/J$. To the contrary, the Ising interaction in FSP is diagonal in the Fock bases, which preserves, rather than destroys, spatial structures in $|\boldsymbol{s}_0\rangle$ via the DW constraints.

\subsection{Analytical derivation of full and subspace thermalized wave packet distribution}

In this section, we derive the distribution of a thermalized wave packet in Fock space, which serves as a reference point to understand and benchmark the results obtained in the main text.  We would consider two types of thermalization. One is the ``full thermal'' wave functions being ergodic throughout the entire Hilbert space, corresponding to the ``thermal'' case in our main text. The other is the ``subspace thermal'' wave function, which is only thermalized within a specific domain wall subspace. This corresponds to the FSP regime, where the light cone has spread across the entire system.

The distribution of Fock-basis amplitudes for a fully thermalized quantum state $|\psi\rangle$ (i.e., not only in a given DW sector through lightcone propagations, but being ergodic throughout the entire Hilbert space) obeys Porter-Thomas distribution~\cite{Porter1956, Boixo2018, Claeys2025},
\begin{align}
	a(\boldsymbol{s}) = |\langle \boldsymbol{s}|\psi\rangle|^2,
	\qquad
	P_{\text{PT}}(a) = {\cal H} e^{- {\cal H} a},
    \qquad 
    {\cal H} = 2^L,
\end{align}
where ${\cal H}$ is the Hilbert space dimension for a chain of $L$ qubits. This distribution is derived under the generic assumption that the wave function can be expanded into Fock bases with coefficients being random numbers $|\psi\rangle =\sum_{\boldsymbol{s}} [c_1(\boldsymbol{s})+{\rm i}c_2(\boldsymbol{s})]|\boldsymbol{s}\rangle, c_1,c_2\in\mathbb{R}$, which are only constrained by the normalization condition $\sum_{\boldsymbol{s}} a(\boldsymbol{s}) = \sum_{\boldsymbol{s}} [c_1^2(\boldsymbol{s}) + c_2^2(\boldsymbol{s})] = 1$.

\begin{figure}
	[h]
	\includegraphics[width=16cm]{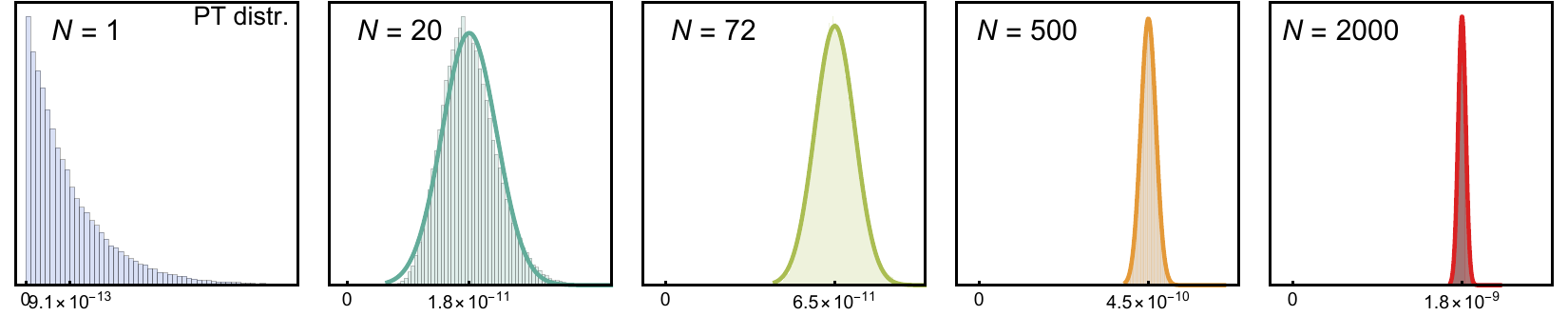}
	\caption{ \label{fig:a1} Summation over $N$ random variables, each having Porter-Thomas (PT) distribution ($N=1$ case), converges to normal distributions with the mean $\mu = N/\mathcal{H}$. Here $L=40$ and $\mathcal{H}=2^{40}\approx 1.1\times10^{12}$. Shades are numerically generated data while solid lines are Eq.~\eqref{eq:clt_pt}.}
\end{figure}
In our experiments, we measure the wave function in terms of the Hamming distance $d$ or the total domain wall numbers $w$. Each $d$ or $w$ involve summations over the amplitudes $a(\boldsymbol{s})$ of large numbers of Fock bases $|\boldsymbol{s}\rangle$. Thus, we could invoke the central limit theorem to obtain the probability distributions of wave functions in $d$ or $w$ coordinates. Specifically, the mean and deviation of Porter-Thomas distribution read
\begin{align}
	\mu = \int_0^\infty aP_{\text{PT}}(a) da = \frac{1}{{\cal H}},
	\qquad
	\sigma^2 = \int_0^\infty (a-\mu)^2 P_{\text{PT}}(a)da = \frac{1}{{\cal H}^2}.
\end{align}
Then, for a summation over $N$ such variables, we have the probability distribution
\begin{align} \label{eq:clt_pt}
	y = \sum_{j=1}^N a(\boldsymbol{s}_j),
	\qquad
	P(y;N) = \frac{1}{\sqrt{2\pi N}\sigma} e^{- \frac{(y - N\mu)^2}{2N\sigma^2}}
	=\frac{{\cal H}}{\sqrt{2\pi N}} e^{- \frac{{\cal H}^2(y - N/{\cal H})^2}{2N}}.
\end{align}
A straightforward verification in Fig.~\ref{fig:a1} shows that summing over $20$ variables already converges well to such normal distributions. In our experiment with 72 qubits, the dominantly populated $d$ and $w$ sectors have $N$ orders of magnitude larger than what we have tested, so it is a good approximation to adopt the normal distribution given by the central limit theorem.

\begin{figure}
	[h]
	\includegraphics[width=8cm]{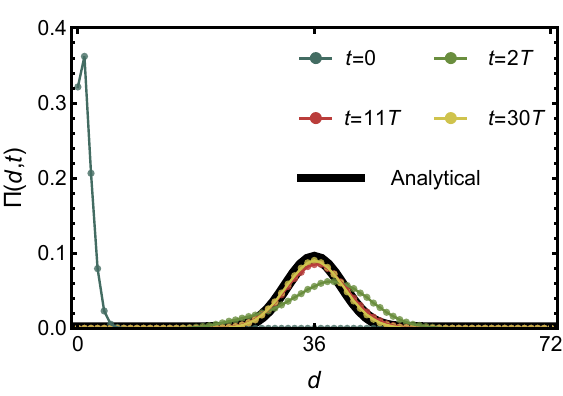}
	\includegraphics[width=8cm]{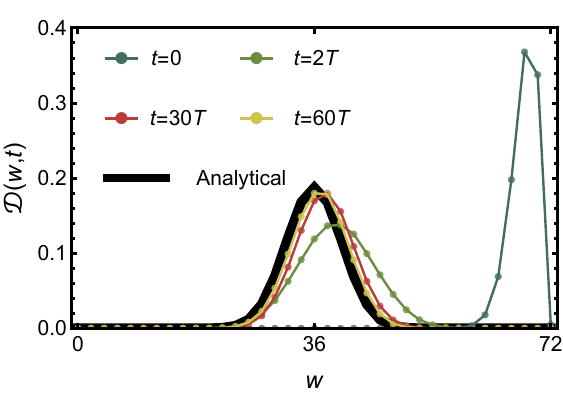}
	\caption{ \label{fig:a2} 
		Distribution of a thermalized wave packet in Fock space, plotted in the coordinates of (left) Hamming distance $d$, corresponding to Fig.~2{\bf a}, {\bf c} of the main text, and (right) domain wall number $w$, corresponding to Fig.~4{\bf a} of the main text. In both cases, the experimentally measured wave packet converges to analytical predictions.}
\end{figure}

Importantly, the normal distribution gives the average value
\begin{align}
	\langle y\rangle = \int_0^\infty yP(y;N) dy = N\mu = \frac{N}{{\cal H}}.
\end{align}
Thus, in treating the distributions in terms of Hamming distance and domain wall numbers, we can effectively take the thermalized wave packet as if it were evenly distributed over all ${\cal H}=2^L$ Fock bases with equal probability $1/{\cal H} = 2^{-L}$, and the total probability at each $d$ and $w$ is proportional to the respective density of states $N$. Thus, we have the probability distribution in Hamming distance $d$ and domain wall numbers $w$ coordinates for a fully thermalized wave packet as
\begin{align} \label{eq:P_d_w}
	\Pi_{\text{ft}}(d) \approx \frac{1}{2^L}\frac{L!}{d!(L-d)!} 
	\xrightarrow{L\gg1} \frac{1}{\sqrt{\pi L/2}} e^{-\frac{(d-L/2)^2}{L/2}},
	\qquad
	\mathcal{D}_{\text{ft}}(w) \approx \frac{1}{2^{L-1}}\frac{L!}{w!(L-w)!}
	\xrightarrow{L\gg1} \frac{2}{\sqrt{\pi L/2}} e^{-\frac{(w-L/2)^2}{L/2}}.
\end{align}
Here, the different locations of $d$ flipped qubit and $w$ domain walls give the factorial terms, while the normalization differs by a factor of 2 as the globally flipped pattern gives the same DW counts. We find the agreement of analytical result in Eq.~\eqref{eq:P_d_w} with experimental data in Fig.~\ref{fig:a2}, verifying the ergodicity of the wave packet in the thermal case $\lambda_1=1.2$ of our main text.

Based on the form of Eq.~\eqref{eq:P_d_w}, we can directly see that the scaling of wave packet width in Fock space $\Delta x$ scales in the fully thermalized limit as
\begin{align} \label{eq:fulltherm}
	\text{Full thermal (i.e. $\lambda_1=2\lambda_2=1.2$):}\qquad
	\Delta x = \sqrt{\sum_{d=0}^L (d-L/2)^2 \Pi_{\text{ft}}(d)}
	=\frac{\sqrt{L}}{2} \propto \sqrt{L}.
\end{align}
Such a scaling is verified in Fig.~4{\bf e} of our main text with large $\lambda_1$, where data of $\Delta x/\sqrt{L}$ collapses.

On the other hand, for our purpose of benchmarking the FSP regime, we also consider the situation where thermalization only occurs within a certain DW sector. That would correspond to the situation where $|\psi(t)\rangle$ is exponentially localized onto a certain DW $w$ sector, while the lightcone has spread across the whole system of size $L$ so all Fock bases within the $W(\boldsymbol{s})=w$ are almost evenly populated. 

The derivation is most intuitive if we start from the $L$-qubit FM background and flip 1 qubit,
\begin{align}
	|\boldsymbol{s}_0\rangle = |10000\dots00\rangle.
\end{align}
This initial state carries 2 domain walls. It belongs to the subspace $\{\boldsymbol{s}| W(\boldsymbol{s})=2\}$ with $2C_{L}^{2} = L(L-1)$ Fock bases, which can be enumerated in the following way,
\begin{align}\nonumber
	&10000\dots0000, 
	&& {\color{RoyalBlue}\boldsymbol{01}}000\dots0000, 
	&& {\color{RoyalBlue}\boldsymbol{0}}0{\color{RoyalBlue}\boldsymbol{1}}00\dots 0000, 
	&&
	{\color{RoyalBlue}\boldsymbol{0}}00{\color{RoyalBlue}\boldsymbol{1}}0\dots0000,
	\quad\dots\quad 
	{\color{RoyalBlue}\boldsymbol{0}}0000\dots 00{\color{RoyalBlue}\boldsymbol{1}}0 
	&& {\color{RoyalBlue}\boldsymbol{0}}0000\dots 000{\color{RoyalBlue}\boldsymbol{1}};
	\\
	\nonumber
	&1{\color{RoyalBlue}\boldsymbol{1}}000\dots0000, 
	&& {\color{RoyalBlue}\boldsymbol{011}}00\dots0000, 
	&& {\color{RoyalBlue}\boldsymbol{0}}0{\color{RoyalBlue}\boldsymbol{11}}0\dots 0000, 
	&&
	{\color{RoyalBlue}\boldsymbol{0}}00{\color{RoyalBlue}\boldsymbol{11}}\dots0000,
	\quad\dots\quad 
	{\color{RoyalBlue}\boldsymbol{0}}0000\dots 00{\color{RoyalBlue}\boldsymbol{11}} && 
	10000\dots 000{\color{RoyalBlue}\boldsymbol{1}};
	\\
	\nonumber
	&1{\color{RoyalBlue}\boldsymbol{11}}00\dots0000,
	&& {\color{RoyalBlue}\boldsymbol{0111}}0\dots0000,
	&&
	{\color{RoyalBlue}\boldsymbol{0}}0{\color{RoyalBlue}\boldsymbol{111}}0\dots 000,
	&&
	{\color{RoyalBlue}\boldsymbol{0}}00{\color{RoyalBlue}\boldsymbol{111}}0\dots 00,
	\quad\dots\quad
	10000\dots00{\color{RoyalBlue}\boldsymbol{11}},
	&&
	1{\color{RoyalBlue}\boldsymbol{1}}000\dots000{\color{RoyalBlue}\boldsymbol{1}};
	\\
	\nonumber
	&\dots\dots
	\\ \label{eq:fmenumerate}
	&1{\color{RoyalBlue}\boldsymbol{1111}}\dots{\color{RoyalBlue}\boldsymbol{111}}0,
	&&
	{\color{RoyalBlue}\boldsymbol{01111}}\dots{\color{RoyalBlue}\boldsymbol{1111}},
	&&
	10{\color{RoyalBlue}\boldsymbol{111}}\dots{\color{RoyalBlue}\boldsymbol{1111}},
	&&
	1{\color{RoyalBlue}\boldsymbol{1}}0{\color{RoyalBlue}\boldsymbol{11}}\dots{\color{RoyalBlue}\boldsymbol{1111}},
	\quad\dots\quad
	1{\color{RoyalBlue}\boldsymbol{1111}}\dots{\color{RoyalBlue}\boldsymbol{1}}0{\color{RoyalBlue}\boldsymbol{11}},
	&&
	1{\color{RoyalBlue}\boldsymbol{1111}}\dots{\color{RoyalBlue}\boldsymbol{11}}0{\color{RoyalBlue}\boldsymbol{1}}
\end{align}
There are $L$ columns in the array related by translations of the pattern, and $L-1$ rows for different numbers of qubits being $1$, which span the dimension $L(L-1)$ of the subspace. Flipped qubits relative to $\boldsymbol{s}_0$ are highlighted by bold blue numbers. We can intuitively see that the qubit patterns in the $n$-th row have 
\begin{align}\nonumber
	(L-n)\times \text{ Fock states $\boldsymbol{s}$:\quad  }
	&
	D(\boldsymbol{s},\boldsymbol{s}_0) = n+1;
	\\
	n\times \text{ Fock states $\boldsymbol{s}$:\quad  }
	&
	D(\boldsymbol{s},\boldsymbol{s}_0) = n-1.
\end{align}
Now, consider the very late time limit for the FSP regime. The light-cone has propagated across the whole system, such that all Fock states within this subspace can be approximately regarded as being evenly populated. The domain wall separation indicated by Fig.~\ref{fig:eigs}{\bf a} and Fig.~\ref{fig:u1_fsp_eigs}{\bf b}, {\bf d} means Fock bases in other domain wall subspaces have exponentially small amplitudes and can be approximately neglected. Under such assumptions, we can derive the wave packet width in the ``subspace thermal'' case as
\begin{align}\nonumber
	\text{Subspace thermal (FSP at $t\rightarrow\infty$):}
	\qquad
	&
	\langle x\rangle = \frac{1}{L(L-1)}\sum_{n=1}^{L-1} \left[
	(L-n)(n+1) + n(n-1)
	\right] = \frac{L}{2},
	\\ \nonumber
	&
	\Delta x = \sqrt{\frac{1}{L(L-1)} 
		\sum_{n=1}^{L-1} \left[
		(L-n) \left(n+1-\frac{L}{2} \right)^2
		+
		n\left(n-1-\frac{L}{2}\right)^2
		\right]
	}
	\\ \label{eq:subtherm}
	&=
	\sqrt{\frac{L^2-6L+20}{12}}
	\xrightarrow{L\gg1} \propto L.
\end{align}
For our experimental scaling analysis in Fig.~4{\bf d} of the main text, we use the ``1FM'' initial state
\begin{align}
	|\boldsymbol{s}_0\rangle = |{\color{BrickRed}\boldsymbol{0}}01010\dots1010\rangle,
\end{align}
where we have flipped the qubit 1 and created the mini-FM region at qubits $(L,1,2)$. This initial state carries $L-2$ domain walls.
An enumeration similar to Eq.~\eqref{eq:fmenumerate} can be done for the Fock states in the $W(\boldsymbol{s})=L-2$ subspace,
\begin{align} \nonumber
	&001010\dots1010, 
	&& {\color{RoyalBlue}\boldsymbol{11}}1010\dots1010,
	&&
	{\color{RoyalBlue}\boldsymbol{1}}0{\color{RoyalBlue}\boldsymbol{0}}010\dots1010,
	&&
	{\color{RoyalBlue}\boldsymbol{1}}01{\color{RoyalBlue}\boldsymbol{1}}10\dots1010,
	\quad\dots\quad
	{\color{RoyalBlue}\boldsymbol{1}}01010\dots10{\color{RoyalBlue}\boldsymbol{0}}0,
	&&
	{\color{RoyalBlue}\boldsymbol{1}}01010\dots101{\color{RoyalBlue}\boldsymbol{1}};
	\\ \nonumber
	&0{\color{RoyalBlue}\boldsymbol{1}}1010\dots1010,
	&&
	{\color{RoyalBlue}\boldsymbol{110}}010\dots1010,
	&&
	{\color{RoyalBlue}\boldsymbol{1}}0{\color{RoyalBlue}\boldsymbol{01}}10\dots1010,
	&&
	{\color{RoyalBlue}\boldsymbol{1}}01{\color{RoyalBlue}\boldsymbol{10}}0\dots1010, 
	\quad\dots\quad 
	{\color{RoyalBlue}\boldsymbol{1}}01010\dots10{\color{RoyalBlue}\boldsymbol{01}},
	&&
	001010\dots101{\color{RoyalBlue}\boldsymbol{1}};
	\\ \nonumber
	&0{\color{RoyalBlue}\boldsymbol{10}}010\dots1010,
	&& 
	{\color{RoyalBlue}\boldsymbol{1101}}10\dots1010,
	&&
	{\color{RoyalBlue}\boldsymbol{1}}0{\color{RoyalBlue}\boldsymbol{010}}010\dots10,
	&&
	{\color{RoyalBlue}\boldsymbol{1}}01{\color{RoyalBlue}\boldsymbol{101}}10\dots10,
	\quad\dots\quad 
	001010\dots 10{\color{RoyalBlue}\boldsymbol{01}},
	&&
	0{\color{RoyalBlue}\boldsymbol{1}}1010\dots101{\color{RoyalBlue}\boldsymbol{1}};
	\\ \nonumber
	& \dots\dots 
	\\  \label{eq:afmenumerate}
	&
	0{\color{RoyalBlue}\boldsymbol{10101}}\dots{\color{RoyalBlue}\boldsymbol{010}}0,
	&&
	{\color{RoyalBlue}\boldsymbol{110101}}\dots{\color{RoyalBlue}\boldsymbol{0101}},
	&&
	00{\color{RoyalBlue}\boldsymbol{0101}}\dots{\color{RoyalBlue}\boldsymbol{0101}},
	&&
	0{\color{RoyalBlue}\boldsymbol{1}}1{\color{RoyalBlue}\boldsymbol{101}}\dots{\color{RoyalBlue}\boldsymbol{0101}},
	\quad\dots\quad 
	0{\color{RoyalBlue}\boldsymbol{10101}}\dots{\color{RoyalBlue}\boldsymbol{0}}0{\color{RoyalBlue}\boldsymbol{01}},
	&&
	0{\color{RoyalBlue}\boldsymbol{10101}}\dots{\color{RoyalBlue}\boldsymbol{01}}1{\color{RoyalBlue}\boldsymbol{1}}.
\end{align}
We note that flipping all odd qubits in Eq.~\eqref{eq:fmenumerate} with $W(\boldsymbol{s})=2$ turns those Fock bases into Eq.~\eqref{eq:afmenumerate} with $W(\boldsymbol{s})=L-2$. Then, a parallel derivation after Eq.~\eqref{eq:fmenumerate} leads to the same conclusion Eq.~\eqref{eq:subtherm} for the ``1FM'' initial state.

With the result $\Delta x\propto \sqrt{L}$ for the fully thermalized case in Eq.~\eqref{eq:fulltherm} and $\Delta x\propto L$ for the domain-wall subspace thermalized case in Eq.~\eqref{eq:subtherm}, we can qualitatively understand the scaling results in Fig.~4{\bf c}, {\bf e} of the main text. 
\begin{itemize}
	\item In the FSP regime ($\lambda_1=2\lambda_2\sim0.1$), focusing on the experimentally accessible timescale, the light-cone has not propagated throughout the whole system (see Fig.~3{\bf a} and {\bf b} of the main text). Suppose the light-cone has fully thermalized $N$ qubits ($N<L$), we can follow the above derivation for the domain wall sector $w=L-2$ for a hypothetical $N$-qubit system, and have $\Delta x\propto N$, which is actually independent of the system size $L$. However, the external noise would cause depolarization effects, bringing a certain fraction of the density matrix into the infinite-temperature one, resulting in $\sqrt{L}$ scaling as in the ``full-thermal" case Eq.~\eqref{eq:fulltherm}. Thus, overall, we would see 
	\begin{align}
		\Delta x\propto \sqrt{L}
	\end{align}
	scaling in the experimental data.
	
	\item In the critical regime ($\lambda_1=2\lambda_2\sim0.4$), the relatively large perturbation has allowed the light-cone to propagate across the whole system of $L$ qubits within the experimental timescale. While in this case, different domain wall sectors are significantly mixed, they are still unevenly populated, so the scaling of 
	\begin{align}
		\Delta x \propto L^\nu, \qquad \nu\in(0.5,1)
	\end{align}
	is expected to be sandwiched between $\nu=0.5$ for the full thermalized case and $\nu=1$ for subspace thermalization within a fixed domain wall sector.
	
	\item Finally, in the thermal regime ($\lambda_1=2\lambda_2\sim1.2$), both the external noise and internal thermalizing evolution would lead to the ``full thermal" result
	\begin{align}
		\Delta x \propto \sqrt{L}
	\end{align}
	in Eq.~\eqref{eq:fulltherm}.
\end{itemize}

In summary, we show that the scaling of wave packet width in Fock space $\Delta x\propto L^\nu$ can distinguish the FSP, critical, and thermal regimes in our system, intriguingly similar to its functionality in many-body localization (MBL) in distinguishing the MBL-critical-thermal regimes~\cite{DeTomasi2021,Yao2023}. For an idealized noise-free situation, we would have $\Delta x\sim L^0$ for FSP, $\Delta x\propto L^{\nu\in(0.5,1)}$ for criticality, and $\Delta x\propto L^{0.5}$ for thermal situation. Taken into account noise with depolarization effects, the exponent $\nu\geqslant0.5$ is bounded from below, and we would have $\Delta x\propto L^{0.5}$ deep inside FSP and thermal regimes, which are divided by an anomalously wide $\Delta x\propto L^{\nu\in(0.5,1)}$ intermediate critical regime.

\section{Numerical results}

\begin{figure}[h]
    \includegraphics[width=18cm]{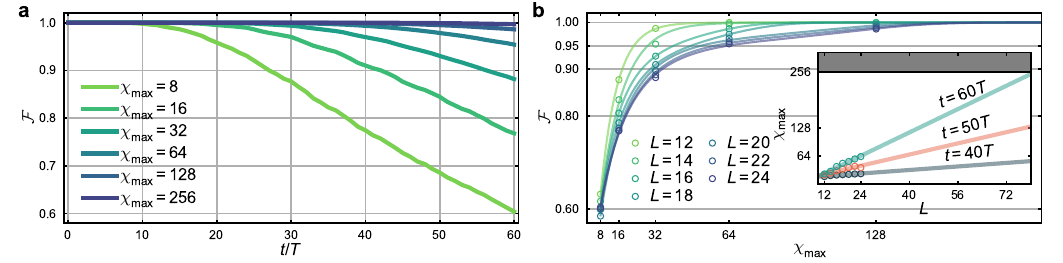}
    \caption{{\bf MPS simulation with different maximum bond dimension.} 
    {\bf a,} MPS simulation fidelity $\mathcal{F}$ as a function of Floquet unitary cycle for different maximum bond dimensions $\chi_{\text{max}}$. These results are simulated for a 24-qubit system. Unless otherwise specified, all results presented below are obtained with $\lambda_1=2\lambda_2 = 0.2 $ and averaged over 10 random $\varphi_2$ realizations, mirroring experimental conditions.
    {\bf b,} $\mathcal{F}$ as a function of $\chi_{\text{max}}$ for different system sizes at $t = 60T$. Inset displays a linear fit of $\chi_{\text{max}}$ required to satisfy $\mathcal{F} = 0.95$ threshold as a function of system size $L$.}
    \label{fig:sim_fid}
\end{figure}

\subsection{Benchmarking matrix product state (MPS) method}

To elucidate the FSP mechanism and validate our analytical predictions, we perform numerical simulations for systems up to $L = 72$ qubits, employing extrapolation techniques. The exponential memory growth of exact diagonalization (ED) renders it infeasible for system sizes comparable to those in our experiment. Instead, we leverage the matrix product state (MPS) method, which is ideally suited for one-dimensional systems and efficiently captures area-law growing entanglement. This allows us to simulate large systems for a practical number of Floquet cycles.

\begin{figure}[t]
    \includegraphics[width=18cm]{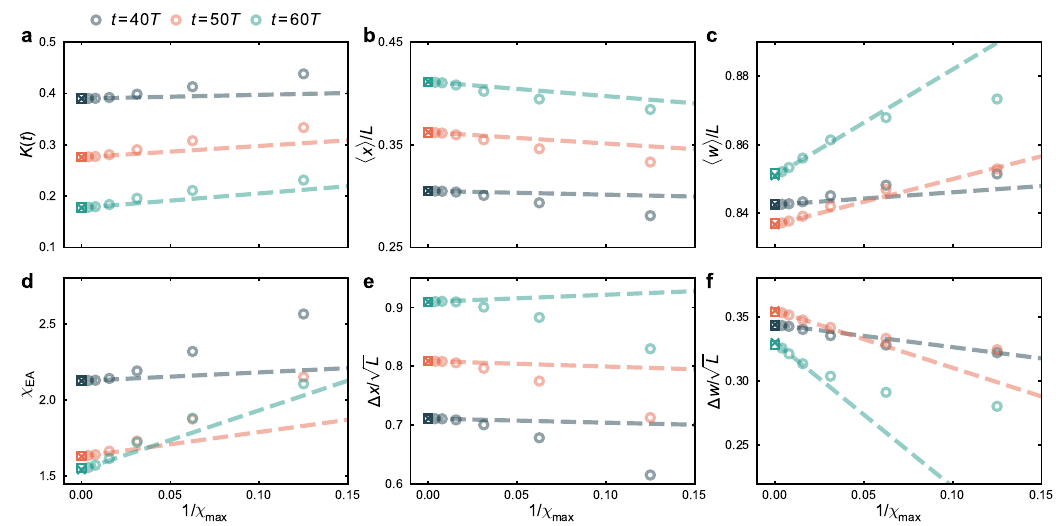}
    \caption{{\bf Extrapolations of MPS results.} Numerical results for the 24-qubit system as a function of the inverse maximum bond dimension $1/\chi_{\rm max}$. Colored dots represent MPS results at specific time slices, as indicated in the legend. Dashed lines show linear fits to the data points at the last two smallest values of $1/\chi_{\text{max}}$. Crosses denote extrapolation results from the linear fit to the limit of $1/\chi_{\text{max}} \rightarrow 0$, while squares represent values obtained via the ED method. In all cases, the crosses overlap with the squares, indicating that MPS results provide a reliable approximation to the exact ED values, particularly as $\chi_{\rm max}$ increases.}
    \label{fig:extra}
\end{figure}

In practice, simulations are conducted at the quantum circuit level using Qiskit~\cite{qiskit2024}. We begin with a Fock state $|\psi_0\rangle$, initialized as a statevector of the form $|\boldsymbol{s}\rangle = \bigotimes_{j=1}^{L} |s_j\rangle$, where $\boldsymbol{s}$ is a bitstring, and $|s_j\rangle$ is the basis state of qubit $j$ with $s_j \in \{0, 1\}$. For the statevector simulation method, the initial state evolves in time to $t = nT$ by applying the Floquet unitary operator $n$ times, yielding $|\psi_{t=nT}\rangle = U_{\rm F}^n |\psi_0\rangle = (U_2 U_1)^n |\psi_0\rangle$. In the MPS method, as introduced by Vidal \cite{Vidal2003PRL}, quantum states in the computational basis are expressed as $|\psi\rangle = \sum_{s_1=0}^{1} \dots \sum_{s_L=0}^{1} c_{s_{1} \dots s_L} |s_1\rangle \otimes \dots \otimes |s_L\rangle$. The coefficients can be decomposed into products of matrices as
\begin{align}
    c_{s_{1} \dots s_L} = \sum_{\alpha_1, \dots, \alpha_{L-1}} \Gamma_{\alpha_1}^{[1] s_1} \Lambda_{\alpha_1}^{[1]} \Gamma_{\alpha_1 \alpha_2}^{[2] s_2} \Lambda_{\alpha_2}^{[2]} \Gamma_{\alpha_2 \alpha_3}^{[3] s_3} \dots \Gamma_{\alpha_{L-1}}^{[L] s_L}
\end{align}
up to a bond dimension $\chi$, where $s_j$ and $\alpha_j$ take values in $\{0, 1\}$ and $\{1, \dots, \chi\}$, respectively. Each tensor $\Gamma^{[j]}$ is assigned to qubit $j$, and single-qubit gates are applied to this local tensor. Each vector $\Lambda^{[j]}$ stores the Schmidt coefficients between qubits $j$ and $j+1$. In our model, two-qubit gates are applied only between adjacent qubits, e.g., between qubit $j$ and $j+1$; hence, only $\Gamma^{[j]}$, $\Lambda^{[j]}$, and $\Gamma^{[j+1]}$ need to be updated.

Theorem~1 in Ref.~\cite{Vidal2003PRL} establishes that the number of parameters for an exact description of a quantum state scales linearly with system size but exponentially with entanglement. As our target quantum circuit deepens or the system evolves toward ergodicity, entanglement growth can necessitate a prohibitively large number of parameters for an exact MPS representation. To manage computational constraints, we approximate the MPS by imposing a maximum bond dimension, $\chi_{\text{max}}$, and subsequently extrapolate our results to the $\chi \rightarrow \infty$ limit. Specifically, we simulate the cycle-by-cycle evolution of an initial Fock state $|\psi_0\rangle = |\boldsymbol{s}\rangle$ up to $n$ cycles for a series of increasing bond dimensions $\{\chi_1, \chi_2, \ldots, \chi_m\}$. This procedure yields a set of approximated states $ \{|\psi_{t=0}^{\chi_j}\rangle, |\psi_{t=T}^{\chi_j}\rangle, \ldots, |\psi_{t=nT}^{\chi_j}\rangle\}_{j=1}^m$ for analysis.

\begin{figure}[t]
    \includegraphics[width=18cm]{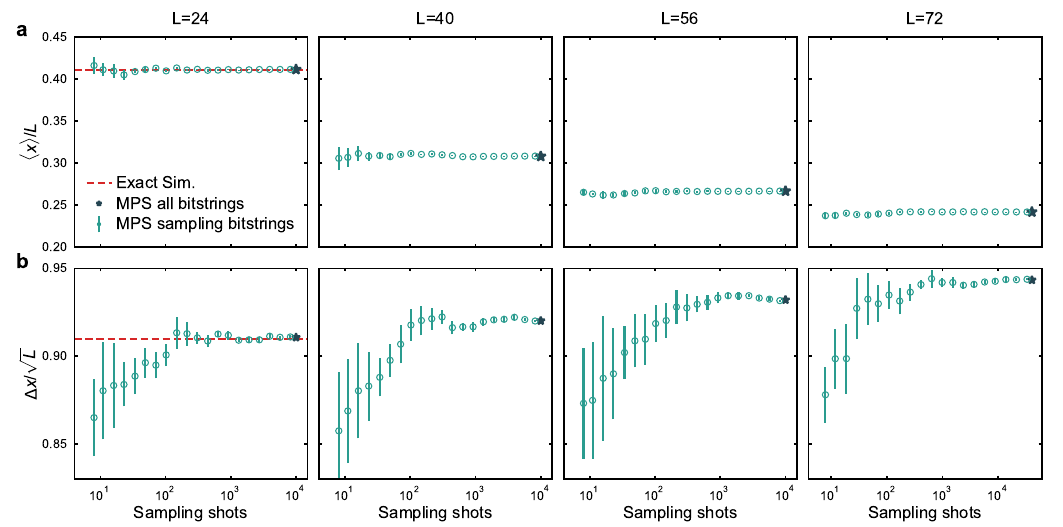}
    \caption{{\bf Convergence of observables as a function of sampling shots in MPS results.} 
    Fock-space observables plotted as a function of the number of sampling shots for different system sizes, with the maximum bond dimension fixed at $\chi_{\text{max}} = 256$ and time set to $t = 60T$. Red dashed lines denote results from exact simulations. Stars correspond to values calculated using all sampling shots ($10,000$ shots for $L \leq 56$ and $40,000$ shots for $L = 72$), while cycles with error bars denote values calculated from 5 random sampling trials at a specified number of shots.
    }
    \label{fig:mps_sampling}
\end{figure}

The state vector simulation is effective when the system size and circuit depth are not too large. Thus, we can estimate the accuracy of MPS simulation by comparing it with the exact state vector simulation for $L = 24$ and $ \lambda_1=2\lambda_2 = 0.2$. The accuracy is quantified via extending the MPS state into a state vector and calculating the fidelity $\mathcal{F}_{t=nT}^{\chi_j} = |\langle\psi_{t=nT}^{\chi_j} | U_{\rm F}^n |\psi_0\rangle|^2$. 
As shown in Fig.~\ref{fig:sim_fid}{\bf a}, $\mathcal{F}_{t}^{\chi_j}$ decays much more slowly when we increase the maximum bond dimension $\chi_{\text{max}}$. By fixing $t=60T$, we compare $\mathcal{F}_{t=60 T}^{\chi_j}$ for different system sizes in Fig.~\ref{fig:sim_fid}{\bf b}. We choose $\mathcal{F} = 0.95$ as a threshold and present the required $\chi_{\text{max}}$
as a function of system size $L$. A linear fit suggests that the $\mathcal{F}$ remains above $0.95$ for even for $L=72$ and $t=60T$. Therefore, we use $\chi_{\text{max}} = 256$ throughout the MPS simulations in this work. 

Now, we estimate the bias introduced by truncation errors for physical observables. We calculate them for different $\chi_{\text{max}}$ and perform extrapolations. This examination includes not only local observations, such as the temporal auto-correlator $K(t)$ [Eq.~\eqref{eq:autocorr}] and the Edwards-Anderson parameter $\chi_{\text{EA}} = \frac{1}{L-1} \sum_{j \neq k} \langle \sigma^z_j \sigma^z_k \rangle^2$, but also considers global observables, such as Fock and domain wall space distributions. As illustrated in Fig.~\ref{fig:extra}, a linear fit of the last two points to $1/\chi_{\text{max}}\rightarrow 0$ limit (crosses) aligns well with the results obtained from the exact method (squares).

\begin{figure}[!h]
    \includegraphics[width=18cm]{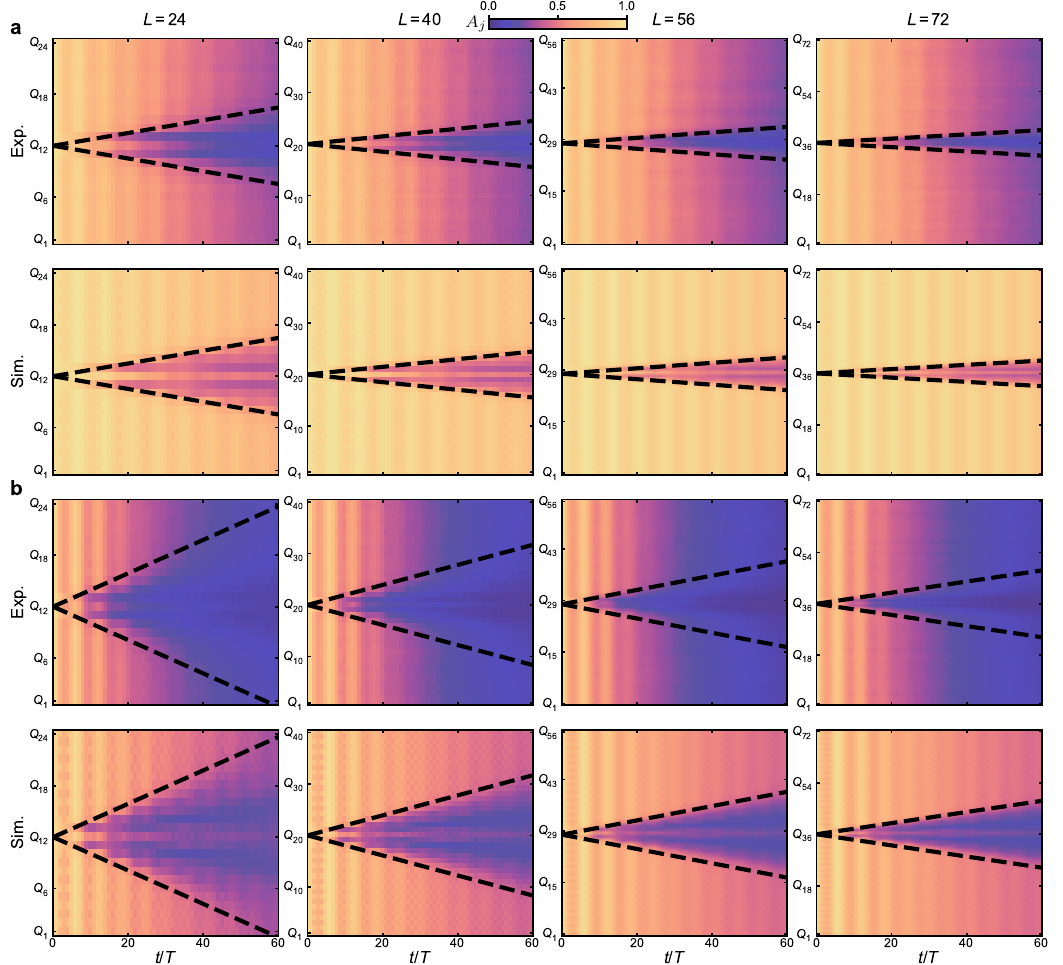}
    \caption{Experimental and numerical results of site-resolved equal-time correlator $A_{j}$ for $\lambda_1=2\lambda_2 = 0.1$ ({\bf a}) and $\lambda_1=2\lambda_2 = 0.2$ ({\bf b}), with a shared color bar located at the top. Each column corresponds to a specific system size. Dashed lines represent analytically calculated propagation speed for comparison.}
    \label{fig:c1j}
\end{figure}

We note that while local observables, e.g., temporal auto-correlator $K(t)$ and Edwards-Anderson parameter $\chi_{\text{EA}}$, can be computed directly from MPS states, some quantities, such as Fock space distributions, are challenging to calculate from MPS states due to the need for a comprehensive statistical analysis of all bitstrings, which becomes impractical for larger system sizes. This raises the question of whether we can obtain a coarse estimate of the distribution by taking a sufficient number of samples.
We find that scaling the number of random samples allows us to achieve a rough estimate for the Fock space wave packet. We collect bitstring samples from MPS states and compare them with exact simulation results for $L = 24$, which are shown in the leftmost panels of Fig.~\ref{fig:mps_sampling}. Even as the system size grows up to 72, we observe good convergence with approximately $10,000$ shots. For all results presented below, we employ $10,000$ shots for $L \leq 56$ and $40,000$ shots for $L = 72$. 

\subsection{MPS results and additional experimental data}

Using the simulation methods outlined above, we get a set of simulated results for further analysis. All results presented below are averaged over 10 random realizations of $\varphi_2$ to match experimental conditions. For system sizes $L \leq 24$, results are obtained through exact simulations; for larger system sizes ($L > 24$), results are estimated from extrapolations using MPS results.

As a numerical reproduction of Fig.~3{\bf b} in the main text, we further compare experimentally measured with simulated site-resolved equal-time correlator $A_j$ in Fig.~\ref{fig:c1j}. In all cases, the analytically predicted light cone, represented by the dashed black lines, aligns well with both experimental and simulated results. We identify two distinct prethermalization mechanisms: first, the light cone rapidly diminishes $A_j$ as it propagates to nearby sites; second, the dynamics outside the light cone remain stable. Notably, for the $L=24$ system at $t=60T$ with $\lambda_1=2\lambda_2 = 0.2$, the light cone extends to nearly the entire system. Generally, larger perturbations lead to faster propagation of the light cone. It can be estimated that for sufficiently large perturbations (e.g., $\lambda_1=2\lambda_2=0.4$), the light cone could spread to approximately $60$ sites after $t=60T$, which exceeds the capability of exact simulation. While experimental results exhibit an additional exponential decay not present in the simulations, they still preserve oscillatory dynamics and a similar light cone structure. This motivates us to investigate the role of noise in current quantum devices and the robustness of the FSP-to-thermal transition. In the next section, we further perform noisy simulations to benchmark experimental errors.

\subsection{Noisy simulation}

\begin{figure}[t]
    \includegraphics[width=18cm]{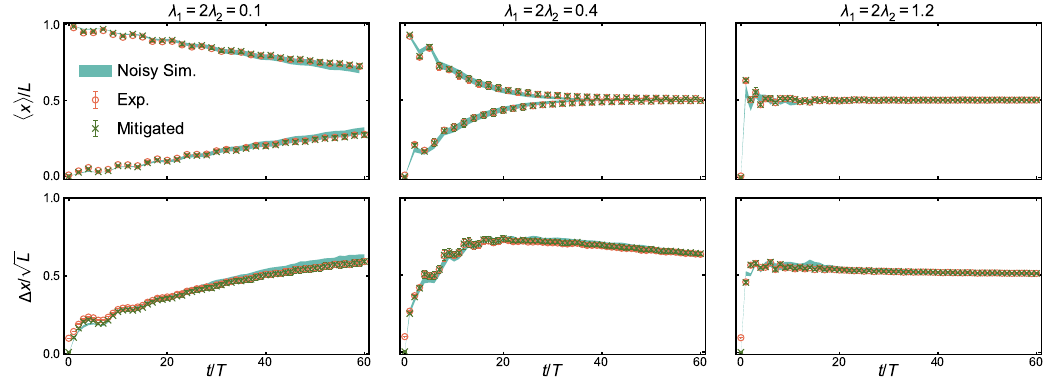}
    \caption{{\bf Experimental data versus noisy simulations for 24-qubit system.} Time evolution of Fock-space observables under varying perturbations. The green stripes represent the results of noisy simulations, with the width indicating the standard deviation of the mean derived from 10 random $\varphi_2$ realizations. The raw experimental data are denoted by red circles, and the results after readout error mitigation are marked with green crosses.} 
    \label{fig:noisy_sim}
\end{figure}

In the current NISQ era, noise has a significant impact on experimental outcomes, making its characterization and benchmarking essential. We classify this noise into two primary sources: readout error and circuit error.

Readout errors arise during qubit measurement. We model these errors as acting independently on each qubit. Readout error mitigation can be efficiently applied to local or well-structured observables. We use it to suppress readout errors for the equal-time correlator $C_{jk}(t)$ and correlation magnitude $A_j$ in the main text. However, it is challenging to do the same for global quantities such as Fock-space wave packet~\cite{Yao2023}. Here, we will show that Fock-space observables are robust to small readout errors. Assuming a uniform readout error rate $\epsilon$ (approximately $0.01$ as experimentally measured in Fig.~\ref{fig:dev_info}), each sampling shot for a qubit has a probability of $\epsilon$ to yield incorrect results. The readout process with measurement errors for a group of qubits can be represented as mapping a bitstring $\boldsymbol{s}$ from a single shot to a distribution of bitstrings in Fock space with $\Pi(d)=C_{L}^d\epsilon^d(1-\epsilon)^{L-d}$, in which the Hamming distance $d$ is defined by $d=D(\boldsymbol{s}_{\text{error}},\boldsymbol{s})$. Consequently, the mean Hamming distance to the original bitstring $\boldsymbol{s}$ can be expressed as
\begin{align}
    \langle \delta x \rangle = \sum_{d=0}^{L} d \cdot C_{L}^{d} \epsilon^{d} (1-\epsilon)^{L-d} = L \epsilon.
\end{align}
Using the triangle inequality, we obtain the normalized mean Hamming distance $|\langle x_{\text{error}} \rangle - \langle x \rangle|/L\leq \langle \delta x \rangle/L = \epsilon$. Similarly, the biased wave packet width is bounded by $|\Delta x_{\text{error}} - \Delta x|/\sqrt{L} \leq \sqrt{\epsilon(1 - \epsilon)} \approx \sqrt{\epsilon}$. 
To show the robustness of Fock-space wave packet properties against readout errors, we compare experimentally measured $\langle x \rangle, \langle\Delta x \rangle$ after readout error mitigation using the procedure in Ref.~\cite{Yao2023}(crosses) with the raw data (circles)  in Fig.~\ref{fig:noisy_sim} for a 24-qubit system. Despite slight differences in packet widths at very early times ($t = 0$), the data are close to each other thereafter. 

The second noise source is circuit error, which includes depolarizing and decoherence errors. These errors typically result in a deterministic decay of the signal over time. To quantify this effect, we perform noisy simulations using the Monte Carlo wavefunction method~\cite{Mlmer1993MonteCarlo} implemented in MindSpore Quantum~\cite{xu2024mindspore} framework. Our noise model incorporates energy relaxation and dephasing (parameterized by $T_1$ and $T_{\phi}$) as well as single- and two-qubit gate errors described by a depolarizing noise model. The numerical results for a 24-qubit system are shown in Fig.~\ref{fig:noisy_sim}, in which we use experimentally calibrated device performance for the 24-qubit system in Fig.~\ref{fig:dev_info} ($T_1=128.1       ~\mu$s, $e_{P}^{\text{SQ}}=0.048\%$, $e_{P}^{\text{TQ}}=0.40\%$). Since the dephasing time is usually longer when the qubits interact with each other, we set $T_2=2T_2^{\text{SE}}=32.2~\mu$s in our simulation (see Ref.~\cite{Song2019Science,xu2018prl} for the choice of $T_2$). For each circuit, we run 100 random noisy realizations with 1,000 shots per circuit. The strong agreement with experimental data validates the effectiveness of our error model.

\begin{figure}[t]
    \includegraphics[width=18cm]{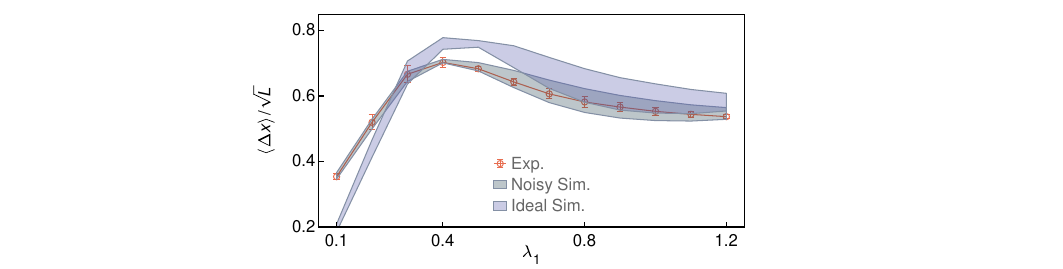}
    \caption{{\bf Noisy simulation of FSP-thermalization crossover.} Time-averaged normalized wave packet width over the time window $t\in[10T,30T]$, plotted against perturbation strength $\lambda_1$ for a 24-qubit system. Error bars show the standard deviation of the mean from 10 realizations of $\varphi_2$.} 
    \label{fig:compare_noisy}
\end{figure}

Finally, we assess how noise distorts the experimental results, particularly its impact on the scaling behavior. We simulate the data in Fig. 4 of the main text for the 24-qubit system (Fig.~\ref{fig:compare_noisy}). Although the amplitudes are different between ideal and noisy simulations, both of them indicate the same crossover regime, confirming that the observed phenomena are still reliable in the presence of noise.

\addcontentsline{toc}{section}{Reference} 
\let\oldaddcontentsline\addcontentsline
\renewcommand{\addcontentsline}[3]{}
\bibliography{TC.bib}
\let\addcontentsline\oldaddcontentsline